\documentclass[11pt,a4paper]{article}
\usepackage{amsmath}
\usepackage{subcaption}
\usepackage{afterpage}
\usepackage{amsfonts}
\usepackage{amssymb}
\usepackage{mathtools}
\usepackage{physics}
\usepackage{simpler-wick}
\usepackage{empheq}
\usepackage{stackengine}
\usepackage{mathrsfs}
\usepackage{comment}
\usepackage{graphicx}
\usepackage[usenames,dvipsnames]{xcolor} 
\usepackage[colorlinks=true, linkcolor=RoyalBlue, citecolor=Maroon, urlcolor=Maroon,linktoc=page]{hyperref}
\allowdisplaybreaks[4] 
\usepackage{jheppub}
\usepackage{mathtools}
\usepackage{float}
\usepackage{tikz-cd}
\usepackage{amsmath}
\usepackage{multicol}
\usepackage{enumitem}

\date{}

\newcommand{\mc}{\mathcal}
\newcommand{\z}{\left}
\newcommand{\y}{\right}
\newcommand{\p}{\partial}
\newcommand{\cd}{\cdots}
\newcommand{\bp}{\bar{\partial}}
\newcommand{\SCLS}{{\text{S$\mathbb{C}$LS}}}
\newcommand{\hSCLS}{{\widehat{\text{S$\mathbb{C}$LS}}}}

\newcommand{\mV}{{\mathsf{V}}}
\newcommand{\mA}{{\mathsf{A}}}
\newcommand{\hmA}{{\widehat{\mathsf{A}}}}
\newcommand{\Li}{{\text{Li}}}
\newcommand{\mx}{{\mathsf{x}}}
\newcommand{\my}{{\mathsf{y}}}
\newcommand{\hmx}{{\widehat{\mathsf{x}}}}
\newcommand{\hmy}{{\widehat{\mathsf{y}}}}
\newcommand{\MO}{{\mathcal{O}}}
\newcommand{\ma}{{\mathsf{a}}}
\newcommand{\hma}{{\widehat{\mathsf{a}}}}

\newcommand{\MV}{{\mathscr{V}}}
\newcommand{\MR}{{\mathscr{R}}}
 \usepackage[normalem]{ulem}
\usepackage[textsize=scriptsize,textwidth=2.5cm]{todonotes}

\let\exporig\exp
\usepackage{physics}
\usepackage{enumitem}
\let\exp\exporig

\affiliation[\ensuremath{\gamma}]{Yau Mathematical Sciences Center, Tsinghua University, Beijing 100084, China}
\affiliation[\ensuremath{\tau}]{Department of Mathematical Sciences, Tsinghua University, Beijing 100084, China}
\affiliation[\ensuremath{\sigma}]{Center for the Fundamental Laws of Nature, Harvard University, Cambridge, MA, USA}
\affiliation[\ensuremath{\lambda}]{Kavli Institute for Theoretical Sciences (KITS), University of Chinese Academy of Sciences, Beijing 100190, China}

\usepackage{xspace}

\title{$\mathcal{N}=1$ super complex Liouville string}

\author{Zhengyuan Du$^{\gamma,\tau,\sigma}$, Kangning Liu$^{\gamma,\tau}$, Zhe-fei Yu$^{\lambda}$}

\emailAdd{duzy21@mails.tsinghua.edu.cn}
\emailAdd{lkn22@mails.tsinghua.edu.cn}
\emailAdd{yuzhefei@ucas.ac.cn}

\abstract{
We study the (type 0B) $\mathcal{N}=1$ supersymmetric complex Liouville string (S$\mathbb{C}$LS), a supersymmetric extension of the bosonic complex Liouville string ($\mathbb{C}$LS). We compute the sphere three-point amplitudes (including NS-NS-NS and NS-R-R types) and find they share the same form as the sphere three-point amplitude of the bosonic $\mathbb{C}$LS. Analysis of the analytic structure of the NS-NS-NS-NS four-point amplitude and the higher equations of motion also yields results identical to the bosonic case. Based on these  findings, we propose that the dual matrix model for the S$\mathbb{C}$LS is the same as that for the bosonic $\mathbb{C}$LS. We also investigate a related theory $\widehat{\text{S}\mathbb{C}\text{LS}}$, which differs in the gauged worldsheet supersymmetry. A parallel analysis is performed for $\widehat{\text{S}\mathbb{C}\text{LS}}$, and a candidate for its dual matrix model is proposed. We then carry out a partial numerical evaluation of the moduli space integral, which provides further evidence for both the proposals of the dual matrix model  regarding  S$\mathbb{C}$LS and $\widehat{\text{S}\mathbb{C}\text{LS}}$.
}

\begin{document}
\maketitle
\setlength{\parskip}{.3\baselineskip}
\section{Introduction}
Two-dimensional string theory, despite being a simplified model compared to its higher-dimensional counterparts, holds great significance in the realm of theoretical physics (cf. \cite{Klebanov:1991qa,Ginsparg:1993is,Jevicki:1993qn,Polchinski:1994mb,Martinec:2004td} for reviews). On one hand, it serves as a highly valuable testing ground for exploring fundamental aspects of string theory—due to its relative simplicity, 2D string theory is solvable in many respects; on the other hand, it provides a platform to understand the holographic principle in quantum theories of gravity. For example, the two-dimensional $c=1$ string theory (or type 0A/0B $\hat{c}=1$
superstring theories) admits a simpler description in terms of a double-scaled matrix quantum mechanics \cite{Klebanov:1991qa,Ginsparg:1993is,Jevicki:1993qn,Polchinski:1994mb,Martinec:2004td,Moore:1991zv, Takayanagi:2003sm,Douglas:2003up,Balthazar:2017mxh}. In this string/matrix model duality, non-perturbative effects (e.g., D-instanton-mediated phenomena \cite{Balthazar:2019rnh,Balthazar:2019ypi,Sen:2019qqg,Sen:2020cef,Sen:2020oqr,Sen:2020ruy,Sen:2020eck,Sen:2021qdk,DeWolfe:2003qf,Balthazar:2022apu,Chakravarty:2022cgj,Sen:2022clw,Eniceicu:2022xvk} and time-dependent stringy dynamics of rolling tachyons \cite{Klebanov:2003km,McGreevy:2003kb,McGreevy:2003ep}) persist, and crucially, are more computationally tractable than their counterparts in higher-dimensional string theories. Another well-known class of examples includes the $(2,p)$ and $(p,q)$ minimal strings (see e.g., \cite{Seiberg:2003nm,Seiberg:2004at}). For these theories, the corresponding worldsheet theories—excluding ghost fields—are constructed by combining a Liouville conformal field theory with $(2,p)$ and $(p,q)$ minimal models, respectively. It has long been recognized that both types of minimal strings admit dual descriptions in terms of (double-scaled) matrix models \cite{David:1984tx,Kazakov:1985ds,Kazakov:1985ea,Ambjorn:1985az,Douglas:1989ve,Gross:1989vs,Douglas:1989dd,Brezin:1990rb}. A key distinction, however, lies in the $(p,q)$ minimal string case: its dual theory requires not one but two matrix models, unlike the simpler single-matrix model dual typically associated with the $(2,p)$ variant.

Recently, irrational analogs of the $(2,p)$ and $(p,q)$ minimal string was introduced in \cite{Collier:2023cyw} and \cite{Collier:2024kmo} (see \cite{Collier:2024kwt,Collier:2024lys,Collier:2024mlg,Collier:2025pbm} for extended discussions) respectively. The former is named the Virasoro minimal string (VMS), defined by coupling Liouville CFT with central charge $c\geq 25$ to ``timelike Liouville CFT'' with central charge $\widehat{c}=26-c$ on the worldsheet:
\begin{equation}
 \begin{array}{c}
 \text{Liouville CFT}\\
\text{$c\geq 25$} \
\end{array}\oplus 
 \begin{array}{c}
 \text{Timelike Liouville CFT}\\
\text{$\widehat{c}=26-c\leq 1$} \
\end{array}
\oplus \begin{array}{c}
 \text{$\mathfrak{bc}$-ghosts}\\
\text{$c_{\mathfrak{bc}}=-26$} \
\end{array}
\end{equation}
Similar to the $(2,p)$ minimal string, this string theory is shown to have a dual description as a double-scaled
matrix integral;
The latter is named the complex Liouville string ($\text{$\mathbb{C}$LS}$), defined at the level of the worldsheet CFT by coupling two copies of
Liouville CFT—each with a complex central charge, where the two central charges are complex conjugates of one another:
\begin{equation}
 \begin{array}{c}
 \text{Liouville CFT}\\
\text{$c^+=13+i\lambda$} \
\end{array}\oplus 
 \begin{array}{c}
 \text{(Liouville CFT)*}\\
\text{$c^-=13-i\lambda$} \
\end{array}
\oplus \begin{array}{c}
 \text{$\mathfrak{bc}$-ghosts}\\
\text{$c_{\mathfrak{bc}}=-26$}, \\
\end{array}\qquad \text{with $\lambda\in\mathbb{R}$}.
\end{equation}
This string theory, analogous to the $(p,q)$ minimal string, admits a reformulation as a double-scaled two-matrix integral. These irrational variants of minimal strings turn to be particularly intriguing: beyond offering examples of holographic dualities that admit rigorous computational treatment, these models also prove to be connected to lower-dimensional gravity in (A)dS spacetimes. It is shown in \cite{Collier:2023cyw} that the VMS is a stringy realization of AdS$_2$ Jackiw-Teitelboim (JT) gravity, which is also known to dual to a double-scaled matrix integral\footnote{That is, JT gravity emerges as a semiclassical limit of the VMS. Note that JT gravity can also be embedded in $(2,p)$ minimal string with $p\to\infty$ \cite{Seiberg:2019unp}.}. Even more notably, the $\text{$\mathbb{C}$LS}$ admits a reformulation in terms of two-dimensional dilaton gravity, which admits both AdS$_2$
and dS$_2$ vacua as classical solutions \cite{Collier:2024kmo,Collier:2025pbm}. Therefore, it provides
an exact version of holography involving two-dimensional de Sitter (dS)
spacetimes. Moreover, the string amplitudes of $\text{$\mathbb{C}$LS}$ can 
be reinterpreted as cosmological correlators of 3d pure
de Sitter gravity with non-dynamical massive particles, thereby furnishing a precise de Sitter/matrix model correspondence \cite{Collier:2024kmo,Collier:2025lux}. It is also found that $\text{$\mathbb{C}$LS}$ is related to  the  double-scaled SYK model \cite{Blommaert:2025eps}.
In the present paper, we generalize the complex Liouville string to incorporate worldsheet $\mathcal{N}=1$ supersymmetry and investigate their potential dual description in terms of matrix integrals\footnote{Very recently, foundational work toward generalizing the Virasoro minimal string to its $\mathcal{N}=1$ supersymmetric counterpart has been laid out in \cite{Johnson:2024fkm,Muhlmann:2025ngz,Rangamani:2025wfa,Johnson:2025vyz}.}.

String amplitudes, defined by integrating worldsheet correlation functions over the moduli space of Riemann surfaces, are generally challenging to compute. For two-dimensional string theories, these integrals could only be evaluated numerically \cite{Balthazar:2017mxh,Balthazar:2018qdv,Balthazar:2022atu,Rodriguez:2023kkl,Rodriguez:2023wun,Collier:2024kmo}. Recently, in \cite{Collier:2024kmo,Collier:2024kwt}, the authors have initiated a novel bootstrap program for string amplitudes in the complex Liouville string, which leverages the analytic structure and other constraints from the worldsheet to determine the string amplitudes. This program successfully yields several (equivalent) analytic closed-form expressions for the sphere four-point amplitude\footnote{Then the torus one-point amplitude is also known, since in $\text{$\mathbb{C}$LS}$, it can be obtained directly from the sphere four-point amplitude \cite{Collier:2024kwt}.}, which are consistent with the results from numerical evaluation of the moduli space integral. Furthermore, the proposed dual matrix model not only successfully reproduces this sphere four-point amplitude but also enables predictions of all higher-point and higher-genus amplitudes via the mechanism known as topological recursion \cite{Collier:2024kmo,Collier:2024lys}. Subsequently, this string/matrix model correspondence was also tested and verified at the non-perturbative level \cite{Collier:2024mlg}.

One of the main aims of this paper is to generalize this bootstrap program to the $\mathcal{N}=1$ supersymmetric case and thereby determine the amplitudes in the (type 0B) $\mathcal{N}=1$ super complex Liouville string ($\text{S$\mathbb{C}$LS}$). The primary
ingredient for the construction of the worldsheet theory is $\mathcal{N}=1$  super-Liouville theory. The structure constants of this CFT were bootstrapped in \cite{Rashkov:1996np,Poghossian:1996agj}, similar to the derivation of the DOZZ formula in the bosonic Liouville theory. Building on these structure constants, we will compute sphere three-point amplitudes involving  both  NS and R vertices (it should be noted that NS vertices exist in both the type 0A and type 0B theories, whereas  R vertices exist only in the type 0B theory). Interestingly, both of the NS-NS-NS and NS-R-R three-point amplitudes take exactly the same form as those in the bosonic $\text{$\mathbb{C}$LS}$. This is quite non-trivial since there is no known explicit relation between the DOZZ structure constant and its $\mathcal{N}=1$ super counterpart. In fact, to achieve this agreement, picture changing plays an essential role. We then focus on bootstrapping the sphere four-point amplitudes of four NS vertex operators. The bootstrap conditions that satisfied by the amplitude are:
\begin{itemize}
 \item The analytic structure of the amplitudes.
 \item The higher equations of motion.
\end{itemize}
Somewhat surprisingly, these bootstrap conditions take exactly the same form as those in the bosonic case. Thus, we propose that the NS-NS-NS-NS sphere four-point amplitude again coincides with the four-point amplitude in the bosonic $\text{$\mathbb{C}$LS}$. Technically, this fact is remarkable, as the moduli space integral requires summing over 8 terms, involving 8 distinct conformal block expansions\footnote{This is because of the picture changing. Besides, in $\mathcal{N}=1$ superconformal field theories, there are in total 8 distinct types of conformal block expansions (in the NS sector). It happens that all of them emerge in our computation of the moduli space integral. }—unlike the bosonic case, where only one term needs to be considered.  Considering these findings,   we further propose that the type 0B $\text{S$\mathbb{C}$LS}$ admits a dual description in terms of the same two-matrix model dual to the bosonic $\text{$\mathbb{C}$LS}$. 

Apart from the $\text{S$\mathbb{C}$LS}$, we also investigate another closely related string theory. This counterpart, denoted as $\hSCLS$, is constructed from the same pair of super-Liouville theories on the worldsheet, with the key distinction lying in the different definition of the total supercurrent\footnote{Concretely, $\hSCLS$ is related to $\SCLS$ by reversing the sign of the supercurrent 
 $G$ in one of the two super-Liouville theories (while $\bar{G}$ remains unchanged). Note that a similar construction also appeared in the recent works on super-Virasoro minimal string \cite{Muhlmann:2025ngz,Rangamani:2025wfa}.}. We perform a parallel analysis of this theory relative to the $\SCLS$, involving the computation of the three-point amplitude, analysis of the four-point amplitude’s analytic structure, derivation of higher equations of motion and numerical evaluation of the four-point moduli integral. From them, we proposed its two-matrix model dual, which (at least) could reproduce the  3-point amplitudes with NS vertices. The associated spectral curve differs from that of the two-matrix model dual to the $\SCLS$.
 
We denote by $\mA^{(b)}_{g,n_{NS},n_{R}}$ and $\hmA^{(b)}_{g,n_{NS},n_{R}}$ the genus-$g$   amplitudes with    $n_{NS}$ NS legs and $n_{R}$ R legs in $\text{S$\mathbb{C}$LS}$ and $\widehat{\text{S$\mathbb{C}$LS}}$,  respectively. Collectively, we find:
\allowdisplaybreaks[4] 
\begin{subequations}
    \begin{align}
        &\mA^{(b)}_{0,3,0}(P_1,P_2,P_3)=2b\sum_{k\in \mathbb{Z}_{\geq 1}} \frac{\prod_{i=1}^{3}\sin (2kb\pi P_i)}{\sin(k\pi b(b+b^{-1}))}\label{3,0pt}\\ &\hmA^{(b)}_{0,3,0}(P_1,P_2,P_3)=2b\sum_{k\in \mathbb{Z}_{\geq 1}-\frac{1}{2}}\frac{\prod^3_{i=1}\sin\z(2kb\pi P_i\y)}{\sin\z(k\pi b(b+b^{-1})\y)}\label{h3,0pt}\\
        &\mA^{(b)}_{0,1,2}(P_1,P_2;P_3)=2b\sum_{k\in \mathbb{Z}_{\geq 1}}^\infty \frac{\prod_{i=1}^{3}\sin (2kb\pi P_i)}{\sin(k\pi b(b+b^{-1}))}\label{1,2pt}\\
 &\hmA^{(b)}_{0,1,2}(P_1,P_2;P_3)=2b\sum_{k\in \mathbb{Z}_{\geq 1}-\frac{1}{2}}^{\infty}\frac{\sin\z(2kb\pi P_3\y)\prod^2_{i=1}\cos\z(2kb\pi P_i\y)}{\sin\z(k\pi b(b+b^{-1})\y)}\label{h1,2pt}\\
        &\mA^{(b)}_{0,4,0}(P_1,P_2,P_3,P_4)=\sum_{m\in \mathbb{Z}_{\geq 1}}\frac{2b^2 \mathsf{V}_{0,4}^{(b)}(iP_1,iP_2,iP_3,iP_4)\prod_{j=1}^4\sin(2\pi m b P_j)}{\sin(\pi m b(b+b^{-1}))^2}\nonumber \\
  &\hspace{3cm}-\sum_{m_1,m_2\in \mathbb{Z}_{\geq 1}}\frac{\sin(2\pi m_1bP_1)\sin(2\pi m_1bP_2)\sin(2\pi m_2bP_3)\sin(2\pi m_2bP_4)}{\pi^2 \sin(\pi m_1b(b+b^{-1}))\sin(\pi m_2b(b+b^{-1}))}\nonumber \\
  &\hspace{3cm}\times \z(\frac{1}{(m_1+m_2)^2}-\frac{\delta_{m_1\neq m_2}}{(m_1-m_2)^2}\y)+2\text{ perms}\label{4,0pt}\\
  &\hmA^{(b)}_{0,4,0}(P_1,P_2,P_3,P_4)=\sum_{m\in \mathbb{Z}_{\geq 1}-\frac{1}{2}}\frac{2b^2 \mathsf{V}_{0,4}^{(b)}(iP_1,iP_2,iP_3,iP_4)\prod_{j=1}^4\sin(2\pi m b P_j)}{\sin(\pi m b(b+b^{-1}))^2}\nonumber \\
  &\hspace{3cm}-\sum_{m_1,m_2\in\mathbb{Z}_{\geq 1}-\frac{1}{2}} \frac{\sin(2\pi m_1bP_1)\sin(2\pi m_1bP_2)\sin(2\pi m_2bP_3)\sin(2\pi m_2bP_4)}{\pi^2 \sin(\pi m_1b(b+b^{-1}))\sin(\pi m_2b(b+b^{-1}))}\nonumber \\
  &\hspace{3cm}\times \z(\frac{1}{(m_1+m_2)^2}-\frac{\delta_{m_1\neq m_2}}{(m_1-m_2)^2}\y)+2\text{ perms}\label{h4,0pt}
    \end{align}
\end{subequations}
For the above expressions of sphere 4-point amplitudes $\mA^{(b)}_{0,4,0}$ and $\hmA^{(b)}_{0,4,0}$, we further perform a direct numerical evaluation of the moduli space integral, finding consistent agreement\footnote{We have evaluated only half of the full moduli space integral, and the result matches half of our proposed analytic expression (see section \ref{Numerical}). The complete numerical computation will be addressed in future work.}.

\subsection*{Outline}
The paper is organized as follows:

In Section \ref{worldsheet}, we first review the fundamentals of $\mathcal{N}=1$ super-Liouville theory. On this basis, we define the worldsheet theory for the (type 0A/0B) super complex Liouville string $\SCLS$  and construct physical vertex operators in both the NS and R sectors.  By gauging distinct worldsheet supercurrents, we introduce another closely related theory and denote it as $\hSCLS$. In
Section \ref{3-point}, we  compute the sphere 3-point  amplitudes in the two aforementioned theories. We find the amplitudes $\mA^{(b)}_{0,3,0}(\boldsymbol{P})$ and $\mA^{(b)}_{0,1,2}(\boldsymbol{P})$ of $\SCLS$ coincide with the 3-point  amplitude in bosonic $\mathbb{C}$LS  exactly, whereas the corresponding amplitudes $\hmA^{(b)}_{0,3,0}(\boldsymbol{P})$ and $\hmA^{(b)}_{0,1,2}(\boldsymbol{P})$ of $\hSCLS$  differ.  
In Section \ref{Analyticstructure}, we analyze the analytic structures of sphere 4-point  amplitudes $\mA^{(b)}_{0,4,0}$ and $\hmA^{(b)}_{0,4,0}$ in $\SCLS$ and $\hSCLS$ respectively. Leveraging established results from super-Liouville theory, we explicitly derive key analytic features of these amplitudes, including their discontinuities and momentum poles.
In Section \ref{GRHEOM}, we derive the higher equations of motion for the sphere 4-point amplitudes. These equations establish a direct connection between four-point amplitudes with degenerate external momenta and the three-point amplitudes discussed in the previous section. 
In Section \ref{2Matrix},  after a brief review of the two-matrix model, we propose dual matrix integrals for type 0B $\SCLS$ (identical to that of the bosonic $\mathbb{C}$LS) and $\hSCLS$ (a distinct one). They reproduce  the sphere 3-point  amplitudes on the string side and give the analytic closed expressing for the  sphere 4-point  amplitudes. 
In Section \ref{Check}, we  verifies that these 4-point amplitudes satisfy all worldsheet-derived constraints. We also  present a partial numerical evaluation of the moduli space integral and find further consistency. There are several appendices that collect some further details and computations.

\section{The worldsheet theory}\label{worldsheet}
In this section, we present the definition of the $\mathcal{N}=1$ super complex Liouville string ($\text{S$\mathbb{C}$LS}$), which shares many similarities with its bosonic counterpart in \cite{Collier:2024kwt}. In the RNS formalism, the worldsheet theory of the $\text{S$\mathbb{C}$LS}$ is constructed by coupling two copies of $\mathcal{N}=1$ super-Liouville theories with complex central charges:
\begin{equation}\label{worldsheettheory}
 \begin{array}{c}
 \text{$\mathcal{N}=1$ super-Liouville}\\
\text{$c^+=\frac{15}{2}+i\lambda$} \
\end{array}\oplus 
 \begin{array}{c}
 (\text{$\mathcal{N}=1$ super-Liouville})^*\\
\text{$c^-=\frac{15}{2}-i\lambda$} \
\end{array}
\oplus  \begin{array}{c}
\text{$\mathfrak{bc}$-ghosts}\\
c_{\mathfrak{bc}}=-26 \
\end{array} \oplus \begin{array}{c}
\text{$\beta \gamma$-ghosts}\\
c_{\beta \gamma}=11 \
\end{array}  
\end{equation}
where $\lambda \in \mathbb{R}_+$. 

\subsection{Super-Liouville field theory}

Before proceeding to a concrete discussion of the $\text{S$\mathbb{C}$LS}$, we first review the fundamentals of $\mathcal{N}=1$ super-Liouville field theory (SLFT), see e.g., \cite{Nakayama:2004vk}.

The action of $\mathcal{N}=1$ super-Liouville field theory is:
\begin{equation}\label{LSuperL}
	S=\frac{1}{4\pi}\int d^2z\z(\p\phi\bp\phi+\frac{1}{4}QR\phi+\bar{\psi}\p\bar{\psi}+\psi\bp\psi\y)+\int d^2z\z[4\pi\mu^2b^2:e^{b\phi}::e^{b\phi}:+2i\mu b^2\psi\bar{\psi}e^{b\phi}\y]
\end{equation}
where $\phi$ and $\psi$ $(\bar{\psi})$ are the Liouville field and its superpartner respectively. The parameter $\mu$ is the cosmological constant and $Q=b^{-1}+b$ is the Liouville charge. 
The stress energy tensor and the superconformal current are:
\begin{equation}
\begin{aligned}
  T&=-\frac{1}{2}(\partial\phi\partial\phi-Q\partial^2\phi+\psi\partial\psi)\\
  G&=i(\psi\partial\phi-Q\partial\psi).
\end{aligned}
\end{equation}
These currents generate the $\mathcal{N}=1$ superconformal algebra with central charge
\begin{equation}
 c=\frac{3}{2}+3Q^2.
\end{equation}
The spectrum of SLFT comprises the NS and R sectors, which we describe below.

\subsection*{The NS sector}
In the NS sector, we can construct the super-primary operator $V_P$ (labeled by $P$):
\begin{equation}
 V_P\propto e^{\alpha \phi},\quad \alpha=\frac{Q}{2}+P.
\end{equation}
This super primary is the bottom component of a superfield, which also includes three other Virasoro primaries:
\begin{equation}
 \Lambda_P =[G_{-\frac{1}{2}},V_P],\quad \bar{\Lambda}_P =[\bar{G}_{-\frac{1}{2}},V_P],\quad W_P=\{G_{-\frac{1}{2}},[\bar{G}_{-\frac{1}{2}},V_P]\}.
\end{equation}
Note that these 3 operators are not super-primaries.
The weights $(h_P,\bar{h}_P)$ of the super-primary $V_P$ are 
\begin{equation}\label{Theweights}
 h_P=\bar{h}_P=\frac{\alpha}{2}\z(Q-\alpha\y)=\frac{1}{8}\z(Q^2-4P^2\y).
\end{equation}
Accordingly, the weights of the other 3 operators $\Lambda_P$, $\bar{\Lambda}_P$, $W_P$ are:
\begin{equation}
 \z(h_P+\frac{1}{2},\bar{h}_P\y),\quad \z(h_P,\bar{h}_P+\frac{1}{2}\y),\quad \z(h_P+\frac{1}{2},\bar{h}_P+\frac{1}{2}\y), 
\end{equation}
respectively.

In the NS sector, there are two independent structure constants, which can be derived in an analogous manner to the DOZZ formula in bosonic Liouville theory \cite{Rashkov:1996np,Poghossian:1996agj,Fukuda:2002bv} (see appendix \ref{Bmn} for their standard form).
Following the convention adopted for the bosonic $\text{$\mathbb{C}$LS}$ in \cite{Collier:2024kwt},  we here set $V_P=V_{-P}$. Consequently, the 2 structure constants (3-point functions) are reflection-symmetric and take the form (see \cite{Muhlmann:2025ngz,Rangamani:2025wfa}):
\begin{equation}
 \begin{aligned}
  &\langle V_{P_1}(z_1)V_{P_2}(z_2)V_{P_3}(z_3)\rangle =\frac{C_b(P_{1},P_2,P_3)}{|z_{12}|^{2(h_{P_1}+h_{P_2}-h_{P_3})}|z_{13}|^{2(h_{P_1}+h_{P_3}-h_{P_2})}|z_{23}|^{2(h_{P_2}+h_{P_3}-h_{P_1})}}\\
  &\langle W_{P_1}(z_1)V_{P_2}(z_2)V_{P_3}(z_3)\rangle =\frac{\widetilde{C}_b(P_1,P_2,P_3)}{|z_{12}|^{2(h_1+h_2-h_3+\frac{1}{2})}|z_{13}|^{2(h_1+h_3-h_2+\frac{1}{2})}|z_{23}|^{2(h_2+h_3-h_1-\frac{1}{2})}}
 \end{aligned}
\end{equation}
where the structure constants are
\begin{equation}\label{SC}
 \begin{aligned}
  &C_b(P_1,P_2,P_3)=\frac{\Gamma_b^{NS}(2Q)}{2\Gamma_b^{NS}(Q)^3}\frac{\Gamma^{NS}_b(\frac{Q}{2}\pm P_1\pm P_2\pm P_3)}{\prod_{i=1}^3\Gamma_b^{NS}(Q\pm 2P_i)}\\
  &\widetilde{C}_b(P_1,P_2,P_3)=i\frac{\Gamma_b^{NS}(2Q)}{\Gamma_b^{NS}(Q)^3}\frac{\Gamma^{R}_b(\frac{Q}{2}\pm P_1\pm P_2\pm P_3)}{\prod_{i=1}^3\Gamma_b^{NS}(Q\pm 2P_i)}.
 \end{aligned}
\end{equation}
In the above, the product is over all the permutations of the signs in the numerator. The function $\Gamma^{NS}_b$ and $\Gamma^{R}_b$ are defined as (see appendix \ref{doublegamma} for definition and properties of the relevant special functions)
\begin{equation}
 	\begin{aligned}
		&\Gamma_b^{NS}(x)\equiv\Gamma_b\z(\frac{x}{2}\y)\Gamma_b\z(\frac{x+b+b^{-1}}{2}\y)\\
		&\Gamma^{R}_b(x)\equiv\Gamma_b\z(\frac{x+b}{2}\y)\Gamma_b\z(\frac{x+b^{-1}}{2}\y)
	\end{aligned}
\end{equation}
Note that the structure constants \eqref{SC} are proportional to those obtained in \cite{Rashkov:1996np,Poghossian:1996agj,Fukuda:2002bv}; the precise relation is given in appendix \ref{Bmn} (see also \cite{Rangamani:2025wfa}).

The two point function of $V_P$, or the normalization, could be read from the above three point function. 
Notice that:
\begin{equation}
  \lim_{P_3\to \frac{Q}{2}}C_b(P_1,P_2,P_3)=\frac{1}{\rho_{NS}^{(b)}(P_1)}(\delta(P_1-P_2)+\delta(P_1+P_2))
\end{equation}
Then the two-point function of two $V_P$'s is:
\begin{equation}
     \langle V_{P_1}(z) V_{P_2}(0)\rangle=\frac{1}{\rho^{(b)}_{NS}(P_1)}\frac{\z(\delta(P_1+P_2)+\delta(P_1-P_2)\y)}{|z|^{2(h_{P_1}+h_{P_2})}}
\end{equation}
where $\rho_{NS}^{(b)}$ is the spectral density in the NS sector:
\begin{equation}\label{measurerho}
 \rho_{NS}^{(b)}(P)=-4\sin\z(\pi bP\y)\sin\z(\pi b^{-1}P\y).
\end{equation}

\subsection*{The R sector}
In the R sector, we can construct two super-primary operators $R_P^{(\pm)}$:
\begin{equation}
	R_P^{(+)}\sim \mu e^{\alpha \phi},\quad R_P^{(-)}\sim\sigma e^{\alpha \phi},\quad \alpha =\frac{Q}{2}+P
\end{equation}
where $\mu$ and $\sigma$ are the disorder and spin field.
 They have the same conformal weights
\begin{equation}
 \z(h_P+\frac{1}{16},\bar{h}_P+\frac{1}{16}\y).
\end{equation}
We adopt the convention in which the two-point functions are \cite{Muhlmann:2025ngz,Rangamani:2025wfa}
\begin{equation}\label{norR}
 \langle R^{(\pm)}_{P_1}(z) R^{(\pm)}_{P_2}(0)\rangle=\frac{1}{\rho^{(b)}_{R}(P_1)}\frac{\z(\delta(P_1-P_2)\pm\delta(P_1+P_2)\y)}{|z|^{2(h_{P_1}+h_{P_1})+\frac{1}{4}}},
\end{equation}
where $\rho_R^{(b)}$ is the R sector spectral density
\begin{equation}
 \rho_{R}^{(b)}(P)=2\sqrt{2}\cos\z(\pi bP\y)\cos\z(\pi b^{-1}P\y).
\end{equation}
 We have two additional independent structure constants that involving  R fields:
\begin{equation}
 \begin{aligned}
 &\langle R^{(\pm)}_{P_1}(z_1)R^{(\pm)}_{P_2}(z_2) V_{P_3}(z_3)\rangle=\frac{C^{e}_b(P_1,P_2;P_3)\pm C^{o}_b(P_1,P_2;P_3)}{|z_{12}|^{2(h_{P_1}+h_{P_2}-h_{P_3}+\frac{1}{8})}|z_{13}|^{2(h_{P_1}+h_{P_3}-h_{P_2})}|z_{23}|^{2(h_{P_2}+h_{P_3}-h_{P_1})}},
 \end{aligned}
\end{equation}
where
\begin{equation}\label{RSC}
 \begin{aligned}
  &C^{e}_b(P_1,P_2;P_3)=\frac{\Gamma_b^{NS}(2Q)}{\sqrt{2}\Gamma_b^{NS}(Q)^3}\frac{\Gamma^{R}_b(\frac{Q}{2}\pm (P_1+ P_2)\pm P_3)\Gamma^{NS}_b(\frac{Q}{2}\pm (P_1- P_2)\pm P_3)}{\Gamma_b^{R}(Q\pm 2P_1)\Gamma_b^{R}(Q\pm 2P_2)\Gamma_b^{NS}(Q\pm 2P_3)}\\
  &C^{o}_b(P_1,P_2;P_3)=\frac{\Gamma_b^{NS}(2Q)}{\sqrt{2}\Gamma_b^{NS}(Q)^3}\frac{\Gamma^{NS}_b(\frac{Q}{2}\pm (P_1+ P_2)\pm P_3)\Gamma^{R}_b(\frac{Q}{2}\pm (P_1- P_2)\pm P_3)}{\Gamma_b^{R}(Q\pm 2P_1)\Gamma_b^{R}(Q\pm 2P_2)\Gamma_b^{NS}(Q\pm 2P_3)}.
 \end{aligned}
\end{equation}
 Note that our structure constant conventions follow \cite{Muhlmann:2025ngz,Rangamani:2025wfa}; their correspondence with those in \cite{Rashkov:1996np,Poghossian:1996agj,Fukuda:2002bv} is given in \cite{Rangamani:2025wfa}.
These structure constants are consistent with the
normalizations in \eqref{norR}, since: 
\begin{equation}
 \begin{aligned}
  &\lim_{P_3\to \frac{Q}{2}}C^{e}_b(P_1,P_2;P_3)=\frac{1}{\rho_{R}^{(b)}(P_1)}\delta(P_1-P_2)\\
  &\lim_{P_3\to \frac{Q}{2}}C^{o}_b(P_1,P_2;P_3)=\frac{1}{\rho_{R}^{(b)}(P_1)}\delta(P_1+P_2)
 \end{aligned}
\end{equation}

\subsection{The spectrum and vertex operators}
We now discuss the spectrum of the $\text{S$\mathbb{C}$LS}$ and construct the associated vertex operators. To do so, we first elaborate on the definition of the worldsheet theory.

\subsection*{Analytic continuation}
The definition of the worldsheet theory  \eqref{worldsheettheory} involves the analytic continuation of super-Liouville CFT.
To obtain the required central charge, we consider 
\begin{equation}
 b=b^+\in e^{\frac{i\pi}{4}}\mathbb{R}_+
\end{equation}
which then gives
\begin{equation}
 c^+=\frac{15}{2}+3b_+^2+3b^{-2}_+\in \frac{15}{2}+i\mathbb{R}_+.
\end{equation}
Axiomatically, super-Liouville theory with complex $c$ can be defined by analytically continuing the OPE data away from real values of $b$ while preserving crossing symmetry of correlation functions\footnote{See \cite{Harlow:2011ny} for a detailed discussion of the analytic continuation of the bosonic Liouville theory. It is expected that a similar treatment applies to super-Liouville theory. Besides, in \cite{Ribault:2015sxa}, it is numerically verified that crossing symmetry holds in  analytic continuated Liouville theory with a complex central charge. We have checked numerically that the same holds true for analytic continuated super-Liouville theory.}.

\subsection*{Reality conditions}
The worldsheet theory \eqref{worldsheettheory} combines the complex Liouville theory with its complex conjugate, thereby rendering the resulting theory real-valued.
We denote the super-Viraosoro generators of the two super-Liouville theories by $\{L_m^+,G_n^+\}$ and $\{L_m^-,G_n^-\}$. Then we can define the total super-Virasoro generators on the worldsheet as: 
\begin{equation}
 L_m\equiv L^+_m+L^-_m, \qquad G_n\equiv G^+_n+G^-_n
\end{equation}
The reality condition implies the following relation
\begin{equation}
 (L^+_m)^\dagger=L_{-m}^-, \qquad (G^+_m)^\dagger=G_{-m}^-.
\end{equation}
Recall that the central charges are 
\begin{equation}
 c^\pm=\frac{15}{2}+3(b^\pm)^{-2}+3(b^\pm)^2,\quad b^+\in e^{\frac{i\pi}{4}}\mathbb{R}_+,\quad b^-\in e^{-\frac{i\pi}{4}}\mathbb{R}_+.
\end{equation}
The reality condition on the worldsheet implies that $(c^+)^*=c^-$, so we have 
\begin{eqnarray}
 b^-=(b^+)^*\implies b^-=-ib^+
\end{eqnarray}
Therefore, we can choose $b^+\in e^{\frac{i\pi}{4}}\mathbb{R}$ and $b^-\in e^{-\frac{i\pi}{4}}\mathbb{R}$.

\subsection*{Vertex Operator}
In the NS sector, physical operators can be constructed by combining the super-primary fields $V^+_{P^+}$ and $V^-_{P^-}$ (from the two super-Liouville theories with $c=c^+$ and $c=c^-$, respectively). Let their conformal weights be denoted by $h^+$ and $h^-$. Given  $(L_0^+)^\dagger=L_0^-$, it follows that $(h^{+})^*=h^-$, and they can be expressed as  (from \eqref{Theweights})
\begin{equation}
 h^\pm=\frac{(Q^\pm)^2}{8}-\frac{(P^\pm)^2}{2}=\frac{c^\pm-\frac{3}{2}}{24}-\frac{(P^\pm)^2}{2},\quad (P^+)^*=\pm P^-.
\end{equation}
Then the on-shell condition 
\begin{equation}
 h^++h^-=\frac{1}{2}
\end{equation}
implies
\begin{equation}
 (P^\pm)^2\in i\mathbb{R},\quad P^-=\pm iP^+.
\end{equation}
Thus, we can choose the convention
\begin{equation}
 P^+\in e^{-\frac{\pi i}{4}}\mathbb{R},\quad P^-=iP^+\in e^{\frac{\pi i}{4}}\mathbb{R},
\end{equation}
which ensures
\begin{equation}
 b^+ P^+=b^- P^-\in\mathbb{R}.
\end{equation}
 Finally, we can write the physical vertex operator in the NS sector by combining the two vertex operators $V^+_{P^+}$ and $V^-_{P^-}$ (and with the ghosts): 
\begin{equation}\label{NSp-1}
 \MV_{P}^{(-1)}=\mathcal{N}_b(P)\mathfrak{c}\bar{\mathfrak{c}}e^{-\Phi-\bar{\Phi}}V^+_PV^-_{iP}.
\end{equation}
Here and in what follows, we consistently use the notation $P\equiv P^+$. The so-called leg-factor $\mathcal{N}_b(P)$ corresponds to a change of normalization of the vertex operator, chosen as in \eqref{normalizations}. Note that the above vertex operator is in picture $(-1,-1)$, so we use a superscript ``$(-1)$''.
To compute the amplitudes, we also need the vertex operator in picture $(0,0)$: 
\begin{equation}\label{NSp0}
 \begin{aligned}
 \MV_{P}^{(0)}\equiv \mathcal{N}_b(P) \mathfrak{c}\bar{\mathfrak{c}} G_{-\frac{1}{2}}\bar{G}_{-\frac{1}{2}}V^+_PV^-_{iP}&= \mathcal{N}_b(P)\mathfrak{c}\bar{\mathfrak{c}}\z(G_{-\frac{1}{2}}^++G_{-\frac{1}{2}}^-\y)\z(\bar{G}_{-\frac{1}{2}}^++\bar{G}_{-\frac{1}{2}}^-\y)V^+_PV^-_{iP}\\
 &=\mathcal{N}_b(P)\mathfrak{c}\bar{\mathfrak{c}}(W^+_PV^-_{iP}+V^+_PW^-_{iP}-\bar{\Lambda}^+_P\Lambda^-_{iP}+\Lambda^+_P\bar{\Lambda}^-_{iP})
 \end{aligned}
\end{equation}
where we have chosen the worldsheet supercurrents that enter the BRST algebra to be
\begin{equation}\label{supercurrentworldsheet}
    G \equiv G^+ + G^-, \quad \bar G\equiv \bar G^+ + \bar G^-
\end{equation}
In the R sector, physical vertex operators  take the following form  (in picture $(-\frac{1}{2},-\frac{1}{2})$):
\begin{equation}
 \MR_{P}^{(-\frac{1}{2})}=\mathcal{R}_b (P)\mathfrak{c}\bar{\mathfrak{c}}e^{-\frac{\Phi}{2}-\frac{\bar{\Phi}}{2}}\z(R_{P}^{+(-)} R_{iP}^{-(+)}+iR_{P}^{+(+)} R_{iP}^{-(-)}\y),
\end{equation}
where $\mathcal{R}_b$ is the leg-factor of the R sector operator, chosen as in \eqref{normalizations}.
After the GSO projection, we obtain type 0A and type 0B superstrings. In type 0A theory,  all R-sector vertex operators are projected out, leaving only those in the NS sector. In type 0B theory, by contrast, both NS-sector and R-sector vertex operators are present in the spectrum.

\subsection{An alternative theory}
A notable novel feature is that an alternative string theory can be constructed from the same two super-Liouville theories on the worldsheet.
When combining the two super-Liouville theories to construct the worldsheet CFT, we can, in fact, define another total supercurrent. This is due to an automorphism of the super-Virasoro algebra, which reverses the sign of the supercurrent ($G\to -G$) while preserving other generators. Thus, the alternative worldsheet supercurrent (denoted as $\widehat{G},\widehat{\bar{G}}$) can be defined as\footnote{In the 10d superstring theory in flat spacetime, we have spacetime Lorentzian symmetries that can mix worldsheet fermions in different directions. So there is only one choice similar to \eqref{supercurrentworldsheet} that preserves spacetime Lorentzian symmetry. However, here we do not have spacetime Lorentzian symmetry, and the alternative choice of BRST quantization is consistent. }:
\begin{equation}
 \widehat{G}\equiv -G^++G^-, \qquad \widehat{\bar{G}}\equiv\bar{G}=\bar{G}^++\bar{G}^-
\end{equation}
Gauging this worldsheet superconformal symmetry yields a distinct string theory.
 
 In this alternative theory, the R vertex operator in picture $(-\frac{1}{2},-\frac{1}{2})$ has a different form:
\begin{equation}
 \widehat{\mathscr{R}}_{P}^{(-\frac{1}{2})}=\mathcal{R}_b(P)\mathfrak{c}\bar{\mathfrak{c}}e^{-\frac{\Phi}{2}-\frac{\bar{\Phi}}{2}}(R_P^{+(+)}R_{iP}^{-(+)}+R_P^{+(-)}R_{iP}^{-(-)})
\end{equation}
For the NS vertex operator, while it takes the same form as \eqref{NSp-1} in picture $(-1,-1)$, its form differs from \eqref{NSp0} in picture $(0,0)$:
\begin{equation}
 \begin{aligned}
 \widehat{\MV}_{P}^{(0)}= \mathcal{N}_b(P) \mathfrak{c}\bar{\mathfrak{c}}\widehat{G}_{-\frac{1}{2}}\widehat{\bar{G}}_{-\frac{1}{2}}V^+_PV^-_{iP}&= \mathcal{N}_b(P) \mathfrak{c}\bar{\mathfrak{c}}\z(-G_{-\frac{1}{2}}^++G_{-\frac{1}{2}}^-\y)\z(\bar{G}_{-\frac{1}{2}}^++\bar{G}_{-\frac{1}{2}}^-\y)V^+_PV^-_{iP}\\
 &= \mathcal{N}_b(P) \mathfrak{c}\bar{\mathfrak{c}}(-W^+_PV^-_{iP}+V^+_PW^-_{iP}-\bar{\Lambda}^+_P\Lambda^-_{iP}-\Lambda^+_P\bar{\Lambda}^-_{iP}).
 \end{aligned}
\end{equation}
Hereafter, we  refer to this alternative super complex Liouville string  as $\hSCLS$.

\subsection{String amplitudes}
We denote the genus-$g$ string amplitude in the $\SCLS$ with $n_{NS}$ NS states and $n_R$ R states  by $\mA^{(b)}_{g,n_{NS},n_R}$
 (the corresponding amplitudes in the  $\hSCLS$ are denoted as $\hmA^{(b)}_{g,n_{NS},n_{R}}$). They can be calculated by integrating the worldsheet correlators over the moduli space ($\hmA^{(b)}_{g,n_{NS},n_R}$ can be calculated similarly): 
\begin{equation}
 \mA_{g,n_{NS},n_{R}}^{(b)}(\boldsymbol{P})\equiv C^{(b)}_{\Sigma_g}\int_{\mathcal{M}_{g,n_{NS},n_R}} \left\langle \prod^{N_{p}}_{k=1}\chi_k\bar{\chi}_k\prod_{t=1}^{N_B}\mathcal{B}_{t}\bar{\mathcal{B}}_t\prod_{i=1}^{n_R}\mathscr{R}^{(-\frac{1}{2})}_{P^{R}_j}\prod_{j=1}^{n_{NS}}\mathscr{V}^{(-1)}_{P^{NS}_j}\right\rangle_{\Sigma_{g,n_{NS},n_{R}}}
\end{equation}
 Here, $\boldsymbol{P}=\{\boldsymbol{P}^{R};\boldsymbol{P}^{NS}\}$  denotes the momenta of the legs, where $\boldsymbol{P}^{NS}$ represents the momentum of NS legs and $\boldsymbol{P}^R$  that of R legs. $N_{p}=2g-2+n_{NS}+\frac{1}{2}n_{R}$ is the number of picture changing operator (PCO) insertion and $N_B=3g-3+n_{NS}+n_{R}$ is the number of $\mathfrak{b}$-ghost insertion. To render the expressions for the string amplitudes more elegant and simplify the mapping to the dual matrix integral, we adopt the normalization factor (which is purely conventional) as follows:
 \begin{align}\label{normalizations}
 &\mathcal{N}_b(P)=\frac{i(b^2-b^{-2})\sin(\pi b P)\sin(\pi b^{-1} P)}{2\pi P\cos\z(\frac{\pi b^2}{2}\y)\cos\z(\frac{\pi b^{-2}}{2}\y)}\nonumber\\
 &\mathcal{R}_b(P)=\frac{e^{\frac{i\pi}{4}}(b^2-b^{-2})\cos(\pi b P)\cos(\pi b^{-1} P)}{2\sqrt{2}\pi \cos\z(\frac{\pi b^2}{2}\y)\cos\z(\frac{\pi b^{-2}}{2}\y)}\\
 &C_{S^2}^{(b)}=-32\pi^4\z(\frac{\cos\z(\frac{\pi b^2}{2}\y)\cos\z(\frac{\pi b^{-2}}{2}\y)}{(b^2-b^{-2})}\y)^2\nonumber
 \end{align}
There are some simple properties of the string amplitudes inherited from the duality of super-Liouville theory.
\subsection*{Parity in $P$}
Because the super-Liouville structure constant is even under $P\to -P$, the parity of the amplitude depends on the parity of the leg factor. Consequently, the amplitude is odd under $P\to -P$ for the NS-leg and even for the R-leg.
\subsection*{Duality and swap symmetry}
The worldsheet theoies have $b\to -b$ and $b\to b^{-1}$. So under such a transformation, the amplitudes will transform
\begin{equation}
 \begin{aligned}
  &\mA^{(-b)}_{g,n_{NS},g_R}(\boldsymbol{P})=\prod_{j=1}^{n_{NS}}\frac{\mathcal{N}_{-b}(P_j)}{\mathcal{N}_b(P_j)}\prod_{i=1}^{n_{R}}\frac{\mathcal{R}_{-b}(P_i)}{\mathcal{R}_b(P_i)}\mA^{(b)}_{g,n_{NS},n_R}(\boldsymbol{P})=\mA^{(b)}_{g,n_{NS},g_R}(\boldsymbol{P})\\
  &\mA^{(b^{-1})}_{g,n_{NS},g_R}(\boldsymbol{P})=\prod_{j=1}^{n_{NS}}\frac{\mathcal{N}_{b^{-1}}(P_j)}{\mathcal{N}_b(P_j)}\prod_{i=1}^{n_{R}}\frac{\mathcal{R}_{b^{-1}}(P_i)}{\mathcal{R}_b(P_i)}\mA^{(b)}_{g,n_{NS},n_R}(\boldsymbol{P})=(-1)^{n_{NS}+n_R}\mA^{(b)}_{g,n_{NS},g_R}(\boldsymbol{P})
 \end{aligned}
\end{equation}
The same relation will be satisfied by $\hmA^{(b)}_{g,n_{NS},n_R}$. Finally, we can exchange the two super-Liouville theories on the worldsheet, which means we do $b\to -ib$ and $\boldsymbol{P}\to i\boldsymbol{P}$ transformation. We can see under swapping, the R-vertex will transform with an additional sign
\begin{equation}
\mathscr{R}_{P_1}^{-\frac{1}{2}}\mathscr{R}_{P_2}^{-\frac{1}{2}}\to- \mathscr{R}_{P_1}^{-\frac{1}{2}}\mathscr{R}_{P_2}^{-\frac{1}{2}}  
\end{equation}
We get
\begin{equation}
\begin{aligned}
  \mA_{g,n_{NS},n_R}^{(-ib)}(i\boldsymbol{P})&=(-1)^{\frac{n_R}{2}}\prod_{j=1}^{n_{NS}}\frac{\mathcal{N}_{-ib}(iP_j)}{\mathcal{N}_{b}(P_j)}\prod_{j=1}^{n_{NS}}\frac{\mathcal{R}_{-ib}(iP_i)}{\mathcal{R}_{b}(P_i)}\mA_{g,n_{NS},n_{R}}^{(b)}(\boldsymbol{P})\\
 &=(-i)^{n_{NS}+n_R}\mA^{(b)}_{g,n_{NS},n_R}(\boldsymbol{P})
\end{aligned}
\end{equation}
For $\hSCLS$, we can see the R-vertex is invariant under swapping
\begin{equation}
 \widehat{\mathscr{R}}_{P_1}^{(-\frac{1}{2})}\widehat{\mathscr{R}}_{P_2}^{(-\frac{1}{2})}\to \widehat{\mathscr{R}}_{P_1}^{(-\frac{1}{2})}\widehat{\mathscr{R}}_{P_2}^{(-\frac{1}{2})}
\end{equation}
We also need to consider the transformation of PCO $\chi\to - \chi$. Finally we have 
\begin{equation}
\begin{aligned}
  \hmA_{g,n_{NS},n_R}^{(-ib)}(i\boldsymbol{P})&=(-1)^{N_P}\prod_{j=1}^{n_{NS}}\frac{\mathcal{N}_{-ib}(iP_j)}{\mathcal{N}_{b}(P_j)}\prod_{j=1}^{n_{NS}}\frac{\mathcal{R}_{-ib}(iP_i)}{\mathcal{R}_{b}(P_i)}\hmA_{g,n_{NS},n_{R}}^{(b)}(\boldsymbol{P})\\
 &=i^{n_{NS}+n_R}\hmA^{(b)}_{g,n_{NS},n_R}(\boldsymbol{P})
\end{aligned}
\end{equation}
This symmetry is called swap symmetry.
\subsection*{Trivial zeros}
The leg factors $\mathcal{N}_b(P)$ and $\mathcal{R}_b(P)$ introduce trivial zeros to the amplitudes
\begin{equation}
 \begin{aligned}
  &\mA_{g,n_{NS},n_R}^{(b)}\Big|_{P^{NS}_j=mb\text{ or } mb^{-1}}=\hmA_{g,n_{NS},n_R}^{(b)}\Big|_{P^{NS}_j=mb\text{ or } mb^{-1}}=0\quad m\in \mathbb{Z},\\
  &\mA_{g,n_{NS},n_R}^{(b)}\Big|_{P^{R}_j=(m+\frac{1}{2})b\text{ or } (m+\frac{1}{2})b^{-1}}=\hmA_{g,n_{NS},n_R}^{(b)}\Big|_{P^{R}_j=(m+\frac{1}{2})b\text{ or } (m+\frac{1}{2})b^{-1}}=0\quad m\in \mathbb{Z}.
 \end{aligned}
\end{equation}

\section{Sphere 3-point Amplitudes}\label{3-point}
In this section, we analyze the sphere 3-point amplitudes in the $\SCLS$ and $\hSCLS$ theories, including both the NS-NS-NS and NS-R-R types. These amplitudes are expressed as infinite sums. We will show that the amplitudes $\mA^{(b)}_{0,3,0}$ and $\mA^{(b)}_{0,1,2}$ in $\SCLS$ are identical to the three-point amplitude in the bosonic $\mathbb{C}\text{LS}$, while the amplitudes $\hmA^{(b)}_{0,3,0}$ and $\hmA^{(b)}_{0,1,2}$ in $\hSCLS$ are different.

\subsection{NS-NS-NS amplitude}
Since there are no  moduli integral in the sphere 3-point amplitudes, one can directly read their expressions from the structure constants \eqref{SC} in SLFT. For the  NS-NS-NS amplitude, we have 
\begin{equation}\label{A030}
 \begin{aligned}
  \mA_{0,3,0}^{(b)}(P_1,P_2,P_3)&=C_{S^2}^{(b)}\left\langle \mathscr{V}_{P_1}^{(-1)}\mathscr{V}_{P_2}^{(-1)}\mathscr{V}_{P_3}^{(0)}\right\rangle_{S^2} \\
  &=\mA^{(b)+}_{0,3,0}(P_1,P_2,P_3)+\mA^{(b)-}_{0,3,0}(P_1,P_2,P_3).
 \end{aligned}
\end{equation}
where
\begin{equation}
 \begin{aligned}
 & \mA_{0,3,0}^{(b)+}(P_1,P_2,P_3)\equiv C_{S^2}^{(b)}\z(\prod_{i=1}^3 \mathcal{N}_b(P_i)\y)C_{b}(P_1,P_2,P_3)\widetilde{C}_{-ib}(iP_1,iP_2,iP_3)\\
 &\mA_{0,3,0}^{(b)-}(P_1,P_2,P_3)\equiv C_{S^2}^{(b)}\z(\prod^3_i \mathcal{N}_b(P_i)\y)\widetilde{C}_{b}(P_1,P_2,P_3)C_{-ib}(iP_1,iP_2,iP_3).
 \end{aligned}
\end{equation}
Introducing these two factors are  convenient, as $\hmA_{0,3,0}^{(b)}$ can now be written as:
\begin{equation}\label{hA030}
 \hmA_{0,3,0}^{(b)}(P_1,P_2,P_3)=\mA^{(b)+}_{0,3,0}(P_1,P_2,P_3)-\mA^{(b)-}_{0,3,0}(P_1,P_2,P_3)
\end{equation}
The two factors $\mA^{(b)\pm}_{0,3,0}$  exhibit different analytic structures. By exploiting the poles, zeros  and  double periodic properties of $\mA^{(b)\pm}_{0,3,0}$, we can obtain  various expressions of the amplitudes $\mA_{0,3,0}^{(b)}$ and $\hmA_{0,3,0}^{(b)}$.

\subsection*{Zeros and poles of $\mA_{0,3,0}^{(b)+}(P_1,P_2,P_3)$ and $\mA_{0,3,0}^{(b)-}(P_1,P_2,P_3)$}
The super DOZZ structure constants $C_{b}$ and $\widetilde{C}_{b}$ possess infinitely many  poles and zeros: 

$C_b(P_1,P_2,P_3)$ has
\begin{itemize}
 \item Zeros: when $P_k=\pm \frac{1}{2}\z(rb+sb^{-1}\y),\quad r,s \in \mathbb{Z}_{\geq 1},\quad r+s=2\mathbb{Z}$
 \item Poles: when $\pm P_1\pm P_2\pm P_3=(r-\frac{1}{2})b+(s-\frac{1}{2})b^{-1},\quad r,s\in\mathbb{Z}_{\geq 1},\quad r+s\in 2\mathbb{Z}$
\end{itemize}

$\widetilde{C}_b(P_1,P_2,P_3)$ has
\begin{itemize}
 \item Zeros: when $P_k=\pm \frac{1}{2}\z(rb+sb^{-1}\y),\quad r,s \in \mathbb{Z}_{\geq 1},\quad r+s=2\mathbb{Z}$
 \item Poles: when $\pm P_1\pm P_2\pm P_3=(r-\frac{1}{2})b+(s-\frac{1}{2})b^{-1},\quad r,s\in\mathbb{Z}_{\geq 1},\quad r+s\in 2\mathbb{Z}+1$
\end{itemize}
Combining the poles and zeros of the super DOZZ structure constants and zeros of the leg factors, we  get the analytic structure of $\mA^{(b)\pm}_{0,3,0}$:

$\mA_{0,3,0}^{(b)+}(P_1,P_2,P_3)$ has
\begin{itemize}
 \item Zeros: when $P_k= \frac{r}{2}b+\frac{s}{2}b^{-1},\quad r,s \in \mathbb{Z}$, $r+s\in 2\mathbb{Z}$
 \item Poles: $P_1\pm P_2\pm P_3=(r-\frac{1}{2})b+(s-\frac{1}{2})b^{-1},\quad r,s\in \mathbb{Z},\ r+s\in2\mathbb{Z}$
\end{itemize}

$\mA_{0,3,0}^{(b)-}(P_1,P_2,P_3)$ has
\begin{itemize}
 \item Zeros: when $P_k= \frac{r}{2}b+\frac{s}{2}b^{-1},\quad r,s \in \mathbb{Z}$, $r+s\in 2\mathbb{Z}$
 \item Poles: $P_1\pm P_2\pm P_3=(r-\frac{1}{2})b+(s-\frac{1}{2})b^{-1},\quad r,s\in \mathbb{Z},\ r+s\in 2\mathbb{Z}+1$
\end{itemize}
Combining them together, we find:

$\mA_{0,3,0}^{(b)}(P_1,P_2,P_3)$ and $\hmA_{0,3,0}^{(b)}(P_1,P_2,P_3)$ have
\begin{itemize}
 \item Zeros: when $P_k= \frac{r}{2}b+\frac{s}{2}b^{-1},\quad r,s \in \mathbb{Z}$, $r+s\in 2\mathbb{Z}$
 \item Poles: $P_1\pm P_2\pm P_3=(r+\frac{1}{2})b+(s+\frac{1}{2})b^{-1},\quad r,s\in \mathbb{Z}$
\end{itemize}
Note that both  $\mA^{(b)}_{0,3,0}$ and $\hmA_{0,3,0}^{(b)}$  share the same poles as the sphere 3-point  amplitude in the bosonic $\mathbb{C}$LS \cite{Collier:2024kwt}. In contrast, their zeros superficially appear to differ from (less than) those of the sphere 3-point  amplitude in the bosonic $\mathbb{C}$LS, where there are no constraint on the parity of $r+s$. However, for  $\mA^{(b)}_{0,3,0}$,  we will show that emergent zeros exist—such that its zero set is exactly identical to that of the bosonic 3-point amplitude..

\subsection*{Periodic conditions}
Employing the shift identities \eqref{shiftGamma} of double gamma function $\Gamma_b$, we find that the amplitudes $\mA^{(b)\pm}_{0,3,0}$ satisfy  the periodic conditions:
\begin{equation}\label{NSb}
 \begin{aligned}
 & \mA^{(b)\pm}_{0,3,0}(P_1+b^{-1},P_2,P_3)=\mA^{(b)\mp}_{0,3,0}(P_1,P_2,P_3),\\
 & \mA^{(b)\pm}_{0,3,0}(P_1+b,P_2,P_3)=\mA^{(b)\mp}_{0,3,0}(P_1,P_2,P_3).
 \end{aligned}
\end{equation}
This means $\mA^{(b)\pm}_{0,3,0}$ are  double-periodic and the period is $(2b,2b^{-1})$. It follows that  $\mA^{(b)}_{0,3,0}$ is double-periodic and $\hmA^{(b)}_{0,3,0}$ is anti-double-periodic—both with respect to the period  $(b,b^{-1})$.

Now we can show the emergent zeros of $\mA^{(b)}_{0,3,0}$. From the above  periodic properties, we have:
\begin{equation}\label{cancel}
 \begin{aligned}
 \mA^{(b)\pm}_{0,3,0}\z(\frac{rb}{2}+\frac{s}{2b},P_2,P_3\y)&=-\mA^{(b)\pm}_{0,3,0}\z(-\frac{rb}{2}-\frac{s}{2b},P_2,P_3\y)\\
 &=-\mA^{(b)\pm}_{0,3,0}\z(\frac{rb}{2}+\frac{s}{2b}-(rb+sb^{-1}),P_2,P_3\y)\\
 &=-\mA^{(b)\mp}_{0,3,0}\z(\frac{rb}{2}+\frac{s}{2b},P_2,P_3\y),\quad \text{for $r+s\in 2\mathbb{Z}+1$}
 \end{aligned}
\end{equation}
The above equation means that $\mA^{(b)}_{0,3,0}(P_1,P_2,P_3)$ has emergent zeros at $P_k=\frac{r}{2}b+\frac{s}{2}b^{-1}$,  arising from the cancellation between $\mA^{(b)+}_{0,3,0}$ and $\mA^{(b)-}_{0,3,0}$. Thus, both the poles and zeros of $\mA^{(b)}_{0,3,0}$ are the same as those of the  bosonic 3-point amplitude,   implying that these two amplitudes themselves are very likely identical.  In what follows, we provide several explicit expressions for $\mA^{(b)\pm}_{0,3,0}$ and demonstrate that this is indeed the case.

\subsection*{Theta function representation}

Using \eqref{TwoGamma}, we obtain the theta function representations for $\mA^{(b)\pm}_{0,3,0}$
\begin{equation}\label{TA}
 \begin{aligned}
 &\mA^{(b)+}_{0,3,0}(P_1,P_2,P_3)=\frac{e^{-\frac{i\pi b^2}{2}}b\vartheta_3(0|b^2)\eta(b^2)^3\prod_i\vartheta_1(bP_i|b^2) \vartheta_3(bP_i|b^2)}{4\vartheta_3\z(\frac{1+b^2}{4}\pm \frac{bP_1}{2}\pm \frac{bP_2}{2}\pm \frac{bP_3}{2}|b^2\y)}\\
 &\mA^{(b)-}_{0,3,0}(P_1,P_2,P_3)=-\frac{e^{-\frac{i\pi b^2}{2}}b\vartheta_3(0|b^2)\eta(b^2)^3\prod_i\vartheta_1(bP_i|b^2) \vartheta_3(bP_i|b^2)}{4\vartheta_3\z(\frac{b^2-1}{4}\pm \frac{bP_1}{2}\pm \frac{bP_2}{2}\pm \frac{bP_3}{2}|b^2\y)}.
 \end{aligned}
\end{equation}
The properties of $\vartheta_i$ and $\eta$ functions are summarized in Appendix \ref{thetafunctions}. Then from \eqref{A030} and \eqref{hA030} we get the theta function representations for $\mA^{(b)}_{0,3,0}$ and $\hmA^{(b)}_{0,3,0}$. It is not immediately obvious that summing the two equations in \eqref{TA} produces the same theta function representation for the bosonic three-point amplitude as given in \cite{Collier:2024kwt}; however, this can be established by comparing their respective poles, zeros, and double periods, as demonstrated in the subsequent part of this section.

\subsection*{Infinite sum representation}
Notably, $\mA^{(b)\pm}_{0,3,0}$ can also be expressed as follows:
 \begin{align}\label{sA}
 \mA^{(b)+}_{0,3,0}(P_1,P_2,P_3)&=\frac{b}{4}\sum_{r=1}^\infty\sum_{\sigma_1,\sigma_2,\sigma_3=\pm}\frac{\sigma_1\sigma_2\sigma_3}{1+e^{\pi ib(\sigma_1P_1+\sigma_2P_2+\sigma_3P_3+2rb-\frac{1}{2}(b-b^{-1}))}}\nonumber\\
 &+\frac{b}{4}\sum_{r=0}^\infty\sum_{\sigma_1,\sigma_2,\sigma_3=\pm}\frac{\sigma_1\sigma_2\sigma_3}{1+e^{\pi ib(\sigma_1P_1+\sigma_2P_2+\sigma_3P_3+2rb+\frac{1}{2}(b-b^{-1}))}}\nonumber\\
  \mA^{(b)-}_{0,3,0}(P_1,P_2,P_3)&=\frac{b}{4}\sum_{r=1}^\infty\sum_{\sigma_1,\sigma_2,\sigma_3=\pm}\frac{\sigma_1\sigma_2\sigma_3}{1+e^{\pi ib(\sigma_1P_1+\sigma_2P_2+\sigma_3P_3+2rb-\frac{1}{2}(b+b^{-1}))}}\\
 &+\frac{b}{4}\sum_{r=0}^\infty\sum_{\sigma_1,\sigma_2,\sigma_3=\pm}\frac{\sigma_1\sigma_2\sigma_3}{1+e^{\pi ib(\sigma_1P_1+\sigma_2P_2+\sigma_3P_3+2rb+\frac{1}{2}(b+b^{-1}))}}\nonumber
 \end{align}
These sums will converge everywhere if we assume $\Im(b^2)>0$. We now demonstrate that the expressions  in \eqref{sA} are indeed correct. Firstly, it is easy to check that the expressions in \eqref{sA} have the  correct poles. Besides, they also satisfy the periodic relation \eqref{NSb}.
Since the expressions in \eqref{sA} are odd functions with respect to the momentum $P_k$,  we have
\begin{equation}
 \begin{aligned}
 \mA^{(b)\pm}_{0,3,0}\z(\frac{rb}{2}+\frac{s}{2b},P_2,P_3\y)&=-\mA^{(b)\pm}_{0,3,0}\z(-\frac{rb}{2}-\frac{s}{2b},P_2,P_3\y)\\
 &=-\mA^{(b)\pm}_{0,3,0}\z(\frac{rb}{2}+\frac{s}{2b}-(rb+sb^{-1}),P_2,P_3\y)\\
 &=-\mA^{(b)\pm}_{0,3,0}\z(\frac{rb}{2}+\frac{s}{2b},P_2,P_3\y),\quad \text{for $r+s\in 2\mathbb{Z}$}
 \end{aligned}
\end{equation}
This shows $\mA^{(b)\pm}_{0,3,0}\z(P_1,P_2,P_3\y)$ have (the correct) zeros at $P_k=\frac{1}{2}\z(rb+sb^{-1}\y),\ r+s\in 2\mathbb{Z}$. Then \eqref{sA} and \eqref{TA}—which are valid expressions for $\mA^{(b)\pm}_{0,3,0}$ -share the same poles and zeros and both are doubly periodic with the same period.  Their ratio should therefore be a constant depending on $b$; this constant can be determined by comparing the residues at their poles, and is found to be 1. Consequently, the expressions in \eqref{sA} are correct.
From them, we can  write  infinite sum representations   for the NS-NS-NS amplitudes explicitly:
\begin{equation}
 \begin{aligned}
 \mA^{(b)}_{0,3,0}(P_1,P_2,P_3)&=\frac{b}{2}\sum_{r=0}^\infty\sum_{\sigma_1,\sigma_2,\sigma_3=\pm}\frac{\sigma_1\sigma_2\sigma_3}{1+e^{2\pi ib(\sigma_1P_1+\sigma_2P_2+\sigma_3P_3+(r+\frac{1}{2})b)}}\\
 \hmA^{(b)}_{0,3,0}(P_1,P_2,P_3)&=\frac{ib}{2}\sum_{r=0}^\infty\sum_{\sigma_1,\sigma_2,\sigma_3=\pm}\frac{(-1)^re^{\pi i b(\sigma_1P_1+\sigma_2P_2+\sigma_3P_3+(r+\frac{1}{2})b)}\sigma_1\sigma_2\sigma_3}{1+e^{2\pi ib(\sigma_1P_1+\sigma_2P_2+\sigma_3P_3+(r+\frac{1}{2})b)}}
 \end{aligned}
\end{equation}
The equation for $\mA^{(b)}_{0,3,0}$ above is exactly the same as (one of) the infinite sum representation for the bosonic 3-point amplitude \cite{Collier:2024kwt}.

\subsection*{Second infinite sum representation}
We can also express the amplitudes as the following infinite sums:
\begin{equation}
 \begin{aligned}
 &\mA^{(b)+}_{0,3,0}(P_1,P_2,P_3)=2b\sum_{m=1}^\infty (-1)^m\frac{\cos\z(\frac{1}{2}m\pi b\z(b+b^{-1}\y)\y)\prod_{i=1}^3\sin (bm\pi P_i)}{\sin(m\pi b^2)}\\
 &\mA^{(b)-}_{0,3,0}(P_1,P_2,P_3)=2b\sum_{m=1}^\infty \frac{\cos\z(\frac{1}{2}m\pi b(b+b^{-1})\y)\prod_{i=1}^3\sin (bm\pi P_i)}{\sin(m\pi b^2)}
 \end{aligned}
\end{equation}
These sums converge when $|\Im(2b(\pm P_1\pm P_2\pm P_3))|<\Im(b^2)$. Then the 3-point amplitudes are 
\begin{equation}\label{SinA3}
 \begin{aligned}
  &\mA^{(b)}_{0,3,0}(P_1,P_2,P_3)=2b\sum_{k\in \mathbb{Z}_{\geq 1}} \frac{\prod_{i=1}^{3}\sin (2kb\pi P_i)}{\sin(k\pi b(b+b^{-1}))}\\ &\hmA^{(b)}_{0,3,0}(P_1,P_2,P_3)=2b\sum_{k\in \mathbb{Z}_{\geq 1}-\frac{1}{2}}\frac{\prod^3_{i=1}\sin\z(2kb\pi P_i\y)}{\sin\z(k\pi b(b+b^{-1})\y)}
 \end{aligned}
\end{equation}
These are the forms that we introduced in the
introduction \eqref{3,0pt}, \eqref{h3,0pt}.
The key difference between the two equations in \eqref{SinA3} lies in their summation ranges: for $\mA_{0,3,0}^{(b)}$,  runs over positive integers, while for 
 $\hmA^{(b)}_{0,3,0}$  it runs over positive half-integers. From the perspective of the dual matrix integral, this discrepancy arises from the distinct sets of branch points of the spectral curves. We will elaborate on this point in section \ref{2Matrix}.

\subsection{NS-R-R amplitudes}
The NS-R-R amplitudes can be directly expressed in terms of the structure constants \eqref{RSC} of SLFT:
\begin{equation}\label{A012}
 \begin{aligned}
  \mA_{0,1,2}^{(b)}(P_1,P_2,P_3)&=C_{S^2}^{(b)}\left\langle \MR_{P_1}^{(-\frac{1}{2})}\MR_{P_2}^{(-\frac{1}{2})}\mathscr{V}_{P_3}^{(-1)}\right\rangle_{S^2} \\
  &=\mA^{(b)+}_{0,1,2}(P_1,P_2;P_3)+\mA^{(b)-}_{0,1,2}(P_1,P_2,P_3)
 \end{aligned}
\end{equation}
where
\begin{equation}
 \begin{aligned}
  &\mA^{(b)+}_{0,1,2}(P_1,P_2;P_3)\equiv2C_{S^3}^{(b)}\mathcal{N}_b(P_3)\mathcal{R}_b(P_1)\mathcal{R}_b(P_2)C_b^{e}(P_1,P_2;P_3)C_{-ib}^{o}(iP_1,iP_2;iP_3)\\
  &\mA^{(b)-}_{0,1,2}(P_1,P_2;P_3)\equiv-2C_{S^3}^{(b)}\mathcal{N}_b(P_3)\mathcal{R}_b(P_1)\mathcal{R}_b(P_2)C_b^{o}(P_1,P_2;P_3)C_{-ib}^{e}(iP_1,iP_2;iP_3).
 \end{aligned}
\end{equation}
It follows that $\hmA^{(b)}_{0,1,2}$ can  be written as:
\begin{equation}\label{hA012}
 \hmA^{(b)}_{0,1,2}(P_1;P_2,P_3)=\mA^{(b)+}_{0,1,2}(P_1;P_2,P_3)-\mA^{(b)-}_{0,1,2}(P_1;P_2,P_3)
\end{equation}
The two factors $\mA^{(b)\pm}_{0,1,2}$ also exhibit distinct  analytic structures. We proceed to perform an analysis similar to that for the NS-NS-NS amplitudes in the previous subsection, yielding various expressions for  $\mA^{(b)}_{0,1,2}$ and $\hmA^{(b)}_{0,1,2}$.

\subsection*{Zeros and poles of $\mA_{0,1,2}^{(b)+}(P_1,P_2;P_3)$ and $\mA_{0,1,2}^{(b)-}(P_1,P_2;P_3)$}
 First, we present the poles and zeros of the  structure constants $C_b^e$ and $C^o_b$ in \eqref{RSC}:

$C_b^e(P_1,P_2;P_3)$ has
\begin{itemize}
 \item Zeros: when $P_{1,2}=\pm \frac{1}{2}\z(rb+sb^{-1}\y),\quad r,s \in \mathbb{Z}_{\geq 1},\quad r+s\in 2\mathbb{Z}+1$\\
 $P_3=\pm\frac{1}{2}\z(rb+sb^{-1}\y),\quad r,s\in \mathbb{Z}_{\geq 1},\quad r+s=2\mathbb{Z}$
 \item Poles: when $\pm (P_1-P_2) \pm P_3=(r-\frac{1}{2})b+(s-\frac{1}{2})b^{-1} ,\quad r,s\in \mathbb{Z}_{\geq 1},\quad r+s\in 2\mathbb{Z}$\\
 $\pm (P_1+P_2) \pm P_3=(r-\frac{1}{2})b+(s-\frac{1}{2})b^{-1},\quad r,s\in \mathbb{Z}_{\geq 1},\quad r+s\in 2\mathbb{Z}+1$
\end{itemize}

$C_b^o(P_1,P_2;P_3)$ has
\begin{itemize}
 \item Zeros: when $P_{1,2}=\pm \frac{1}{2}\z(rb+sb^{-1}\y),\quad r,s \in \mathbb{Z}_{\geq 1},\quad r+s\in2\mathbb{Z}+1 $\\
 $P_3=\pm\frac{1}{2}\z(rb+sb^{-1}\y),\quad r,s\in \mathbb{Z}_{\geq 1},\quad r+s=2\mathbb{Z}$
 \item Poles: when $\pm (P_1-P_2) \pm P_3=(r-\frac{1}{2})b+(s-\frac{1}{2})b^{-1} ,\quad r,s\in \mathbb{Z}_{\geq 1},\quad r+s\in 2\mathbb{Z}+1$\\
 $\pm (P_1+P_2) \pm P_3=(r-\frac{1}{2})b+(s-\frac{1}{2})b^{-1},\quad r,s\in \mathbb{Z}_{\geq 1},\quad r+s\in 2\mathbb{Z}$
\end{itemize}
Combining the poles and zeros of the super DOZZ structure constants and the zeros of the leg factors, we get the analytic structure of $\mA^{(b)\pm}_{0,1,2}$: 

$\mA_{0,1,2}^{(b)+}(P_1,P_2;P_3)$ has
\begin{itemize}
 \item Zeros: when $P_{1,2}= \frac{r}{2}b+\frac{s}{2}b^{-1},\quad r,s \in \mathbb{Z}$, $r+s\in 2\mathbb{Z}+1$\\
 $P_3= \frac{r}{2}b+\frac{s}{2}b^{-1},\quad r,s \in \mathbb{Z}$, $r+s\in 2\mathbb{Z}$
 \item Poles: $P_1- P_2\pm P_3=(r-\frac{1}{2})b+(s-\frac{1}{2})b^{-1},\quad r,s\in \mathbb{Z},\ r+s\in 2\mathbb{Z}$\\
 $P_1+ P_2\pm P_3=(r-\frac{1}{2})b+(s-\frac{1}{2})b^{-1},\quad r,s\in \mathbb{Z},\ r+s\in 2\mathbb{Z}+1$
\end{itemize}

$\mA_{0,1,2}^{(b)-}(P_1,P_2;P_3)$ has
\begin{itemize}
 \item Zeros: when $P_{1,2}= \frac{r}{2}b+\frac{s}{2}b^{-1},\quad r,s \in \mathbb{Z}$, $r+s\in 2\mathbb{Z}+1$\\
 $P_3= \frac{r}{2}b+\frac{s}{2}b^{-1},\quad r,s \in \mathbb{Z}$, $r+s\in 2\mathbb{Z}$
 \item Poles: $P_1- P_2\pm P_3=(r-\frac{1}{2})b+(s-\frac{1}{2})b^{-1},\quad r,s\in \mathbb{Z},\ r+s\in 2\mathbb{Z}+1$\\
 $P_1+ P_2\pm P_3=(r-\frac{1}{2})b+(s-\frac{1}{2})b^{-1},\quad r,s\in \mathbb{Z},\ r+s\in2\mathbb{Z}$
\end{itemize}
Combining them together, we find:

$\mA_{0,1,2}^{(b)}(P_1,P_2;P_3)$ and $\hmA_{0,1,2}^{(b)}(P_1,P_2;P_3)$ have
\begin{itemize}
 \item Zeros: when $P_{1,2}= \frac{r}{2}b+\frac{s}{2}b^{-1},\quad r,s \in \mathbb{Z}$, $r+s\in 2\mathbb{Z}+1$\\
 $P_3= \frac{r}{2}b+\frac{s}{2}b^{-1},\quad r,s \in \mathbb{Z}$, $r+s\in 2\mathbb{Z}$
 \item Poles: $P_1\pm P_2\pm P_3=(r+\frac{1}{2})b+(s+\frac{1}{2})b^{-1},\quad r,s\in \mathbb{Z}$
\end{itemize}
Thus, the NS-R-R amplitudes $\mA^{(b)}_{0,1,2}$ and $\hmA^{(b)}_{0,1,2}$ also share the same poles as the sphere 3-point amplitude in
the bosonic $\mathbb{C}$LS \cite{Collier:2024kwt}, but their zeros seem different from (fewer than) those of the latter at a superficial glance. Similar to $\mA^{(b)}_{0,3,0}$, emergent zeros will also appear in $\mA^{(b)}_{0,1,2}$ (see the discussion on  periodic conditions below) such that the zeros of $\mA^{(b)}_{0,1,2}$  are identical to those of the bosonic 3-point amplitude.

\subsection*{Theta function representation}
We have the theta function representation of the $\mA^{(b)\pm}_{0,1,2}$:
 \begin{align}
  &\mA^{(b)+}_{0,1,2}(P_1,P_2;P_3)=\frac{e^{-\frac{i\pi b^2}{2}}b\vartheta_3(0|b^2)\eta(b^2)^3\vartheta_1(bP_3|b^2)\vartheta_3(bP_3|b^2)\prod_{i=1}^2\vartheta_2(bP_i|b^2)\vartheta_4(bP_i|b^2)}{4\vartheta_3\z(\frac{b^2+1}{4}\pm \frac{b(P_1-P_2)}{2}\pm\frac{bP_3}{2}|b^2\y)\vartheta_3\z(\frac{b^2-1}{4}\pm \frac{b(P_1+P_2)}{2}\pm \frac{bP_3}{2}|b^2\y)}\nonumber\\
  &\mA^{(b)-}_{0,1,2}(P_1,P_2;P_3)=-\frac{e^{-\frac{i\pi b^2}{2}}b\vartheta_3(0|b^2)\eta(b^2)^3\vartheta_1(bP_3|b^2)\vartheta_3(bP_3|b^2)\prod_{i=1}^2\vartheta_2(bP_i|b^2)\vartheta_4(bP_i|b^2)}{4\vartheta_3\z(\frac{b^2+1}{4}\pm \frac{b(P_1+P_2)}{2}\pm \frac{bP_3}{2}|b^2\y)\vartheta_3\z(\frac{b^2-1}{4}\pm \frac{b(P_1-P_2)}{2}\pm \frac{bP_3}{2}|b^2\y)}
 \end{align}
From \eqref{A012} and \eqref{hA012}, we get the theta function representations for $\mA^{(b)}_{0,1,2}$ and $\hmA^{(b)}_{0,1,2}$.

\subsection*{Periodic conditions}
We have the periodic conditions with respect to $P_1$ (same for $P_2$)
\begin{equation}
 \begin{aligned}
 & \mA^{(b)\pm}_{0,1,2}(P_1+b^{-1},P_2;P_3)=\mA^{(b)\mp}_{0,1,2}(P_1,P_2;P_3)\\
 & \mA^{(b)\pm}_{0,1,2}(P_1+b,P_2;P_3)=\mA^{(b)\mp}_{0,1,2}(P_1,P_2;P_3).
 \end{aligned}
\end{equation}
It turns out that the periodic conditions with respect to $P_3$ take the same form:
\begin{equation}
 \begin{aligned}
 & \mA^{(b)\pm}_{0,1,2}(P_1,P_2;P_3+b^{-1})=\mA^{(b)\mp}_{0,1,2}(P_1,P_2;P_3)\\
 & \mA^{(b)\pm}_{0,1,2}(P_1,P_2;P_3+b)=\mA^{(b)\mp}_{0,1,2}(P_1,P_2;P_3).
 \end{aligned}
\end{equation}
Therefore, $\mA^{(b)\pm}_{0,1,2}$ are also double-periodic in all momenta, where the period is $(2b,2b^{-1})$. Then  from \eqref{A012} and \eqref{hA012}, we find that  $\mA^{(b)}_{0,1,2}$ is double periodic with period  $(b,b^{-1})$, while  $\hmA^{(b)}_{0,1,2}$ is double anti-periodic with the same period  $(b,b^{-1})$.  Then, due to the similar procedure as in \eqref{cancel}, there will be emergent zeros so that the zeros of $\mA_{0,1,2}^{(b)}$ are identical to those of the bosonic 3-point amplitude.

\subsection*{Relations among 3-point amplitudes}
There exists an additional transformation that can map $\mA_{0,1,2}^{(b)\pm}$ to $\mA_{0,3,0}^{(b)\pm}$:
\begin{equation}\label{NSRb}
 \begin{aligned}
  &\mA_{0,1,2}^{(b)\pm}\z(P_1+\frac{1}{2}b^{-1},P_2+\frac{1}{2}b^{-1};P_3\y)=\mA_{0,3,0}^{(b)\pm}(P_1,P_2,P_3)\\
  &\mA_{0,1,2}^{(b)\pm}\z(P_1+\frac{1}{2}b,P_2+\frac{1}{2}b;P_3\y)=\mA_{0,3,0}^{(b)\pm}(P_1,P_2,P_3)
  \end{aligned}
\end{equation}
Then we have
\begin{equation}
 \begin{aligned}
  &\mA_{0,1,2}^{(b)}\z(P_1+\frac{1}{2}b^{-1},P_2+\frac{1}{2}b^{-1};P_3\y)=\mA_{0,3,0}^{(b)}(P_1,P_2,P_3)\\
  &\mA_{0,1,2}^{(b)}\z(P_1+\frac{1}{2}b,P_2+\frac{1}{2}b;P_3\y)=\mA_{0,3,0}^{(b)}(P_1,P_2,P_3)
 \end{aligned}
\end{equation}
Similar relations hold for $\hmA^{(b)}_{0,1,2}$.

\subsection*{Infinite sum representation}
The NS-R-R amplitudes can be expressed in another form which manifests the locations of the poles
 \begin{align}
 \mA^{(b)+}_{0,1,2}(P_1,P_2;P_3)&=\frac{b}{4}\sum_{r=1}^\infty\sum_{\sigma_1,\sigma_2,\sigma_3=\pm}\frac{\sigma_1\sigma_2\sigma_3}{1-\sigma_1\sigma_2e^{\pi ib(\sigma_1P_1+\sigma_2P_2+\sigma_3P_3+2rb-\frac{1}{2}(b-b^{-1}))}}\nonumber\\
 &+\frac{b}{4}\sum_{r=0}^\infty\sum_{\sigma_1,\sigma_2,\sigma_3=\pm}\frac{\sigma_1\sigma_2\sigma_3}{1-\sigma_1\sigma_2e^{\pi ib(\sigma_1P_1+\sigma_2P_2+\sigma_3P_3+2rb+\frac{1}{2}(b-b^{-1}))}}\nonumber\\
  \mA^{(b)-}_{0,1,2}(P_1,P_2;P_3)&=\frac{b}{4}\sum_{r=1}^\infty\sum_{\sigma_1,\sigma_2,\sigma_3=\pm}\frac{\sigma_1\sigma_2\sigma_3}{1-\sigma_1\sigma_2e^{\pi ib(\sigma_1P_1+\sigma_2P_2+\sigma_3P_3+2rb-\frac{1}{2}(b+b^{-1}))}}\\
 &+\frac{b}{4}\sum_{r=0}^\infty\sum_{\sigma_1,\sigma_2,\sigma_3=\pm}\frac{\sigma_1\sigma_2\sigma_3}{1-\sigma_1\sigma_2e^{\pi ib(\sigma_1P_1+\sigma_2P_2+\sigma_3P_3+2rb+\frac{1}{2}(b+b^{-1}))}}\nonumber
 \end{align}
From \eqref{A012} and \eqref{hA012}, we  get
\begin{equation}
 \begin{aligned}
 \mA^{(b)}_{0,1,2}(P_1,P_2;P_3)&=\frac{b}{2}\sum_{r=0}^\infty\sum_{\sigma_1,\sigma_2,\sigma_3=\pm}\frac{\sigma_1\sigma_2\sigma_3}{1-\sigma_1\sigma_2e^{2\pi ib(\sigma_1P_1+\sigma_2P_2+\sigma_3P_3+(r+\frac{1}{2})b)}}\\
 \hmA^{(b)}_{0,3,0}(P_1,P_2;P_3)&=\frac{ib}{2}\sum_{r=0}^\infty\sum_{\sigma_1,\sigma_2,\sigma_3=\pm}\frac{(-1)^{r+1}e^{\pi i b(\sigma_1P_1+\sigma_2P_2+\sigma_3P_3+(r+\frac{1}{2})b)}\sigma_3}{1-\sigma_1\sigma_2e^{2\pi ib(\sigma_1P_1+\sigma_2P_2+\sigma_3P_3+(r+\frac{1}{2})b)}}
 \end{aligned}
\end{equation}

\subsection*{Second infinite sum representation}
We further have 
\begin{equation}
 \begin{aligned}
 &\mA^{(b)+}_{0,1,2}(P_1,P_2;P_3)=2b\sum_{m=1}^\infty (-1)^m\frac{\cos\z(\frac{1}{2}m\pi b\z(b+b^{-1}\y)\y)\sin(bm\pi P_3)\prod_{i=1}^2\sin \z(bm\pi P_i+\frac{m\pi}{2}\y)}{\sin(m\pi b^2)}\\
 &\mA^{(b)-}_{0,1,2}(P_1,P_2;P_3)=2b\sum_{m=1}^\infty \frac{\cos\z(\frac{1}{2}m\pi b(b+b^{-1})\y)\sin(bm\pi P_3)\prod_{i=1}^2\sin \z(bm\pi P_i+\frac{m\pi}{2}\y)}{\sin(m\pi b^2)}
 \end{aligned}
\end{equation}
Then the 3-point amplitudes  are 
\begin{equation}
\begin{aligned}
  &\mA^{(b)}_{0,1,2}(P_1,P_2;P_3)=2b\sum_{k\in \mathbb{Z}_{\geq 1}}^\infty \frac{\prod_{i=1}^{3}\sin (2kb\pi P_i)}{\sin(k\pi b(b+b^{-1}))}\\
 &\hmA^{(b)}_{0,1,2}(P_1,P_2;P_3)=2b\sum_{k\in \mathbb{Z}_{\geq 1}-\frac{1}{2}}^{\infty}\frac{\sin\z(2kb\pi P_3\y)\prod^2_{i=1}\cos\z(2kb\pi P_i\y)}{\sin\z(k\pi b(b+b^{-1})\y)}
\end{aligned}
\end{equation}
These are the forms that we introduced in the
introduction \eqref{1,2pt}, \eqref{h1,2pt}.
 In particular, we can see that the NS-NS-NS and NS-R-R 3-point amplitudes in $\SCLS$ are exactly the same as the 3-point amplitude $\mA^{(b)}_{0,3}$ in the bosonic $\mathbb{C}\text{LS}$ \cite{Collier:2024kwt}:
\begin{equation}\label{030=012=bo}
 \mA^{(b)}_{0,3,0}=\mA^{(b)}_{0,1,2}=\mA^{(b)}_{0,3}
\end{equation}
From the point of view of the worldsheet CFT, these equations give  non-trivial relations between  the super DOZZ structure constants and the bosonic DOZZ structure constant. 

We provide further comments on the relation \eqref{030=012=bo}. In fact, analogous relations among amplitudes can be found in the bosonic $c=1$ string and type 0B $\hat{c}=1$ superstring \cite{Balthazar:2022atu}. In that paper, the authors combine the NS and R states in type 0B $\hat{c}=1$ superstring as
\begin{equation}\label{combNSR}
\mathcal{L}_w^{\pm}\equiv \mathcal{T}_w^\pm -\mathcal{A}_w^\pm,\quad \mathcal{R}_w^{\pm}\equiv \mathcal{T}_w^\pm +\mathcal{A}_w^\pm 
\end{equation}
where $\mathcal{T}^{+}_w$ ($\mathcal{T}^{-}_w$) and $\mathcal{A}^{+}_w$ ($\mathcal{A}^{-}_w$) correspond to the in (out) NS and R states respectively (with energy $\omega$). Then it was shown that $\mathcal{L}^{\pm}_w$ and $\mathcal{R}^{\pm}_w$ decouple at the level of perturbation theory (i.e. amplitudes that involve both $\mathcal{L}^{\pm}_w$ and $\mathcal{R}^{\pm}_w$ vanish)\footnote{From the dual matrix model, the decoupling of the ``LHS modes'' $\mathcal{L}_w^{\pm}$ and the ``RHS modes'' $\mathcal{R}_w^{\pm}$ become manifest \cite{Balthazar:2022atu}. It will be interesting to see whether this structure happens also for $\SCLS$ (see section \ref{dualmatrix} for further comments).} and both gives rise to one copy of perturbative amplitudes of the $c=1$ bosonic string theory \cite{Balthazar:2017mxh}. For the 3-point amplitudes, this  in particular implies that both the NS-NS-NS and NS-R-R amplitudes in type 0B $\hat{c}=1$ superstring theory coincide with the 3-point amplitude in the bosonic $c=1$ string theory.

\section{Analytic structure of the sphere 4-point  amplitude}\label{Analyticstructure}
In this section, we analyze the analytic structure of the sphere NS-NS-NS-NS four-point amplitudes $\mA^{(b)}_{0,4,0}$ and $\hmA^{(b)}_{0,4,0}$ in $\SCLS$ and $\hSCLS$, respectively. Their discontinuities and poles are determined by the properties of the underlying super-Liouville correlation functions.

\subsection{Discontinuities of $\mA^{(b)}_{0,4,0}$ and $\hmA^{(b)}_{0,4,0}$}
 We consider the sphere four-point amplitudes with all 4 external operators being super-primaries in the NS sector: $\mathscr{V}_{P_i}(z_i,\bar{z}_i)$ $(i=1,2,3,4)$. Firstly, we fix the positions of three vertex operators to be $(z_1,z_3,z_4)=(0,1,\infty)$. Then the position $z\in\mathbb{C}$ of the last vertex operator $\mathscr{V}_{P_2}(z,\bar{z})$ becomes the modulus of the sphere four-point amplitude.

To get a total picture number of $-2$, we let the first and fourth operators be in picture $-1$ and the second and third operators be in picture $0$. Then  the sphere four-point amplitudes can be expressed as:
\begin{equation}\label{4pt+-}
\begin{aligned}
 & \mA^{(b)}_{0,4,0}(P_1,P_2,P_3,P_4)=\mA^{(b)+}_{0,4,0}(P_1,P_2,P_3,P_4)+\mA^{(b)-}_{0,4,0}(P_1,P_2,P_3,P_4)\\
 & \hmA^{(b)}_{0,4,0}(P_1,P_2,P_3,P_4)=\mA^{(b)+}_{0,4,0}(P_1,P_2,P_3,P_4)-\mA^{(b)-}_{0,4,0}(P_1,P_2,P_3,P_4)
\end{aligned}
\end{equation}
where for convenience (as in the analysis of the 3-point amplitudes), we introduced 
\begin{equation}\label{defApAm}
 \begin{aligned}
  &\mA_{0,4,0}^{(b)+}(P_1,P_2,P_3,P_4)\equiv C^{(b)}_{S^2}\int_{\mathbb{C}}d^2z\mathcal{N}_b(P_1)\mathcal{N}_b(P_2)\mathcal{N}_b(P_3)\mathcal{N}_b(P_4)\left(I_{11}+I_{22}+J_{11}+J_{22}\right)\\
    &\mA_{0,4,0}^{(b)-}(P_1,P_2,P_3,P_4)\equiv C^{(b)}_{S^2}\int_{\mathbb{C}}d^2z\mathcal{N}_b(P_1)\mathcal{N}_b(P_2)\mathcal{N}_b(P_3)\mathcal{N}_b(P_4)\left(I_{12}+I_{21}+J_{12}+J_{21}\right).
 \end{aligned}
\end{equation}
The functions $I_{ab}$ and $J_{ab}$  can be calculated from  4-point functions in the two SLFTs on the worldsheet: 
\begin{equation}
\begin{aligned}
  I_{11}&=\left\langle V^+_{P_1}(0) W^+_{P_2}(z,\bar{z})W^+_{P_3}(1)V^+_{P_4}(\infty)\right\rangle_{g=0} \left\langle V^{-}_{iP_1}(0) V^{-}_{iP_2}(z,\bar{z})V^-_{iP_3}(1)V^-_{iP_4}(\infty)\right\rangle_{g=0}\\
 I_{12}&=\left\langle V^+_{P_1}(0) W^+_{P_2}(z,\bar{z})V^+_{P_3}(1)V^+_{P_4}(\infty)\right\rangle_{g=0} \left\langle V^-_{iP_1}(0) V^-_{iP_2}(z,\bar{z})W^-_{iP_3}(1)V^-_{iP_4}(\infty)\right\rangle_{g=0}\\
 I_{21}&=\left\langle V^+_{P_1}(0) V^+_{P_2}(z,\bar{z})W^+_{P_3}(1)V^+_{P_4}(\infty)\right\rangle_{g=0} \left\langle V^-_{iP_1}(0) W^-_{iP_2}(z,\bar{z})V^-_{iP_3}(1)V^-_{iP_4}(\infty)\right\rangle_{g=0}\\
 I_{22}&=\left\langle V^+_{P_1}(0) V^+_{P_2}(z,\bar{z})V^+_{P_3}(1)V^+_{P_4}(\infty)\right\rangle_{g=0} \left\langle V^-_{iP_1}(0) W^-_{iP_2}(z,\bar{z})W^-_{iP_3}(1)V^-_{iP_4}(\infty)\right\rangle_{g=0}
\end{aligned}
\end{equation}
and
\begin{equation}
\begin{aligned}
  J_{11}&=-\left\langle V^+_{P_1}(0) \Lambda^+_{P_2}(z,\bar{z})\Lambda^+_{P_3}(1)V^+_{P_4}(\infty)\right\rangle_{g=0} \left\langle V^-_{iP_1}(0) \bar{\Lambda}^-_{iP_2}(z,\bar{z})\bar{\Lambda}^-_{iP_3}(1)V^-_{iP_4}(\infty)\right\rangle_{g=0}\\
 J_{12}&=\left\langle V^+_{P_1}(0) \Lambda^+_{P_2}(z,\bar{z})\bar{\Lambda}^+_{P_3}(1)V^+_{P_4}(\infty)\right\rangle_{g=0} \left\langle V^-_{iP_1}(0) \bar{\Lambda}^-_{iP_2}(z,\bar{z})\Lambda^-_{iP_3}(1)V^-_{iP_4}(\infty)\right\rangle_{g=0}\\
 J_{21}&=\left\langle V^+_{P_1}(0) \bar{\Lambda}^+_{P_2}(z,\bar{z})\Lambda^+_{P_3}(1)V^+_{P_4}(\infty)\right\rangle_{g=0} \left\langle V^-_{iP_1}(0) \Lambda^-_{iP_2}(z,\bar{z})\bar{\Lambda}^-_{iP_3}(1)V^-_{iP_4}(\infty)\right\rangle_{g=0}\\
J_{22}&=-\left\langle V^+_{P_1}(0) \bar{\Lambda}^+_{P_2}(z,\bar{z})\bar{\Lambda}^+_{P_3}(1)V^+_{P_4}(\infty)\right\rangle_{g=0} \left\langle V^-_{iP_1}(0) \Lambda^-_{iP_2}(z,\bar{z})\Lambda^-_{iP_3}(1)V^-_{iP_4}(\infty)\right\rangle_{g=0}
\end{aligned}
\end{equation}
\subsection*{Superconformal block decomposition of $\mA^{(b)}_{0,4,0}$ and $\hmA^{(b)}_{0,4,0}$}
All these four-point functions could be written as superconformal block expansions, which will involve 8 NS conformal blocks \cite{Hadasz:2006qb}. We write explicitly the superconformal block expansions for the four-point functions in the SLFT labeled by ``$+$'', the expansions for four-point functions in the conjugate theory (labeled by ``$-$'') are similar. For the 4 correlators that appear in $I_{ab}$, the expansions are (the superscripts ``$+$''  are omitted):
 \begin{align}\label{diagonalblcok}
  &\left\langle V_{P_1}(0) V_{P_2}(z,\bar{z})V_{P_3}(1)V_{P_4}(\infty)\right\rangle_{g=0}\nonumber\\
  &\qquad =\sum_P\Big(C_b(P_4,P_3,P)C_b(P,P_2,P_1) \mathcal{F}^1_{h_P}\left[
 \begin{smallmatrix}
  h_{P_3} & h_{P_2} \\
 h_{P_4} & h_{P_1}
 \end{smallmatrix}
  \right](z)\mathcal{F}^1_{h_P}\left[
 \begin{smallmatrix}
  h_{P_3} & h_{P_2} \\
 h_{P_4} & h_{P_1}
 \end{smallmatrix}
  \right](\bar z) \nonumber\\
  &\qquad -\widetilde{C}_b(P_4,P_3,P)\widetilde{C}_b(P,P_2,P_1) \mathcal{F}^{\frac{1}{2}}_{h_P}\left[
 \begin{smallmatrix}
  h_{P_3} & h_{P_2} \\
 h_{P_4} & h_{P_1}
 \end{smallmatrix}
  \right](z) \mathcal{F}^{\frac{1}{2}}_{h_P}\left[
 \begin{smallmatrix}
  h_{P_3} & h_{P_2} \\
 h_{P_4} & h_{P_1}
 \end{smallmatrix}
  \right](\bar z) \Big)\nonumber\\
  &\left\langle V_{P_1}(0) W_{P_2}(z,\bar{z})V_{P_3}(1)V_{P_4}(\infty)\right\rangle_{g=0}\nonumber\\
  &\qquad =\sum_P\Big(C_b(P_4,P_3,P)\widetilde{C}_b(P,P_2,P_1) \mathcal{F}^1_{h_P}\left[
 \begin{smallmatrix}
  h_{P_3} & *h_{P_2} \\
 h_{P_3} & h_{P_1}
 \end{smallmatrix}
  \right](z) \mathcal{F}^1_{h_P}\left[
 \begin{smallmatrix}
  h_{P_3} & *h_{P_2} \\
 h_{P_3} & h_{P_1}
 \end{smallmatrix}
  \right](\bar z) \nonumber\\
  & \qquad +\widetilde{C}_b(P_4,P_3,P)C_b(P,P_2,P_1) \mathcal{F}^{\frac{1}{2}}_{h_P}\left[
 \begin{smallmatrix}
  h_{P_3} & *h_{P_2} \\
 h_{P_4} & h_{P_1}
 \end{smallmatrix}
  \right](z) \mathcal{F}^{\frac{1}{2}}_{h_P}\left[
 \begin{smallmatrix}
  h_{P_3} & *h_{P_2} \\
 h_{P_4} & h_{P_1}
 \end{smallmatrix}
  \right](\bar z) \Big)\nonumber\\
  & \left\langle V_{P_1}(0) V_{P_2}(z,\bar{z})W_{P_3}(1)V_{P_4}(\infty)\right\rangle_{g=0}\\
  &\qquad = \sum_P\Big(\widetilde{C}_b(P_4,P_3,P)C_b(P,P_2,P_1) \mathcal{F}^1_{h_P}\left[
 \begin{smallmatrix}
  *h_{P_3} & h_{P_2} \\
 h_{P_4} & h_{P_1}
 \end{smallmatrix}
  \right](z) \mathcal{F}^1_{h_P}\left[
 \begin{smallmatrix}
  *h_{P_3} & h_{P_2} \\
 h_{P_4} & h_{P_1}
 \end{smallmatrix}
  \right](\bar z)\nonumber \\
  & \qquad +C_b(P_4,P_3,P)\widetilde{C}_b(P,P_2,P_1) \mathcal{F}^{\frac{1}{2}}_{h_P}\left[
 \begin{smallmatrix}
  *h_{P_3} & h_{P_2} \\
 h_{P_3} & h_{P_1}
 \end{smallmatrix}
  \right](z) \mathcal{F}^{\frac{1}{2}}_{h_P}\left[
 \begin{smallmatrix}
  *h_{P_3} & h_{P_2} \\
 h_{P_3} & h_{P_1}
 \end{smallmatrix}
  \right](\bar z) \Big)\nonumber\\
  & \left\langle V_{P_1}(0) W_{P_2}(z,\bar{z})W_{P_3}(1)V_{P_4}(\infty)\right\rangle_{g=0}\nonumber\\
  &\qquad = \sum_P\Big(\widetilde{C}_b(P_4,P_3,P)\widetilde{C}_b(P,P_2,P_1) \mathcal{F}^1_{h_P}\left[
 \begin{smallmatrix}
  *h_{P_3} & *h_{P_2} \\
 h_{P_4} & h_{P_1}
 \end{smallmatrix}
  \right](z) \mathcal{F}^1_{h_P}\left[
 \begin{smallmatrix}
  *h_{P_3} & *h_{P_2} \\
 h_{P_4} & h_{P_1}
 \end{smallmatrix}
  \right](\bar z) \nonumber\\
  &\qquad -C_b(P_4,P_3,P)C_b(P,P_2,P_1) \mathcal{F}^{\frac{1}{2}}_{h_P}\left[
 \begin{smallmatrix}
  *h_{P_3} & *h_{P_2} \\
 h_{P_4} & h_{P_1}
 \end{smallmatrix}
  \right](z) \mathcal{F}^{\frac{1}{2}}_{h_P}\left[
 \begin{smallmatrix}
  *h_{P_3} & *h_{P_2} \\
 h_{P_4} & h_{P_1}
 \end{smallmatrix}
  \right](\bar z) \Big)\nonumber
 \end{align}
where $*h_i=h_i+\frac{1}{2}$. We also use the symbol $_-h$ to represent either $h$ or $*h$. For the 4 correlators that in appear $J_{ab}$, the superconformal block expansions are:
\begin{equation}\label{nondiagonalblock}
 \begin{aligned}
  &\left\langle V_{P_1}(0) \Lambda_{P_2}(z,\bar{z})\Lambda_{P_3}(1)V_{P_4}(\infty)\right\rangle_{g=0}\\
  &\qquad =-\sum_P\Big(\widetilde{C}_b(P_4,P_3,P)\widetilde{C}_b(P,P_2,P_1)\mathcal{F}^1_{h_P}\left[
 \begin{smallmatrix}
  *h_{P_3} & *h_{P_2} \\
 h_{P_4} & h_{P_1}
 \end{smallmatrix}
  \right](z)\mathcal{F}^{\frac{1}{2}}_{h_P}\left[
 \begin{smallmatrix}
  h_{P_3} & h_{P_2} \\
 h_{P_4} & h_{P_1}
 \end{smallmatrix}
  \right](\bar{z})\\
  &\qquad+C_b(P_4,P_3,P)C_b(P,P_2,P_1)\mathcal{F}^{\frac{1}{2}}_{h_P}\left[
 \begin{smallmatrix}
  *h_{P_3} & *h_{P_2} \\
 h_{P_4} & h_{P_1}
 \end{smallmatrix}
  \right](z)\mathcal{F}^{1}_{h_P}\left[
 \begin{smallmatrix}
  h_{P_3} & h_{P_2} \\
 h_{P_4} & h_{P_1}
 \end{smallmatrix}
  \right](\bar{z})\Big)\\
  &\left\langle V_{P_1}(0) \bar{\Lambda}_{P_2}(z,\bar{z})\bar{\Lambda}_{P_3}(1)V_{P_4}(\infty)\right\rangle_{g=0}\\
  &\qquad =-\sum_P\Big(C_b(P_4,P_3,P)C_b(P,P_2,P_1)\mathcal{F}^1_{h_P}\left[
 \begin{smallmatrix}
  h_{P_3} & h_{P_2} \\
 h_{P_4} & h_{P_1}
 \end{smallmatrix}
  \right](z)\mathcal{F}^{\frac{1}{2}}_{h_P}\left[
 \begin{smallmatrix}
  *h_{P_3} & *h_{P_2} \\
 h_{P_4} & h_{P_1}
 \end{smallmatrix}
  \right](\bar{z})\\
  &\qquad+\widetilde{C}_b(P_4,P_3,P)\widetilde{C}_b(P,P_2,P_1)\mathcal{F}^{\frac{1}{2}}_{h_P}\left[
 \begin{smallmatrix}
  h_{P_3} & h_{P_2} \\
 h_{P_4} & h_{P_1}
 \end{smallmatrix}
  \right](z)\mathcal{F}^{1}_{h_P}\left[
 \begin{smallmatrix}
  *h_{P_3} & *h_{P_2} \\
 h_{P_4} & h_{P_1}
 \end{smallmatrix}
  \right](\bar{z})\Big)\\
  &\left\langle V_{P_1}(0) \Lambda_{P_2}(z,\bar{z})\bar{\Lambda}_{P_3}(1)V_{P_4}(\infty)\right\rangle_{g=0}\\
  &\qquad =\sum_P\Big(C_b(P_4,P_3,P)\widetilde{C}_b(P,P_2,P_1)\mathcal{F}^1_{h_P}\left[
 \begin{smallmatrix}
  h_{P_3} & *h_{P_2} \\
 h_{P_4} & h_{P_1}
 \end{smallmatrix}
  \right](z)\mathcal{F}^{\frac{1}{2}}_{h_P}\left[
 \begin{smallmatrix}
  *h_{P_3} & h_{P_2} \\
 h_{P_4} & h_{P_1}
 \end{smallmatrix}
  \right](\bar{z})\\
  &\qquad-\widetilde{C}_b(P_4,P_3,P)C_b(P,P_2,P_1)\mathcal{F}^{\frac{1}{2}}_{h_P}\left[
 \begin{smallmatrix}
  h_{P_3} & *h_{P_2} \\
 h_{P_4} & h_{P_1}
 \end{smallmatrix}
  \right](z)\mathcal{F}^{1}_{h_P}\left[
 \begin{smallmatrix}
  *h_{P_3} & h_{P_2} \\
 h_{P_4} & h_{P_1}
 \end{smallmatrix}
  \right](\bar{z})\Big)\\
  &\left\langle V_{P_1}(0) \bar{\Lambda}_{P_2}(z,\bar{z})\Lambda_{P_3}(1)V_{P_4}(\infty)\right\rangle_{g=0}\\
 &\qquad =\sum_P\Big(\widetilde{C}_b(P_4,P_3,P)C_b(P,P_2,P_1)\mathcal{F}^1_{h_P}\left[
 \begin{smallmatrix}
  *h_{P_3} & h_{P_2} \\
 h_{P_4} & h_{P_1}
 \end{smallmatrix}
  \right](z)\mathcal{F}^{\frac{1}{2}}_{h_P}\left[
 \begin{smallmatrix}
  h_{P_3} & *h_{P_2} \\
 h_{P_4} & h_{P_1}
 \end{smallmatrix}
  \right](\bar{z})\\
&\qquad-C_b(P_4,P_3,P)\widetilde{C}_b(P,P_2,P_1)\mathcal{F}^{\frac{1}{2}}_{h_P}\left[
 \begin{smallmatrix}
  *h_{P_3} & h_{P_2} \\
 h_{P_4} & h_{P_1}
 \end{smallmatrix}
  \right](z)\mathcal{F}^{1}_{h_P}\left[
 \begin{smallmatrix}
  h_{P_3} & *h_{P_2} \\
 h_{P_4} & h_{P_1}
 \end{smallmatrix}
  \right](\bar{z})\Big)\\
 \end{aligned}
\end{equation}
where various $\mathcal{F}$ are NS conformal blocks. Note that contrary to the  NS conformal blocks in \eqref{diagonalblcok}, the NS conformal blocks in \eqref{nondiagonalblock} are non-diagonal with respect to the left and right moving parts. 
 There are two types of NS conformal blocks: the one with superscript ``$1$'' (``$\frac{1}{2}$'') is called the even (odd) block, which collects contributions of all intermediate operators that lie in the integer (half-integer) level of the NS supermodule \cite{Hadasz:2006qb}. These NS conformal blocks have the following behavior near $z=0$:
\begin{equation}
 \begin{aligned}
 &\mathcal{F}_{h_P}^1\left[\begin{smallmatrix}
  _{-}h_{P_3}&_-h_{P_2}\\
  h_{P_4}&h_{P_1}
 \end{smallmatrix}\right]\sim z^{h_P-_-h_{P_2}-h_{P_1}}\z(1+\mathcal{O}(z)\y)\\
 &\mathcal{F}_{h_P}^{\frac{1}{2}}\left[\begin{smallmatrix}
  _{-}h_{P_3}&_-h_{P_2}\\
  h_{P_4}&h_{P_1}
 \end{smallmatrix}\right]\sim z^{h_P-_-h_{P_2}-h_{P_1}+\frac{1}{2}}F_{h_P}^{\frac{1}{2}}\left[\begin{smallmatrix}
  _-h_{P_3}&_-h_{P_2}\\
  h_{P_4}&h_{P_1}
 \end{smallmatrix}\right]
 \end{aligned}
\end{equation}
where
 \begin{align}
 &F_{h_P}^{\frac{1}{2}}\left[\begin{smallmatrix}
  h_{P_3}&h_{P_2}\\
  h_{P_4}&h_{P_1}
 \end{smallmatrix}\right]=\frac{1}{2h_P}\nonumber\\
 &F_{h_P}^{\frac{1}{2}}\left[\begin{smallmatrix}
  *h_{P_3}&*h_{P_2}\\
  h_{P_4}&h_{P_1}
 \end{smallmatrix}\right]=-\frac{(h_P+h_{P_3}-h_{P_4})(h_P+h_{P_2}-h_{P_1})}{2h_P}\\
 &F_{h_P}^{\frac{1}{2}}\left[\begin{smallmatrix}
  h_{P_3}&*h_{P_2}\\
  h_{P_4}&h_{P_1}
 \end{smallmatrix}\right]=-\frac{h_P+h_{P_2}-h_{P_1}}{2h_P}\nonumber\\
 &F_{h_P}^{\frac{1}{2}}\left[\begin{smallmatrix}
  *h_{P_3}&h_{P_2}\\
  h_{P_4}&h_{P_1}
 \end{smallmatrix}\right]=\frac{h_P+h_{P_3}-h_{P_4}}{2h_P}\nonumber
 \end{align}
In the above expansions, the summations are over all the possible intermediate super-primaries (labeled by $P$), which should be understood as integrals rather than discrete sums:
\begin{equation}
 \sum_P\to -i\int_{i\mathbb{R}_{\geq 0}} dP \rho^{(b)}_{NS}(P)\z(...\y)
\end{equation}

\subsection*{Calculation method}
As an illustration, we provide a detailed derivation of the discontinuities of $\mA^{(b)+}_{0,4,0}$. Let us first take a closer look at $I_{11}$, which is a product of two 4-point correlation functions. Within $I_{11}$, there are four terms, each containing one of the following: $\widetilde{C}\widetilde{C}CC$, $-\widetilde{C}\widetilde{C}\widetilde{C}\widetilde{C}$, $-CCCC$ and $CC\widetilde{C}\widetilde{C}$, where the first two factors are 3-point coefficients of the super-Liouville$^+$ theory while the last two are that of super-Liouville$^-$; the order of the momenta is always like $\widetilde{C}_{43P^+}\widetilde{C}_{P^+ 21}C_{43P^-}C_{P^- 21}$. We denote these terms as, e.g., $I_{11}^{(\widetilde{C}\widetilde{C}CC)}$ (and the same for other $I_{ab}$'s and $J_{ab}$'s), then:
\begin{equation}\label{4.10}
\begin{aligned}
 & \!\int dz^2 I^{(\widetilde{C}\widetilde{C}CC)}_{11}\!\!=\!-i\int_{i\mathbb{R}_{\geq 0}} dP_+ \rho_{NS}^{(b)}(P_+)\widetilde{C}_b(P_4,P_3,P_+)\widetilde{C}_b(P_+,P_2,P_1)\left|\mathcal{F}^1_{h^+_{P_+}}\left[\begin{smallmatrix}
 \frac{1}{2}+h^+_{P_3}&\frac{1}{2}+h^+_{P_2}\\
 h^+_{P_4}&h^+_{P_1}
  \end{smallmatrix}\right]\right|^2
\\
 & \times \int_{i\mathbb{R}} \frac{dP_-}{2i} \rho_{NS}^{(-ib)}(P_-)C_{-ib}(iP_4,iP_3,P_-)C_{-ib}(P_-,iP_2,iP_1)\left|\mathcal{F}^1_{h^-_{P_-}}\left[\begin{smallmatrix}
 \frac{1}{2}-h^+_{P_3}&\frac{1}{2}-h^+_{P_2}\\
 \frac{1}{2}-h^+_{P_4}&\frac{1}{2}-h^+_{P_1}
 \end{smallmatrix}\right]\right|^2
\end{aligned}
\end{equation}
Recall that $\widetilde{C}_b(P_+,P_2,P_1)$ has the following series of poles 
\begin{equation}\label{4.11}
 P_+=P_*=\pm P_1\pm P_2+(s+\frac{1}{2})b+(r+\frac{1}{2})b^{-1},\quad r,s \in \mathbb{Z}_{\geq 0},\quad r+s=2\mathbb{Z}+1
\end{equation}
When we deform the value of $P_1$ or $P_2$, the poles may go across the integral contour $P_+\in i\mathbb{R}$. This happens when
\begin{equation}
 \Re(P_*)<0.
\end{equation}
Similarly, for other branches of the poles of $\widetilde{C}_b(P_+,P_2,P_1)$ and $C_{-ib}(P_-,iP_1,iP_2)$, we get the following conditions
\begin{align}\label{4PoleCase}
 &\Re\z(\pm P_1\pm P_2+(r+\frac{1}{2})b+(s+\frac{1}{2})b^{-1}\y)<0\nonumber\\
 &\Re\z(\pm P_1\pm P_2-(r+\frac{1}{2})b-(s+\frac{1}{2})b^{-1}\y)<0,\quad r,s\in\mathbb{Z}_{\geq 0},\quad r+s\in 2\mathbb{Z}+1\\
 &\Im\z(\pm P_1\pm P_2-(r+\frac{1}{2})b+(s+\frac{1}{2})b^{-1}\y)>0\nonumber\\
 &\Im\z(\pm P_1\pm P_2+(r+\frac{1}{2})b-(s+\frac{1}{2})b^{-1}\y)<0,\quad r,s\in \mathbb{Z}_{\geq 0},\quad r+s\in 2\mathbb{Z}.\nonumber
\end{align} 
Thus, we have: 
\begin{equation}
 \begin{aligned}
&C_{S_2}^{(b)}\prod_{j=1}^4\mathcal{N}_b(P_j)\int_{D^2} d^2z I^{(\widetilde{C}\widetilde{C}CC)}_{11}\\
 & \qquad \supset -2\pi C_{S_2}^{(b)}\prod_{j=1}^4\mathcal{N}_b(P_j)\int_{D^2} d^2z\Big[\underset{P_+=P_*}{\Res}\rho_{NS}^{(b)}(P_+)\widetilde{C}_b(P_+,P_2,P_1)\widetilde{C}_b(P_4,P_3,P_+)\\
 &\qquad \times \int_{i\mathbb{R}} \frac{dP_-}{2i}\rho_{NS}^{(-ib)}(P_-) C_{-ib}(P_-,iP_2,iP_1)C_{-ib}(iP_4,iP_3,P_-)\\
 &\qquad \times |z|^{2-P_-^2-P_*^2}\z(1+\mathcal{O}(z,\bar{z})\y)
 \end{aligned}
\end{equation}
Here $D^2$ is a disk with radius $R$ centered around $z=0$. The integral over  $z$ is convergent for $\Re(P_*^2)<0$. This allows us to first carry out the integral in the regime $\Re(P_*^2)<0$, then extend the solution to $\Re(P_*^2)>0$ via analytical continuation. For the $z$-integral over  the domain $D^2$, we have the following formula:
\begin{equation}
 \int_{D^2} d^2z |z|^{-2-P_*^2-P_-^2}=-\frac{2\pi R^{-P_-^2-P_*^2} }{P_-^2+P_*^2}
\end{equation}
Note that the radius $R$ will not affect  the final result. Because we have assumed $\Re(P_*)<0, \Re(P_*^2)<0$, the poles of $P_-=\pm iP_*$ do not lie on the integral contour of $P_-$. We  can analytic continue $P_1, P_2$  to the region $\Re(P_*)<0, \Re(P_*^2)>0$, which depends on  whether  $P_*$ is continued into this region from above or below. The difference comes from a residue contribution associated with the pole at  $P_-=\pm i P_*$. Thus, we have the following discontinuities of the 4-point amplitude at $P_*=0$:
\begin{equation}\label{4.16}
 \begin{aligned}
 &\underset{P_*=0}{\text{Disc}}\z(C_{S_2}^{(b)}\prod_{j=1}^4\mathcal{N}_b(P_j)\int_{D^2} d^2z I_{11}^{(\widetilde{C}\widetilde{C}CC)}\y)=-4\pi^3 iC^{(b)}_{S_2}\prod_{j=1}^{4}\mathcal{N}_{b}(P_j)\frac{\rho_{NS}^{(b)}(P_*)\rho_{NS}^{(-ib)}(iP_*)}{P_*}\\
 &\times\underset{P=P_*}{\Res}\z(\widetilde{C}_b(P,P_2,P_1)\widetilde{C}_b(P_4,P_3,P)\y)C_{-ib}(iP_*,iP_2,iP_1)C_{-ib}(iP_4,iP_3,iP_*)\\
 &=-4\pi^3 i\frac{\rho_{NS}^{(b)}(P_*)\rho_{NS}^{(-ib)}(iP_*)}{C_{S_2}^{(b)}\mathcal{N}_b(P_*)\mathcal{N}_b(P_*)P_*}\underset{P=P_*}{\Res}\z(\mA^-_{0,3,0}(P_4,P_3,P)\mA^-_{0,3,0}(P,P_2,P_1)\y)\\
 &=8\pi iP_*\underset{P=P_*}{\Res}\z(\mA^-_{0,3,0}(P_4,P_3,P)\mA^-_{0,3,0}(P,P_2,P_1)\y)
 \end{aligned}
\end{equation}
In fact, this is the final answer to the discontinuities of $\mA^{(b)+}_{0,4,0}$ at $P_*=0$, as contributions from other terms cancel with each other. For example, the poles $P_+=P_*$ play a similar role (as above) in the calculation of discontinuities from $I_{11}^{\widetilde{C}\widetilde{C}\widetilde{C}\widetilde{C}}$:
 \begin{align}
 &\underset{P_*= \pm 1 }{\text{Disc}}\z(C_{S_2}^{(b)}\prod_{j=1}^4\mathcal{N}_b(P_j)\int_{D^2} d^2z I_{11}^{(\widetilde{C}\widetilde{C}\widetilde{C}\widetilde{C})}\y)\nonumber\\
 &=\frac{1}{4h^2_{P_*}}4\pi^3 iC^{(b)}_{S_2}\prod_{j=1}^{4}\mathcal{N}_{b}(P_j)\frac{\rho_{NS}^{(b)}(P_*)\rho_{NS}^{(-ib)}(i\sqrt{P_*^2-1})}{\sqrt{P_*^2-1}}\\
 &\times\underset{P=P_*}{\Res}\z(\widetilde{C}_b(P,P_2,P_1)\widetilde{C}_b(P_4,P_3,P)\y)\widetilde{C}_{-ib}(i\sqrt{P_*^2-1},iP_2,iP_1)\widetilde{C}_{-ib}(iP_4,iP_3,i\sqrt{P_*^2-1})\nonumber
 \end{align}
The above term could be canceled by the discontinuities from $J_{11}^{\widetilde{C}\widetilde{C}\widetilde{C}\widetilde{C}}$, which takes the following form
 \begin{align}
 &\underset{P_*=\pm1}{\text{Disc}}\z(\!C_{S_2}^{(b)}\prod_{j=1}^4\mathcal{N}_b(P_j)\int_{D^2} d^2z J_{11}^{\widetilde{C}\widetilde{C}\widetilde{C}\widetilde{C}}\!\y)\nonumber\\
 &= -\frac{1}{4h^2_{P_*}}4\pi^3 iC^{(b)}_{S_2}\prod_{j=1}^{4}\mathcal{N}_{b}(P_j)\frac{\rho_{NS}^{(b)}(P_*)\rho_{NS}^{(-ib)}(i\sqrt{P_*^2-1})}{\sqrt{P_*^2-1}}\\
 &\times\underset{P=P_*}{\Res}\z(\widetilde{C}_b(P,P_2,P_1)\widetilde{C}_b(P_4,P_3,P)\y)\widetilde{C}_{-ib}(i\sqrt{P_*^2-1},iP_2,iP_1)\widetilde{C}_{-ib}(iP_4,iP_3,i\sqrt{P_*^2-1})\nonumber\\
 &=-\underset{P_*=0}{\text{Disc}}\z(C_{S_2}^{(b)}\prod_{j=1}^4\mathcal{N}_b(P_j)\int_{D^2} d^2z I_{11}^{\widetilde{C}\widetilde{C}\widetilde{C}\widetilde{C}}\y)\nonumber.
 \end{align}
Similar calculation shows that discontinuities from $J_{11}^{CCCC}$ and $I_{11}^{CCCC}$ cancel with each other. As for $I_{11}^{ C C \widetilde{C}\widetilde{C}}$, the cancellation turns out to be more complicated: it will be canceled by $I^{ C C \widetilde{C}\widetilde{C}}_{22}$, $J^{ C C \widetilde{C}\widetilde{C}}_{11}$, and $J^{ C C \widetilde{C}\widetilde{C}}_{22}$.  First, note that the leading order of the $z$-expansion of  $J_{11}^{CC\widetilde{C}\widetilde{ C}}$ and $J_{11}^{\widetilde{C}\widetilde{C}CC}$ are not norms of $z$. 
When focusing on discontinuities, we need to integrate $z$ on a neighborhood of $z=0$, and only norms could contribute. So at the leading order of $z$, the above term will not contribute to the discontinuities of the amplitude. However, the next order of the $z$-expansion will be norms, and they help to cancel the discontinuities from $I_{11}^{CC\widetilde{C}\widetilde{C}}$. Furthermore, as argued in \cite{Collier:2024kwt}, the higher-order contribution comes from exchanging the super-Virasoro descendants of the worldsheet SCFT, and they are trivial in the BRST cohomology.  One can explicitly check that these terms do not contribute due to a cancellation from the expansion of conformal blocks.


\subsection*{Summary of the discontinuities of $\mA_{0,4,0}^{(b)}$ and $\hmA_{0,4,0}^{(b)}$}
Now, let us summarize the discontinuities of $\mA_{0,4,0}^{(b)}$ and $\hmA_{0,4,0}^{(b)}$. For $\mA_{0,4,0}^{(b)+}$ we have (the second line is just \eqref{4.16}) 
\begin{equation}
 \begin{aligned}
  \underset{P^{e}_*=0}{\text{Disc}}\z(\mA^+_{0,4,0}(P_1,P_2,P_3,P_4)\y)&=8\pi i P_*^e\underset{P=P^{e}_*}{\Res}\z(\mA^+_{0,3,0}(P_4,P_3,P)\mA^+_{0,3,0}(P,P_2,P_1)\y)\\
  \underset{P^{o}_*=0}{\text{Disc}}\z(\mA^+_{0,4,0}(P_1,P_2,P_3,P_4)\y)&=8\pi i P_*^o\underset{P=P^{o}_*}{\Res}\z(\mA^-_{0,3,0}(P_4,P_3,P)\mA^-_{0,3,0}(P,P_2,P_1)\y)\\
 \end{aligned}
\end{equation}
For $\mA_{0,4,0}^{(b)-}$ we have 
\begin{equation}
 \begin{aligned}
  \underset{P^{e}_*=0}{\text{Disc}}\z(\mA^-_{0,4,0}(P_1,P_2,P_3,P_4)\y)&=8\pi i P_*^e\underset{P=P^{e}_*}{\Res}\z(\mA^-_{0,3,0}(P_4,P_3,P)\mA^+_{0,3,0}(P,P_2,P_1)\y)\\
  \underset{P^{o}_*=0}{\text{Disc}}\z(\mA^-_{0,4,0}(P_1,P_2,P_3,P_4)\y)&=8\pi i P_*^o\underset{P=P^{o}_*}{\Res}\z(\mA^+_{0,3,0}(P_4,P_3,P)\mA^-_{0,3,0}(P,P_2,P_1)\y)\\
 \end{aligned}
\end{equation}
where 
\begin{equation}
 \begin{aligned}
  &P_*^e=\pm P_1\pm P_2+(r+\frac{1}{2})b+(s+\frac{1}{2})b^{-1},\quad r,s\in \mathbb{Z},\quad r+s\in 2\mathbb{Z}\\
  &P_*^o=\pm P_1\pm P_2+(r+\frac{1}{2})b+(s+\frac{1}{2})b^{-1},\quad r,s\in \mathbb{Z},\quad r+s\in2\mathbb{Z}+1
 \end{aligned}
\end{equation}
Note that similar equations hold for any pair of external momenta  $p_i,p_j$. Then from \eqref{4pt+-}, we get:
\begin{equation}\label{DiscA}
\begin{aligned}
 & \underset{P_*=0}{\text{Disc}}\z(\mA^{(b)}_{0,4,0}(P_1,P_2,P_3,P_4)\y)=8\pi i P_*\underset{P=P_*}{\Res}\z(\mA^{(b)}_{0,3,0}(P_4,P_3,P)\mA^{(b)}_{0,3,0}(P,P_2,P_1)\y)\\
 & \underset{P_*=0}{\text{Disc}}\z(\hmA^{(b)}_{0,4,0}(P_1,P_2,P_3,P_4)\y)=8\pi i P_*\underset{P=P_*}{\Res}\z(\hmA^{(b)}_{0,3,0}(P_4,P_3,P)\hmA^{(b)}_{0,3,0}(P,P_2,P_1)\y)
\end{aligned}
\end{equation}
where 
\begin{equation}
 P_*=\pm P_1\pm P_2+(r+\frac{1}{2})b+(s+\frac{1}{2})b^{-1},\quad r,s\in\mathbb{Z}
\end{equation}
The discontinuities of $\mA^{(b)}_{0,4,0}$ and $\hmA^{(b)}_{0,4,0}$ have the same form as $\mathbb{C}\text{LS}$.

\subsection*{General formula}
Our derivation for the NS-NS-NS-NS sphere 4-point amplitudes—where discontinuities are governed by the factorization of 3-point amplitudes—suggests a universal pattern. We therefore conjecture that this relation encapsulates a general factorization property of the theory's analytic structure, extending to higher-point and higher-genus amplitudes.
\begin{equation}
 \begin{aligned}
  \underset{P_*=0}{\text{Disc }} \mA_{g,n,0}^{(b)}(\boldsymbol{P})=2\pi i \Big[\underset{P=P_*}{\Res}\sum_{h\in [0,g],I\subset\{1,...,n\} \text{stable}}&2P\mA^{(b)}_{h,|I|+1,0}(\boldsymbol{P}_I,P)\mA^{(b)}_{h,|I^c|+1,0}(\boldsymbol{P}_{I^c},P)\\
  &+\underset{P=\frac{1}{2}P_*}{\Res}2P_*\mA_{g-1,n+2,0}(\boldsymbol{P},P,P)\Big]\\
    \underset{P_*=0}{\text{Disc }} \hmA_{g,n,0}^{(b)}(\boldsymbol{P})=2\pi i \Big[\underset{P=P_*}{\Res}\sum_{h\in [0,g],I\subset\{1,...,n\} \text{stable}}&2P\hmA^{(b)}_{h,|I|+1,0}(\boldsymbol{P}_I,P)\hmA^{(b)}_{h,|I^c|+1,0}(\boldsymbol{P}_{I^c},P)\\
  &+\underset{P=\frac{1}{2}P_*}{\Res}2P_*\hmA_{g-1,n+2,0}(\boldsymbol{P},P,P)\Big]
 \end{aligned}
\end{equation}

\subsection{Poles of $\mA^{(b)}_{0,4,0}$ and $\hmA^{(b)}_{0,4,0}$}

Regarding the poles of the sphere 4-point amplitudes $\mA^{(b)}_{0,4,0}$ and $\hmA^{(b)}_{0,4,0}$, the corresponding analysis can in fact be straightforwardly generalized to arbitrary general amplitudes. Thus, we start by presenting the poles of general correlation functions in SLFT, defined on an arbitrary Riemann surface.

\subsection*{Poles of correlation functions on $\mathcal{M}_{g,n}$}
 We employ‌ the Coulomb gas formalism to analyse $n$-point correlation functions in SLFT on a genus $g$ Riemann surface $\mathcal{M}_{g,n}$.  To use the Coulomb gas formalism, we can make an expansion of the Lagrangian\footnote{Note that there is a contact term $:e^{b\phi}::e^{b\phi}:$ in the Lagrangian of SLFT \eqref{LSuperL}. However, in practice, we can neglect this contact term \cite{Nakayama:2004vk}.}
\begin{equation}
 \z\langle\prod_{i=1}^{n_V}V_{P_i}(z_i)\prod_{j=1}^{n_W}W_{P_j}(z_j)\prod_{k=1}^{n_\Lambda} \Lambda_{P_k}(z_k)\prod_{t=1}^{n_{\bar{\Lambda}}}\bar{\Lambda}_{P_t}(z_t) \sum_{n=0}^\infty\z(\int d^2z2i\mu b^2 \psi\bar{\psi}e^{b\phi}\y)^n\y\rangle_{free}
\end{equation}
The delta function representing the momentum conservation  is supported at
\begin{equation}
 \sum_{i=1}^{n_V}\alpha_i+\sum_{j=1}^{n_W}\alpha_j+\sum_{k=1}^{n_\Lambda}\alpha_k+\sum_{t=1}^{n_{\bar{\Lambda}}}\alpha_t+rb=(1-g)Q
\end{equation}
where $n_V, n_W, n_{\Lambda}, n_{\bar{\Lambda}}$ are the numbers of $V_P, W_P, \Lambda_P, \bar{\Lambda}_P$, respectively.
In SLFT,  this delta  function is smoothed into a pole. Owing to the $b\to b^{-1}$ duality,  we can incorporate a $sb^{-1}$ term into the above equation, which consequently becomes:
\begin{equation}\label{CFTpole}
 \pm P_1\pm P_2\pm...\pm P_n =\z(\frac{2g-2+n}{2}Q+rb+sb^{-1}\y),\quad r,s \in\mathbb{Z}_{\geq 0}
\end{equation}
 where $n$ is the total number of vertex,  $n\equiv n_V+n_W+n_\Lambda+n_{\bar{\Lambda}}$. 
Because of supersymmetry, the total number of $\Lambda_P$ and $\bar{\Lambda}_P$ should be even, which means $n_\Lambda+n_{\bar{\Lambda}}\in 2\mathbb{Z}$. We also know there should be an even number of $\psi$ and $\bar{\psi}$ so that the correlation function is not zero, so we have 
\begin{equation}
 \begin{aligned}
 &r+s\in 2\mathbb{Z},\quad \text{if }n_W+n_\Lambda\in 2\mathbb{Z}\\
 &r+s\in 2\mathbb{Z}+1,\quad \text{if }n_W+n_\Lambda\in 2\mathbb{Z}+1
 \end{aligned}
\end{equation}

\subsection*{Poles of $\mA^{(b)}_{0,4,0}$ and $\hmA^{(b)}_{0,4,0}$}
For $\mA_{0,4,0}^{(b)+}$, the poles are at 
\begin{equation}\label{polep+}
 P_1\pm P_2\pm P_3\pm P_4=rb+sb^{-1},\quad r,s\in \mathbb{Z}_{\neq 0},\quad r+s\in 2\mathbb{Z}
\end{equation}
For $\mA_{0,4,0}^{(b)-}$, the poles are at 
\begin{equation}\label{polep-}
 P_1\pm P_2\pm P_3\pm P_4=rb+sb^{-1},\quad r,s\in \mathbb{Z}_{\neq 0},\quad r+s\in 2\mathbb{Z}+1
\end{equation}
So  for $\mA_{0,4,0}^{(b)}$ and $\hmA_{0,4,0}^{(b)}$, their poles are at
\begin{equation}
 P_1\pm P_2\pm P_3\pm P_4=rb+sb^{-1},\quad r,s\in \mathbb{Z}_{\neq 0}
\end{equation}
This is the same position
\subsection*{Poles of $\mA^{(b)}_{g,n_{NS},0}$ and $\hmA^{(b)}_{g,n_{NS},0}$}
For general amplitudes $\mA_{g,n_{NS},0}^{(b)}$ and $\mA_{g,n_{NS},0}^{(b)}$, both series of poles in \eqref{CFTpole} will contribute. Finally, the poles of the amplitudes are
\begin{equation}
  \pm P_1\pm P_2\pm...\pm P_n =\z(\frac{2g-2+n}{2}Q+rb+sb^{-1}\y),\quad r,s \in\mathbb{Z}_{\geq 0}
\end{equation}

\section{Ground ring and higher equations of motion}\label{GRHEOM}
Similar to the bosonic case, we use higher equations of motion (HEOM) of super-Liouville field theory \cite{Belavin:2006pv} to evaluate special sphere 4-point 
amplitudes $\mA_{0,4,0}^{(b)}$ and $\hmA_{0,4,0}^{(b)}$. The strategy is as follows (similar to the bosonic case \cite{Collier:2024kwt}):
\begin{itemize}
 \item Set one external vertex operator to a special value.
 \item Write this operator as acting BRST operators $\mathcal Q, \bar{\mathcal Q}$ on a logarithmic operator, which is the logarithmic counterpart of a ground ring operator. This is the (cohomological version of)  higher equations of motion (HEOM).
 \item Use the HEOM to write the moduli integration localizes to the boundary of moduli space. The contributions from the boundary terms can be read from the OPE between the logarithmic operator and the other three operators.
\end{itemize}
We will first review the higher equations of motion in super-Liouville field theory. Then we review the construction of the ground ring and give the higher equations of motion in the string theory (a cohomological version of HEOM). Finally, we use these results to compute the 4-point amplitude.

\subsection{HEOM in SLFTs}
Let us start by reviewing the higher equations of motion in $\mathcal{N}=1$ super-Liouville theory \cite{Belavin:2006pv}. Consider the following operators in SLFT
\begin{equation}
 V'_P=\frac{\partial}{\partial P}V_P
\end{equation}
They are not primary operators, but are logarithmic operators, that is, $L_0$ does not act diagonally on them (the states corresponding to $V'_P$ and $V_P$ are denoted as $|P\rangle'$ and $|P\rangle$ respectively):
\begin{equation}
\begin{aligned}
 L_0|P\rangle&=\frac{1}{8}(Q^2-4P^2)|P\rangle\\
 L_0|P\rangle'&=\frac{\partial}{\partial P}\left(\frac{1}{8}(Q^2-4P^2)|P\rangle\right)=\frac{1}{8}(Q^2-4P^2)|P\rangle'-P|P\rangle
\end{aligned}
\end{equation}
It is easy to see that $L_0$ acts on $(|P\rangle,|P\rangle')$ as a rank 2 Jordan block.

To describe the higher equations of motion, we focus on the logarithmic fields $V'_P$ with $P$ taking  degenerate values. Firstly, recall that  for $P=\frac{mb}{2}+\frac{n}{2b}$ $(m+n\in2\mathbb{Z})$, $V_P$ has a level $\frac{mn}{2}$ null vector. This means that there exists an operator $D_{m,n}$ made out of super-Virasoro modes that annihilates the primary $V_{P=\frac{mb}{2}+\frac{n}{2b}}$:
\begin{equation}
 D_{m,n}V_{P=\frac{mb}{2}+\frac{n}{2b}}=0
\end{equation}
Note that we focus on the null states in the NS sector so $m+n\in2\mathbb{Z}$ (for null states in the R sector, $m+n\in2\mathbb{Z}+1$). The factor $D_{m,n}$ is unique up to an overall normalization, which is fixed by demanding that the term $G_{-\frac{1}{2}}^{mn}$ in $D_{m,n}$ has a unit coefficient. We give the explicit forms of the first several $D_{m,n}$ (note that $G_{-\frac{1}{2}}^2=L_{-1}$):
\begin{equation}
 D_{1,1}=G_{-\frac{1}{2}}, \quad D_{3,1}=L_{-1}G_{-\frac{1}{2}}+b^2G_{-\frac{3}{2}}, \quad D_{2,2}=L_{-1}^2+\frac{(1+b^2)^2}{2b^2}L_{-2}-G_{-\frac{3}{2}}G_{-\frac{1}{2}}
\end{equation}
We can similarly construct the operators $\bar{D}_{m,n}$ for the right-moving modes. Then the higher equations of motion are:
\begin{equation}\label{HEOM}
 D_{m,n}\bar{D}_{m,n}V'_{P=\frac{mb}{2}+\frac{n}{2b}}=B_{m,n}V_{P=\frac{mb}{2}-\frac{n}{2b}}
\end{equation}
where the factor $B_{m,n}$ is determined in \cite{Belavin:2006pv} (see appendix \ref{Bmn}).

\subsection{Ground ring and cohomological HEOM}
Now we want to use the HEOM in SLFT to obtain the HEOM in the super complex Liouville string theory, that is, to obtain a cohomological version of the HEOM. For this, we need to introduce an infinite series of BRST closed vertex operators at ghost number
zero – the ground ring operators. In the super complex Liouville string, they take the following form\footnote{Notice that our normalization of the ground ring operator takes nearly the same form as the bosonic one \cite{Collier:2024kwt}, differing only by a sign.}:
\begin{equation}\label{ground ring normalization}
 \mathcal{O}_{m,n}\equiv \frac{\pi\mathcal{N}_b(\frac{mb}{2}-\frac{n}{2b})}{B_{m,n}}H_{m,n}\bar{H}_{m,n}V^+_{P=\frac{mb}{2}+\frac{n}{2b}}V^-_{P=i(\frac{mb}{2}-\frac{n}{2b})}
\end{equation}
where the operators $H_{m,n}$ are constructed
from  the super-Virasoro generators 
of the two SLFTs and generators of the ghosts, and their action yields descendants of level $\frac{mn-1}{2}$. To fix the normalization of the ground ring operators, we normalize $H_{m,n}$ as:
\begin{equation}
 H_{m,n}=(L_{-1}^+)^{\frac{mn-1}{2}}+...
\end{equation}
For example, the simplest non-trivial case is $H_{3,1}$:
\begin{equation}\label{H31}
 H_{3,1}=L_{-1}^+-L_{-1}^-+G_{-\frac{1}{2}}^-G_{-\frac{1}{2}}^+-b^2\beta_{-\frac{3}{2}}\gamma_{\frac{1}{2}}-2b^2\mathfrak{b}_{-2}c_1+b^2\left(G_{-\frac{1}{2}}^-+G_{-\frac{1}{2}}^+\right)\beta_{-\frac{3}{2}}c_1
\end{equation}
With these normalizations, one can check that the OPEs of the ground ring operators are:
\begin{equation}\label{groundringOPE}
 \mathcal{O}_{m,n}(z)\mathcal{O}_{m',n'}(0)= \sum_{m''\overset{2}{=}|m-m'|+1}^{m+m'-1}\sum_{n''\overset{2}{=}|n-n'|+1}^{n+n'-1}\mathcal{O}_{m'',n''}(0)+\text{BRST exact}.
\end{equation}
In the above, the  OPE coefficient  is exactly $1$.
Importantly,  ordinary physical vertex operators $\mathscr{V}_P$ form modules of the ground ring operators. Thus, up to  BRST exact terms, their OPEs take the following form:
\begin{equation}\label{groundringV}
\begin{aligned}
 \mathcal{O}_{m,n}(z)\mathscr{V}_P^{(-1)}(0)&=\sum_{r\overset{2}{=}1-m}^{m-1}\sum_{s\overset{2}{=}1-n}^{n-1}\mathscr{V}^{(-1)}_{P+\frac{rb}{2}+\frac{s}{2b}}(0)+\text{BRST exact}\\
 \mathcal{O}_{m,n}(z)\mathscr{V}_P^{(0)}(0)&=\sum_{r\overset{2}{=}1-m}^{m-1}\sum_{s\overset{2}{=}1-n}^{n-1}\mathscr{V}^{(0)}_{P+\frac{rb}{2}+\frac{s}{2b}}(0)+\text{BRST exact}
\end{aligned}
\end{equation}
One can check that with the normalization \eqref{ground ring normalization}, the structure constants above are exactly  $1$ (see appendix \ref{Thelegfactor}).

Now we are ready to write the cohomological HEOM.
Since there are two SLFTs on the worldsheet, we can introduce two logarithmic counterparts of $\mathcal{O}_{m,n}$, defined as:
\begin{equation}
\begin{aligned}
  \mathcal{O}_{m,n}'^+\equiv \frac{\pi\mathcal{N}_b(\frac{mb}{2}-\frac{n}{2b})}{B_{m,n}}H_{m,n}\bar{H}_{m,n}V'^+_{P=\frac{mb}{2}+\frac{n}{2b}}V^-_{P=i(\frac{mb}{2}-\frac{n}{2b})}\\
  \mathcal{O}_{m,n}'^-\equiv \frac{\pi\mathcal{N}_b(\frac{mb}{2}-\frac{n}{2b})}{B_{m,n}}H_{m,n}\bar{H}_{m,n}V^+_{P=\frac{mb}{2}+\frac{n}{2b}}V'^-_{P=i(\frac{mb}{2}-\frac{n}{2b})}
\end{aligned}
\end{equation}
Then the equation of motion  \eqref{HEOM} enables the derivation of its cohomological version:
\begin{equation}\label{coHEOM}
\begin{aligned}
  \mathcal{Q}\bar{\mathcal Q}\mathcal{O}_{m,n}'^+&=-\pi\mathscr{V}^{(0)}_{P=\frac{mb}{2}-\frac{n}{2b}}+\text{BRST exact terms}\\
  \mathcal{Q}\bar{\mathcal{Q}}\mathcal{O}_{m,n}'^-&=-\pi i\mathscr{V}^{(0)}_{P=\frac{mb}{2}+\frac{n}{2b}}+\text{BRST exact terms}
\end{aligned}
\end{equation}
where $\mathcal{Q},\bar{\mathcal{Q}}$ are BRST charges and the physical vertex operators on the right-hand side are the ones in picture $0$.

The last essential ingredient for computing string amplitudes is the OPE of $\mathcal{O}_{m,n}'^\pm$ with ordinary physical vertex operators  $\mathscr{V}_P$. Note that since we already know the OPE of $\mathcal{O}_{m,n}^\pm$ with $\mathscr{V}_P$ in \eqref{groundringV},  little work remains to obtain the OPE of 
 $\mathcal{O}_{m,n}'^\pm$ with $\mathscr{V}_P$. Since $\mathcal{O}_{m,n}'^\pm$ is a logarithmic
field, there will be, in particular, a logarithmic term in the OPE, which is the only term relevant to our computation of string amplitudes. Firstly, we have:
\begin{equation}
 \mathcal{Q}\bar{\mathcal{Q}}( \mathcal{O}'^+_{m,n}(z)\mathscr{V}^{(*)}_P(0))=-\pi\mathscr{V}^{(0)}_{P=\frac{mb}{2}-\frac{n}{2b}}(z) \mathscr{V}_P^{(*)}(0),\qquad *=-1,0
\end{equation}
where we have used the HEOM \eqref{coHEOM}. As the right-hand side of the above equation does not contain a logarithmic term, the logarithmic term appearing
in the $\mathcal{O}'^+_{m,n}(z)\mathscr{V}_P(0)$ OPE has to be BRST closed. Considering the degenerate super-Virasoro fusion rules, the only possibility is
\begin{equation}
\begin{aligned}
    \mathcal{O}'^+_{m,n}(z)\mathscr{V}^{(*)}_P(0)=\sum_{r\overset{2}{=}1-m}^{m-1}\sum_{s\overset{2}{=}1-n}^{n-1}&C_{m,n}^{r,s}(P)\mathscr{V}^{(*)}_{P+\frac{rb}{2}+\frac{s}{2b}}(0)\text{log}|z|^2\\
    &+\text{non-logarithmic terms $\&$ BRST exact terms}
\end{aligned}
\end{equation}
where $*=-1,0$ and $C_{m,n}^{r,s}(P)$ is the (to be determined) structure constant. The logarithmic
term can only appear when the $\frac{\partial}{\partial P}$ derivative in the definition of the logarithmic
ground ring operators hit the exponent of the position dependence $|z|$ of the
non-logarithmic OPE. Since the structure constant in the OPE \eqref{groundringV} is $1$, $C_{m,n}^{r,s}(P)$ is given by the derivative of the
$z$-exponent. Thus, we extract  a discrete term in the OPE of the primary fields, which gives rise to a $z$-dependence
\begin{equation}
 |z|^{-2\left(\frac{1}{8}Q^2-\frac{1}{2}P'^2\right)-2\left(\frac{1}{8}Q^2-\frac{1}{2}P^2\right)+2\left(\frac{1}{8}Q^2-\frac{1}{2}(P-P'+\frac{1}{2}(r+m)b+\frac{1}{2}(s+n)b^{-1})^2\right)}
\end{equation}
where $P'$ is the momentum of the super-Liouville vertex operator before specifying to $\frac{mb}{2}+\frac{n}{2b}$. Note that the operator $H_{m,n},\bar{H}_{m,n}$ leads to further $z$-dependence, but it is independent of $P'$. We thus simply have to
take a $P'$ derivative and then put $P'=\frac{mb}{2}+\frac{n}{2b}$, which leads to
\begin{equation}
 C_{m,n}^{r,s}(P)=P+\frac{rb}{2}+\frac{s}{2b}+\frac{mb}{2}+\frac{n}{2b}
\end{equation}
However, there exists a second discrete term with momentum $P+P'+\frac{1}{2}(r-m)b+\frac{1}{2}(s-n)b^{-1}$ that could also have contributed to the logarithm. As explained in the bosonic case \cite{Collier:2024kwt}, to decide which one is the correct answer (and get an unambiguous answer), we
must require that the discrete term appears indeed in the OPE and is more singular than the
continuum.  Then the final unambiguous answer turns out to be:
\begin{equation}\label{Crsmn}
 C_{m,n}^{r,s}(P)=-\sqrt{\left(P+\frac{rb}{2}+\frac{s}{2b}\right)^2}+\frac{mb}{2}+\frac{n}{2b}
\end{equation}
provided that $\text{Re}(p+\frac{rb}{2}+\frac{s}{2b})^2>0$. For $\text{Re}(p+\frac{rb}{2}+\frac{s}{2b})^2<0$, the answer can be analytically continued, but the branch of the
square root is then ambiguous.

\subsection{4-point amplitude with one degenerate external operator}
We can now compute the 4-point amplitude, where one of the external operators is  $\mathscr{V}^{(0)}_{P=\frac{mb}{2}-\frac{n}{2b}}$ (which is  in picture $0$):
\begin{equation}
\begin{aligned}
  \mathsf{A}^{(b)}_{0,4,0}&\left(P_1,P_2,P_3,P_4=\frac{mb}{2}-\frac{n}{2b}\right)\\
  &=-C_{S^2}^{(b)}\int d^2z\left\langle \mathscr{V}^{(-1)}_{P_1}(z_1)\mathscr{V}^{(-1)}_{P_2}(z_2)\mathscr{V}^{(0)}_{P_3}(z_3)b_{-1}\bar{b}_{-1}\mathscr{V}^{(0)}_{P_4=\frac{mb}{2}-\frac{n}{2b}}(z)\right\rangle\\
  &=\frac{1}{\pi}C_{S^2}^{(b)}\int d^2z\left\langle \mathscr{V}^{(-1)}_{P_1}(z_1)\mathscr{V}^{(-1)}_{P_2}(z_2)\mathscr{V}^{(0)}_{P_3}(z_3)b_{-1}\bar{b}_{-1} \mc{Q}\bar{\mc{Q}}\mathcal{O}_{m,n}'^+(z)\right\rangle\\
  &=-\frac{1}{\pi}C_{S^2}^{(b)}\int d^2z\left\langle \mathscr{V}^{(-1)}_{P_1}(z_1)\mathscr{V}^{(-1)}_{P_2}(z_2)\mathscr{V}^{(0)}_{P_3}(z_3)\partial\bar{\partial}\mathcal{O}_{m,n}'^+(z)\right\rangle
\end{aligned}
\end{equation}
where in the second equality, we have used the cohomological HEOM \eqref{coHEOM}.
Then, the integrand is at this point
a total derivative and the moduli integral reduces to boundary contributions from $z_i$ $(i=1,2,3)$ and $\infty$ (since the higher equations of motion only hold with a flat
background metric, we also get a boundary contribution from $\infty$).

We first calculate the contribution from $z_i$. As in the bosonic case, for a logarithmic dependence $f\sim$ log $r^2$, we get $-\pi$ times the prefactor of the logarithm. So the boundary contribution from
 $z_1$ is:
\begin{equation}
\begin{aligned}
  &C_{S^2}^{(b)}\sum_{r\overset{2}{=}1-m}^{m-1}\sum_{s\overset{2}{=}1-n}^{n-1}C_{m,n}^{r,s}(P_1)\left\langle \mathscr{V}^{(-1)}_{P_1+\frac{rb}{2}+\frac{s}{2b}}(z_1)\mathscr{V}^{(-1)}_{P_2}(z_2)\mathscr{V}^{(0)}_{P_3}(z_3)\right\rangle\\
  &=\z(\sum_{r\overset{2}{=}1-m}^{m-1}\sum_{s\overset{2}{=}1-n}^{n-1}C^{r,s}_{m,n}(P_1)\y)\mathsf{A}^{(b)}_{0,3,0}\left(P_1+\frac{(m-1)b}{2}+\frac{n-1}{2b},P_2,P_3\right),
\end{aligned}
\end{equation}
where we have used the double-periodicity of the three-point function $\mathsf{A}^{(b)}_{0,3,0}$.
The contributions from $z_2,z_3$ are similar. For the contribution from $\infty$, we have:
\begin{equation}
 \mathcal{O}'^+_{m,n}(z)\sim (mb+nb^{-1})\mathcal{O}_{m,n}(z)\text{log $|z|^2$}
\end{equation}
as $z\to \infty$. Then the
contribution at infinity is:
\begin{equation}
\begin{aligned}
 &-C_{S^2}^{(b)}(mb+nb^{-1})\left\langle \mathscr{V}^{(-1)}_{P_1}(z_1)\mathscr{V}^{(-1)}_{P_2}(z_2)\mathscr{V}^{(0)}_{P_3}(z_3)\mathcal{O}_{m,n}(\infty)\right\rangle\\
 =&-(mb+nb^{-1})mn\mathsf{A}^{(b)}_{0,3,0}\left(P_1+\frac{(m-1)b}{2}+\frac{n-1}{2b},P_2,P_3\right)
\end{aligned}
\end{equation}
Here we used again double-periodicity of the three-point function $\mathsf{A}^{(b)}_{0,3,0}$.
Finally, we get
\begin{equation}\label{HEOMmA}
\begin{aligned}
 \mathsf{A}^{(b)}_{0,4,0}&\left(P_1,P_2,P_3,P_4=\frac{mb}{2}-\frac{n}{2b}\right)
 =\mathsf{A}^{(b)}_{0,3,0}\left(P_1+\frac{(m-1)b}{2}+\frac{n-1}{2b},P_2,P_3\right)\\&\qquad\times\left[\frac{(mb+nb^{-1})mn}{2}-\sum_{i=1}^3\sum_{r\overset{2}{=}1-m}^{m-1}\sum_{s\overset{2}{=}1-n}^{n-1}\sqrt{\left(P_i+\frac{rb}{2}+\frac{s}{2b}\right)^2}\right]
\end{aligned}
\end{equation}

The upshot of the analysis of HEOM is that the 4-point amplitude with one degenerate external operator reduces the same way as in the bosonic case to the 3-point amplitude \cite{Collier:2024kwt} (with the degenerate external operator removed).

\subsection{Ground ring and HEOM of $\hSCLS$}
Now we derive the ground ring and HEOM of $\hSCLS$. The ground ring operators are defined analogously to \eqref{ground ring normalization}:
\begin{equation}
 \widehat{\mathcal{O}}_{m,n}\equiv \frac{\pi\mathcal{N}_b(\frac{mb}{2}-\frac{n}{2b})}{B_{m,n}}\widehat{H}_{m,n}\widehat{\bar{H}}_{m,n}V^+_{P=\frac{mb}{2}+\frac{n}{2b}}V^-_{P=i(\frac{mb}{2}-\frac{n}{2b})},
\end{equation}
where $\widehat{H}_{m,n}$  has the same form as  $H_{m,n}$, but with $G_{m}^{+}\to -G_{m}^{+}$. Consequently, the OPEs of the ground ring operators become 
\begin{equation}\label{hgroundringOPE}
\begin{aligned}
     \widehat{\mathcal{O}}_{m,n}(z)\widehat{\mathcal{O}}_{m',n'}(0)= \sum_{m''\overset{2}{=}|m-m'|+1}^{m+m'-1}\sum_{n''\overset{2}{=}|n-n'|+1}^{n+n'-1}&(-1)^{\frac{m+n+m'+n'+m''+n''-2}{2}}\widehat{\mathcal{O}}_{m'',n''}(0)\\
     &+\text{BRST exact}.
\end{aligned}
\end{equation}
The ordinary physical vertex operators $\widehat{\mathscr{V}_P}$ form modules of the ground ring operators and their OPEs are:
\begin{equation}
\begin{aligned}
 \widehat{\mathcal{O}}_{m,n}(z)\widehat{\mathscr{V}}_P^{(-1)}(0)&=\sum_{r\overset{2}{=}1-m}^{m-1}\sum_{s\overset{2}{=}1-n}^{n-1}(-1)^{\frac{m+n-r-s-2}{2}}\widehat{\mathscr{V}}^{(-1)}_{P+\frac{rb}{2}+\frac{s}{2b}}(0)+\text{BRST exact}\\
 \widehat{\mathcal{O}}_{m,n}(z)\widehat{\mathscr{V}}_P^{(0)}(0)&=\sum_{r\overset{2}{=}1-m}^{m-1}\sum_{s\overset{2}{=}1-n}^{n-1}(-1)^{\frac{m+n-r-s-2}{2}}\widehat{\mathscr{V}}^{(0)}_{P+\frac{rb}{2}+\frac{s}{2b}}(0)+\text{BRST exact}.
\end{aligned}
\end{equation}
The logarithmic counterparts of $\widehat{\mathcal{O}}_{m,n}$ are defined as:
\begin{equation}
\begin{aligned}
  \widehat{\mathcal{O}}_{m,n}'^+\equiv \frac{\pi\mathcal{N}_b(\frac{mb}{2}-\frac{n}{2b})}{B_{m,n}}\widehat{H}_{m,n}\widehat{\bar{H}}_{m,n}V'^+_{P=\frac{mb}{2}+\frac{n}{2b}}V^-_{P=i(\frac{mb}{2}-\frac{n}{2b})}\\
  \widehat{\mathcal{O}}_{m,n}'^-\equiv \frac{\pi\mathcal{N}_b(\frac{mb}{2}-\frac{n}{2b})}{B_{m,n}}\widehat{H}_{m,n}\widehat{\bar{H}}_{m,n}V^+_{P=\frac{mb}{2}+\frac{n}{2b}}V'^-_{P=i(\frac{mb}{2}-\frac{n}{2b})}.
\end{aligned}
\end{equation}
Then from the HEOM \eqref{HEOM}, we can derive the cohomological version of HEOM:
\begin{equation}
\begin{aligned}
 \widehat{\mathcal{Q}}\widehat{\bar{\mathcal{Q}}}\widehat{\mathcal{O}}_{m,n}'^+&=-(-1)^n\pi\widehat{\mathscr{V}}^{(0)}_{P=\frac{mb}{2}-\frac{n}{2b}}+\text{BRST exact terms}\\
  \widehat{\mathcal{Q}}\widehat{\bar{\mathcal{Q}}}\widehat{\mathcal{O}}_{m,n}'^-&=-\pi i\widehat{\mathscr{V}}^{(0)}_{P=\frac{mb}{2}+\frac{n}{2b}}+\text{BRST exact terms},
\end{aligned}
\end{equation}
where the physical vertex operators on the right-hand side are the ones in picture $0$. Finally, considering the degenerate super-Virasoro fusion rules, we get the OPEs:
\begin{equation}
\begin{aligned}
     \widehat{\mathcal{O}}'^+_{m,n}(z)\widehat{\mathscr{V}}^{(*)}_P(0)=\sum_{r\overset{2}{=}1-m}^{m-1}\sum_{s\overset{2}{=}1-n}^{n-1}&(-1)^{\frac{m+n-r-s-2}{2}}C_{m,n}^{r,s}(P)\widehat{\mathscr{V}}^{(*)}_{P+\frac{rb}{2}+\frac{s}{2b}}(0)\text{log}|z|^2\\
     &+\text{non-logarithmic terms $\&$ BRST exact terms}
\end{aligned}
\end{equation}
where the structure constant $C_{m,n}^{r,s}(P)$ are the same as in \eqref{Crsmn}. Finally, for the 4-point amplitude with one external operator given by $\mathscr{V}^{(0)}_{P=\frac{mb}{2}-\frac{n}{2b}}$, we have:
\begin{equation}
\begin{aligned}
  \hmA^{(b)}_{0,4,0}&\left(P_1,P_2,P_3,P_4=\frac{mb}{2}-\frac{n}{2b}\right)\\
  &=-\frac{(-1)^n}{\pi}C_{S^2}^{(b)}\int d^2z\left\langle \mathscr{V}^{(-1)}_{P_1}(z_1)\mathscr{V}^{(-1)}_{P_2}(z_2)\widehat{\mathscr{V}}^{(0)}_{P_3}(z_3)\partial\bar{\partial}\widehat{\mathcal{O}}_{m,n}'^+(z)\right\rangle
\end{aligned}
\end{equation}
This  moduli integral also reduces to boundary
contributions from $z_i$ and $\infty$. 
The contribution from $z_1$ is:
\begin{equation}
\begin{aligned}
  &(-1)^nC_{S^2}^{(b)}\sum_{r\overset{2}{=}1-m}^{m-1}\sum_{s\overset{2}{=}1-n}^{n-1}(-1)^{\frac{m+n-r-s-2}{2}}C_{m,n}^{r,s}(P_1)\left\langle \mathscr{V}^{(-1)}_{P_1+\frac{rb}{2}+\frac{s}{2b}}(z_1)\mathscr{V}^{(-1)}_{P_2}(z_2)\widehat{\mathscr{V}}^{(0)}_{P_3}(z_3)\right\rangle\\
  &=(-1)^n\z(\sum_{r\overset{2}{=}1-m}^{m-1}\sum_{s\overset{2}{=}1-n}^{n-1}C^{r,s}_{m,n}(P_1)\y)\hmA^{(b)}_{0,3,0}\!\left(P_1+\frac{(m-1)b}{2}+\frac{n-1}{2b},P_2,P_3\right)
\end{aligned}
\end{equation}
where we have used the anti-double-periodicity of the three-point function $\hmA^{(b)}_{0,3,0}$.
Similar results hold for the contributions from $z_2$ and $z_3$. For them, we can further use the relation \eqref{NSb} and \eqref{NSRb} to show that the shift $\frac{(m-1)b}{2}+\frac{n-1}{2b}$ on $P_2$ and $P_3$ can be moved to $P_1$ if $m+n\in 2\mathbb{Z}$\footnote{Note that for $m+n\in 2\mathbb{Z}+1$, this is no longer true. In contrast,  in $\SCLS$,  the 3-point amplitude satisfies  a more constrained relation: $\mA_{0,3,0}^{(b)}(P_1,P_2+\frac{b^{\pm1}}{2},P_3)=\mA_{0,3,0}^{(b)}(P_1+\frac{b^{\pm1}}{2},P_2,P_3)$.}. The
contribution from $\infty$ is:
\begin{equation}
\begin{aligned}
 &-(-1)^nC_{S^2}^{(b)}(mb+nb^{-1})\left\langle \mathscr{V}^{(-1)}_{P_1}(z_1)\mathscr{V}^{(-1)}_{P_2}(z_2)\widehat{\mathscr{V}}^{(0)}_{P_3}(z_3)\widehat{\mathcal{O}}_{m,n}(\infty)\right\rangle\\
 =&-(-1)^n(mb+nb^{-1})mn\widehat{\mathsf{A}}^{(b)}_{0,3,0}\left(P_1+\frac{(m-1)b}{2}+\frac{n-1}{2b},P_2,P_3\right),
\end{aligned}
\end{equation}
where we have used again the anti-double-periodicity of the three-point function $\hmA^{(b)}_{0,3,0}$.
Collecting these 4 contributions, we obtain the final result
\begin{equation}\label{HEOMhmA}
\begin{aligned}
\hmA^{(b)}_{0,4,0}&\left(P_1,P_2,P_3,P_4=\frac{mb}{2}-\frac{n}{2b}\right)
 =(-1)^n\hmA^{(b)}_{0,3,0}\left(P_1+\frac{(m-1)b}{2}+\frac{n-1}{2b},P_2,P_3\right)\\&\qquad\times\left[\frac{(mb+nb^{-1})mn}{2}-\sum_{i=1}^3\sum_{r\overset{2}{=}1-m}^{m-1}\sum_{s\overset{2}{=}1-n}^{n-1}\sqrt{\left(P_i+\frac{rb}{2}+\frac{s}{2b}\right)^2}\right].
\end{aligned}
\end{equation}
Note that compared to the HEOM \eqref{HEOMmA}, \eqref{HEOMhmA} contains an extra factor $(-1)^n$. This factor treats $m$ and $n$ on an equal footing, as $m$ and $n$ always have the same parity.

\section{The  dual two-Matrix integrals}\label{2Matrix}
In this section, we study the dual matrix model. After a brief review of the topological recursion associated with the general two-matrix integral, we proceed to investigate two specific concrete models, which are respective candidates for the dual descriptions of $\SCLS$ and $\hSCLS$. These two two-matrix integrals not only   reproduce the sphere 3-point  amplitudes of $\SCLS$ and $\hSCLS$ respectively, but also yield the sphere 4-point  amplitudes via topological recursion. Further evidence supporting these two proposed dual two-matrix integrals will be presented  in the following section.

\subsection{Review of the two-matrix model}
We give a brief review of the 2-matrix integral, following \cite{Collier:2024lys}.  The 2-matrix integral over $N\times N$ hermitian matrices is defined as 
\begin{equation}\label{matrixpartionfunction}
 \langle \cdot\rangle =\int_{\mathbb{R}^{2N^2}} [dM_1][dM_2](\cdot) e^{-N\tr\z(V_1(M_1)+V_2(M_2)-M_1M_2\y)}.
\end{equation}
In particular,  if we set $V_2(M_2)=\frac{1}{2}M_2^2$, we can integrate out the matrix $M_2$, which reduces the integral to a one-matrix integral. For  single-trace operators $\MO_i \, (i=1,2,...,n)$, their correlation functions are defined as:
\begin{equation}
 \langle \mathcal{\MO}_1...\MO_n\rangle\equiv \int_{\mathbb{R}^{2N^2}}[dM_1][dM_2]\prod^n_{i=1}\MO_i(M_1,M_2)e^{-N\tr\z(V_1(M_1)+V_2(M_2)-M_1M_2\y)}
\end{equation}
 In the large $N$ limit, the connected correlator has a $\frac{1}{N}-$expansion
\begin{equation}
 \langle \MO_1...\MO_n\rangle_c=\sum_{g=0}^\infty \langle \MO_1...\MO_n\rangle_gN^{2-2g-n}.
\end{equation}
The main observables in matrix integral models are resolvents in terms of one matrix, defined as:
\begin{equation}
 R^{(1)}(x)=\tr\frac{1}{x-M_1},\quad R^{(2)}(x)=\tr\frac{1}{x-M_2}.
\end{equation}
One can consider the products of resolvents 
\begin{equation}
 R(x_1,...,x_n)\equiv \prod_{i=1}^n R(x_i)
\end{equation}
then  we have the genus expansion in the large $N$ limit
\begin{equation}
 \langle R(x_1,...,x_n) \rangle_c=\sum^\infty_{g=0}R_{g,n} (x_1,...,x_n)N^{2-2g-n}.
\end{equation}

\subsection*{Loop Equation}
One can solve the matrix integral perturbatively in the $\frac{1}{N}$ expansion by the loop equation \cite{Eynard:2002kg}:
\begin{equation}\label{Loop}
 \begin{aligned}
 \langle P(x,y)R(I)\rangle&=\Big\langle \Big(y-V_1'(x)+\frac{1}{N}R(x)\Big)U(x,y)R(I)\Big\rangle\\
 &+\frac{1}{N}\sum^n_{k=1}\p_{x_k}\Big\langle \frac{U(x,y)-U(x_k,y)}{x-x_k}R(I\setminus x_k)\Big\rangle,
 \end{aligned}
\end{equation}
where
\begin{equation}
 \begin{aligned}
 &U(x,y)=\tr\z(\frac{1}{x-M_1}\frac{V'_2(y)-V'_2(M_2)}{y-M_2}\y)+N(x-V'_2(y))\\
 &P(x,y)=N(V_2'(y)-x)(V_1'(x)-y)-\tr\z(\frac{V_1'(x)-V_1'(M_1)}{x-M_1}\frac{V_2'(y)-V_2'(M_2)}{y-M_2}\y)+N.
 \end{aligned}
\end{equation}

\subsection*{Spectral curve and topological recursion}
Two-matrix integrals exhibit a notable property: their resolvents  can be recursively derived solely from the  knowledge of the spectral curve.
Consider the zero-order term of the loop equation \eqref{Loop}, set $I=\emptyset$, and take the large $N$ limit; the equation then reduces to:
\begin{equation}
 P_0(x,y)=(y-Y(x))U_0(x,y)
\end{equation}
where we have employed  the   subscript ``0'' to denote the genus-0 contribution to $U$ and $P$, and also introduced:
\begin{equation}
    Y(x) = V'_1(x) - R_{0,1}^{(1)}(x).
\end{equation}
Therefore, the equation $P_0(x,Y(x))=0$ gives an implicit parametrization of the spectral curve. We can also choose some direct parametrization $(\mx(z),\my(z))$, where $z\in \Sigma$ is a coordinate on the spectral curve and $\mx,\my:\Sigma\to \mathbb{CP}^1$ are maps that satisfy
\begin{equation}
 P_0(\mx(z),\my(z))=0.
\end{equation}
Now we show that the resolvents can be  determined recursively from the spectral curve. First, we  define
\begin{equation}
 \omega_{g,n}(z,...,z_n)\equiv R_{g,n}(\mx(z_1),...,\mx(z_n)) d\mx(z_1)...d\mx(z_n)
\end{equation}
It is a meromorphic multi-differential on the spectral curve $\Sigma$. The differentials $\omega_{0,1}(z)$ and $\omega_{0,2}(z_1,z_2)$ are defined slightly differently as follows:
\begin{equation}
 \begin{aligned}
  &\omega_{0,1}(z)\equiv(R_{0,1}(\mx(z))-V_1'(\mx(z)))d\mx(z)=-\my(z)d\mx(z)\\
  &\omega_{0,2}(z,z')\equiv\z(R_{0,2}(\mx(z),\mx(z'))+\frac{1}{(\mx(z)-\mx(z'))^2}\y)d\mx(z)d\mx(z')=\frac{dzdz'}{(z-z')^2}
 \end{aligned}
\end{equation}
Then the loop equation will induce a recursion relation—the topological recursion—for $\omega_{g,n}$ with higher $g$ and $n$ values. Now we introduce the recursion kernel, which is a key constituent of the topological recursion formula. Let $z_m^*$  stand for an enumeration of the branch points of $\mx(z)$, i.e.
\begin{equation}
 d\mx(z_m^*)=0
\end{equation}
 We can define the local Galois involution $\sigma_m$ at the branch point $z_m^*$ by the following  two properties:
\begin{equation}
 \mx(z)=\mx(\sigma_m(z)),\quad \sigma_m(z^*_m)=z_m^*
\end{equation}
for $z$ in a neighborhood of $z_m^*$. 
Then the recursion kernel is defined as
\begin{equation}\label{Kernal}
 K_m(z_1,z)\equiv \frac{\frac{1}{2}\int^{z}_{z'=\sigma_m(z)}\omega_{0,2}(z_1,z')}{\omega_{0,1}(z)-\omega_{0,1}(\sigma_m(z))}
\end{equation}
Finally, the topological recursion formula can be written as:
\begin{equation}
 \begin{aligned}
  \omega_{g,n}(z_1,...,z_n)=&\sum_{m \text{ branch points}}\underset{z=z_m^*}{\Res}K_m(z_1,z)\Big(\omega_{g-1,n+1}(z,\sigma_m(z),z_2,...,z_n)\\
  &+\sum_{h=0}^g \sum_{\substack{
 \mathcal{I} \cup \mathcal{J} = \{z_2,..., z_n\} \\
 (h,\mathcal{I}) \neq (0,\emptyset) \\
 (h,\mathcal{J}) \neq (g,\emptyset)
}} \omega_{h,1+|\mathcal{I}|}(z,\mathcal{I})\omega_{g-h,1+|\mathcal{J}|}(\sigma_m(z),\mathcal{J})\Big)
 \end{aligned}
\end{equation}
Any  $\omega_{g,n}$ where  $g$ and $n$ take higher values can be derived from $\omega_{0,1}$ and $\omega_{0,2}$ through the formula provided above.

\subsection{Candidates for the dual matrix models}\label{dualmatrix}
We have already shown that the amplitude $\mA^{(b)}_{0,3,0}$ is the same as the sphere 3-point amplitude in the bosonic $\mathbb{C}$LS. Thus, we propose that the dual matrix model of $\SCLS$ is the same as the one dual to the bosonic $\mathbb{C}$LS, whose spectral curve is \cite{Collier:2024kwt}:
\begin{equation}\label{spetralcurve}
 \mx(z)=-2\cos(\pi b^{-1}\sqrt{z}),\quad \my(z)=2\cos(\pi b \sqrt{z})
\end{equation}
For  $\hSCLS$,  the proposed dual two-matrix integral we put forward has the following spectral curve:
\begin{equation}\label{hspetralcurve}
 \hmx(z)=2\sin(\pi b^{-1}\sqrt{z}),\quad \hmy(z)=2\sin(\pi b \sqrt{z})
\end{equation}
We will show later that this two-matrix integral reproduces the amplitude $\hmA^{(b)}_{0,3,0}$ of $\hSCLS$\footnote{Note that the NS-R-R and NS-NS-NS amplitudes are different in $\hSCLS$.}. Hereafter, we always use $\widehat{\ \ }$ to label the quantities in the two-matrix model \eqref{hspetralcurve}. Although a thorough analysis of the two-matrix integral \eqref{spetralcurve} have already been presented in \cite{Collier:2024kwt}, we nonetheless repeat this analysis to serve as a comparison with that for the two-matrix integral \eqref{hspetralcurve}.

\subsubsection*{Sheets} 
Both the two  spectral curves have infinitely many sheets since $\mx(z)=\mx((\sqrt{z}+2bn)^2)$ and $\hmx(z)=\hmx((\sqrt{z}+2bn)^2)$ for all $n\in \mathbb{Z}$. Notably, there are also infinitely many branch points
\begin{equation}
 \begin{aligned}
  &z_m^*=(mb)^2,\quad m\in\mathbb{Z}_{\geq 1}\\
  &\widehat{z}_m^*=((m+\frac{1}{2})b)^2,\quad m\in\mathbb{Z}_{\geq 0}
 \end{aligned}
\end{equation}
\subsubsection*{Singular points} Both  two spectral curves  have singular points at 
\begin{equation}
 z^\pm_{(r,s)}=\widehat{z}^\pm_{(r,s)}=\z(rb\pm sb^{-1}\y)^2,\quad r,s\in\mathbb{Z}_{\geq 1}
\end{equation}
 Both sign choices  map to the same point under $(\mx(z),\my(z))$ and $(\hmx(z),\hmy(z))$
\begin{equation}
 \begin{aligned}
  &\mx(z^+_{(r,s)})=\mx(z^-_{(r,s)}),\quad \my(z^+_{(r,s)})=\my(z^-_{(r,s)})\\
  &\hmx(\widehat{z}^+_{(r,s)})=\hmx(\widehat{z}^-_{(r,s)}),\quad \hmy(\widehat{z}^+_{(r,s)})=\hmy(\widehat{z}^-_{(r,s)}).
 \end{aligned}
\end{equation}
Consequently, the spectral curves self-intersect at these points, creating nodal singularities.

\subsubsection*{Calculation of $\omega_{0,3}^{(b)}$, $\widehat{\omega}_{0,3}^{(b)}$ and $\ma_{0,3,0}^{(b)}$, $\hma_{0,3,0}^{(b)}$}

For simplicity, we  use the coordinate $w^2=z$. Then we have the first two resolvents for the spectral curves,
\begin{equation}
 \begin{aligned}
  &\omega^{(b)}_{0,1}=-\frac{4\pi\cos(\pi bw)\sin (\pi b^{-1}w)}{b}dw,\quad \omega_{0,2}^{(b)}(w_1,w_2)=\frac{4w_1w_2}{(w^2_1-w^2_2)^2}dw_1dw_2\\
   &\widehat{\omega}^{(b)}_{0,1}=-\frac{4\pi \sin(\pi bw)\cos (\pi b^{-1}w)}{b}dw,\quad\widehat{\omega}_{0,2}^{(b)}(w_1,w_2)=\frac{4w_1w_2}{(w^2_1-w^2_2)^2}dw_1dw_2
 \end{aligned}
\end{equation}
We have the local Galois inversion $\sigma_m(w)=2mb-w$ and $\widehat{\sigma}_m(w)=(2m+1)b-w$. Using \eqref{Kernal}, we get the recursion kernal
\begin{equation}
 \begin{aligned}
 K_m(w_1,w)&=-\frac{bw_1\z(\frac{1}{w_1^2-w^2}-\frac{1}{w_1^2-\sigma_m(w)^2}\y)dw_1}{4\pi\z[\cos(\pi b w)\sin(\pi b^{-1}w)+\cos(\pi b \sigma_m(w))\sin(\pi b^{-1}\sigma_m(w))\y]dw}\\
  \widehat{K}_m(w_1,w)&=-\frac{bw_1\z(\frac{1}{w_1^2-w^2}-\frac{1}{w_1^2-\widehat{\sigma}_m(w)^2}\y)dw_1}{4\pi \z[\sin(\pi bw)\cos(\pi b^{-1}w)+\sin(\pi b\widehat{\sigma}_m(w))\cos(\pi b^{-1}\widehat{\sigma}_m(w))\y]dw}
 \end{aligned}
\end{equation}
Then $\omega_{0,3}$ can be expressed as
\begin{equation}
 \begin{aligned}
  &\omega_{0,3}(w_1,w_2,w_3)=\sum_{m}\underset{w=w_m^*}{\Res}K_m(w_1,w)\Big(\omega_{0,2}(w,w_2)\omega_{0,2}(\sigma_m(w),w_3)+(w_2\leftrightarrow w_3)\Big)
 \end{aligned}
\end{equation}
It follows that:
\begin{equation}
 \begin{aligned}
 \omega^{(b)}_{0,3}(w_1,w_2,w_3)&=-\sum_{m=1}^\infty\frac{16(-1)^mb^4m^3w_1w_2w_3dw_1dw_2dw_3}{\pi^3 \sin(m\pi b^2)(w_1^2-w_m^{*2})^2(w_2^2-w_m^{*2})^2(w_3^2-w_m^{*2})^2}\\
 \widehat{\omega}^{(b)}_{0,3}(w_1,w_2,w_3)&=-\sum_{m=0}^\infty\frac{16(-1)^mb^4(m+\frac{1}{2})^3w_1w_2w_3dw_1dw_2dw_3}{\pi^3\cos\z((m+\frac{1}{2})\pi b^2\y) (w_1^2-\widehat{w}_m^{*2})^2(w_2^2-\widehat{w}_m^{*2})^2(w_3^2-\widehat{w}_m^{*2})^2}
 \end{aligned}
\end{equation}
To relate the amplitudes derived from string theory, we need to use the following dictionary to obtain the amplitudes calculated from the matrix integral  (denoted as $\ma_{g,n,0}$):
\begin{equation}
 \begin{aligned}
 \ma_{g,n,0}(P_1,...,P_n)&\equiv\sum_{m_1,...,m_n } \underset{z_1=z_{m_1}^*}{\Res}...\underset{z_n=z_{m_n}^*}{\Res}\prod_{j=1}^n\frac{\cos(2\pi P_j\sqrt{z_j})}{P_j}\omega_{g,n}(z_1,...,z_n)
 \end{aligned}
\end{equation}
Then, we have 
\begin{equation}
 \begin{aligned}
 &\ma^{(b)}_{0,3,0}=\sum_{m=1}^\infty\frac{2b(-1)^m\prod_j^3\sin(2m\pi bP_j)}{\sin(m\pi b^2)}\\
  &\hma^{(b)}_{0,3,0}=\sum_{m=0}^\infty\frac{2b(-1)^m\prod_j^3\sin((2m+1)\pi bP_j)}{\cos((m+\frac{1}{2})\pi b^2)}
 \end{aligned}
\end{equation}
They match exactly with the string result \eqref{SinA3}.

\subsubsection*{Calculation of $\ma_{0,4,0}^{(b)}$, $\hma_{0,4,0}^{(b)}$}
From the topological recursion, we have 
\begin{equation}
 \begin{aligned}
  &\omega_{0,4}(z_1,...,z_4)\\
  &=\sum_{m}\underset{w=w_m^*}{\Res}K_m(w_1,w)\Big(\omega_{0,2}(w,w_2)\omega_{0,3}(\sigma_m(w),w_3,w_4)+\omega_{0,2}(\sigma_m(w),w_2)\omega_{0,3}(w,w_3,w_4)\Big)\\
  &+(2,3,4 \text{ per})
 \end{aligned}
\end{equation}
Note that there are in fact two infinite summations over $m$  in $\omega_{0,4}$: one is the summation over $m$ in $\omega_{0,4}=\sum_{m}K_{m}(...)$ and the other is the summation  in $\omega_{0,3}=\sum_{m'}(...)$. When $m=m'$, a third-order pole arises, which complicates the form of 
 $\omega_{0,4}$ significantly. Nevertheless‌, the final results for the two amplitudes  turn out to be simple:
 \begin{subequations}
     \begin{align}
  \ma^{(b)}_{0,4,0}(&\boldsymbol{P})=\sum_{m\in \mathbb{Z}_{\geq 1}}\frac{2b^2 \mathsf{V}_{0,4}^{(b)}(iP_1,iP_2,iP_3,iP_4)\prod_{j=1}^4\sin(2\pi m b P_j)}{\sin(\pi m b^2)^2}\nonumber\\
  &-\sum_{m_1,m_2\in \mathbb{Z}_{\geq 1}}\frac{(-1)^{m_1+m_2}\sin(2\pi m_1bP_1)\sin(2\pi m_1bP_2)\sin(2\pi m_2bP_3)\sin(2\pi m_2bP_4)}{\pi^2 \sin(\pi m_1b^2)\sin(\pi m_2b^2)}\nonumber\\
  &\times \z(\frac{1}{(m_1+m_2)^2}-\frac{\delta_{m_1\neq m_2}}{(m_1-m_2)^2}\y)+2\text{ perms}\\
  \hma^{(b)}_{0,4,0}(&\boldsymbol{P})=\sum_{m\in \mathbb{Z}_{\geq 1}-\frac{1}{2}}\frac{2b^2 \mathsf{V}_{0,4}^{(b)}(iP_1,iP_2,iP_3,iP_4)\prod_{j=1}^4\sin(2\pi m b P_j)}{\cos(\pi m b^2)^2}\nonumber\\
  &+\!\!\sum_{m_1,m_2\in\mathbb{Z}_{\geq 1}-\frac{1}{2}}\!\!\!\!\! \frac{(-1)^{m_1+m_2}\sin(2\pi m_1bP_1)\sin(2\pi m_1bP_2)\sin(2\pi m_2bP_3)\sin(2\pi m_2bP_4)}{\pi^2 \cos(\pi m_1b^2)\cos(\pi m_2b^2)}\nonumber\\
  &\times \z(\frac{1}{(m_1+m_2)^2}-\frac{\delta_{m_1\neq m_2}}{(m_1-m_2)^2}\y)+2\text{ perms}
 \end{align}
 \end{subequations}
where $\mathsf{V}^{(b)}_{0,4}(\boldsymbol{P})$ is the analytic continuation of the sphere 4-point amplitude of the Virasoro minimal string
\begin{equation}
 \begin{aligned}
  \mathsf{V}^{(b)}_{0,4}(iP_1,iP_2,iP_3,iP_4)=\frac{b^2+b^{-2}}{4}-\sum_{i}^4P_i^2
 \end{aligned}
\end{equation}
We can further render the forms of these two 4-point amplitudes more consistent:
\begin{subequations}\label{4pointamplitude}
 \begin{align}
  \ma^{(b)}_{0,4,0}(\boldsymbol{P})=&\sum_{m\in \mathbb{Z}_{\geq 1}}\frac{2b^2 \mathsf{V}_{0,4}^{(b)}(iP_1,iP_2,iP_3,iP_4)\prod_{j=1}^4\sin(2\pi m b P_j)}{\sin(\pi m b(b+b^{-1}))^2}\nonumber\\
  &-\sum_{m_1,m_2\in \mathbb{Z}_{\geq 1}}\frac{\sin(2\pi m_1bP_1)\sin(2\pi m_1bP_2)\sin(2\pi m_2bP_3)\sin(2\pi m_2bP_4)}{\pi^2 \sin(\pi m_1b(b+b^{-1}))\sin(\pi m_2b(b+b^{-1}))}\nonumber\\
  &\times \z(\frac{1}{(m_1+m_2)^2}-\frac{\delta_{m_1\neq m_2}}{(m_1-m_2)^2}\y)+2\text{ perms}\\
  \hma^{(b)}_{0,4,0}(\boldsymbol{P})=&\sum_{m\in \mathbb{Z}_{\geq 1}-\frac{1}{2}}\frac{2b^2 \mathsf{V}_{0,4}^{(b)}(iP_1,iP_2,iP_3,iP_4)\prod_{j=1}^4\sin(2\pi m b P_j)}{\sin(\pi m b(b+b^{-1}))^2}\nonumber\\
  &-\sum_{m_1,m_2\in\mathbb{Z}_{\geq 1}-\frac{1}{2}} \frac{\sin(2\pi m_1bP_1)\sin(2\pi m_1bP_2)\sin(2\pi m_2bP_3)\sin(2\pi m_2bP_4)}{\pi^2 \sin(\pi m_1b(b+b^{-1}))\sin(\pi m_2b(b+b^{-1}))}\nonumber\\
  &\times \z(\frac{1}{(m_1+m_2)^2}-\frac{\delta_{m_1\neq m_2}}{(m_1-m_2)^2}\y)+2\text{ perms}
 \end{align}
\end{subequations}
These are the forms that we introduced in the
introduction \eqref{4,0pt}, \eqref{h4,0pt}.
The only difference between $\ma^{(b)}_{0,4,0}$ and $ \hma^{(b)}_{0,4,0}$ resides in the range of the infinite summation.

We close this section with two brief comments. Firstly, the dual matrix model we proposed for the $\SCLS$ could possibly describe all  amplitudes that involve both NS and R vertices (in type 0B theory). In fact, for the three-point amplitudes at least, the structure we observed (eq. \eqref{030=012=bo}) shares similarities with that in the bosonic  $c=1$ and type 0B $\hat{c}=1$ string theories, as noted at the end of section \ref{3-point}. Thus, by similarly defining the ``LHS mode'' and ``RHS mode'' via the combination of NS and R vertices (as in \eqref{combNSR}), we suspect that they will produce two decoupled copies of the amplitude series of the bosonic $\mathbb{C}$LS. This implies, in particular, that the NS-NS-R-R and R-R-R-R 4-point amplitudes will be identical to the NS-NS-NS-NS 4-point amplitude. Secondly, recall that the NS-NS-NS and NS-R-R amplitudes differ in the $\hSCLS$. Currently, we have   only reproduced the NS-NS-NS amplitude  from the proposed dual two-matrix integral.  It will be interesting to investigate whether this two-matrix integral could reproduce amplitudes  involving R  vertices.

\section{Explicit checks of the proposal}\label{Check}
In the previous section, we have proposed the dual matrix integrals for $\SCLS$ and $\hSCLS$, which give closed form formulas of the 4-point amplitudes $\ma^{(b)}_{0,4,0}$ and $ \hma^{(b)}_{0,4,0}$. In the present section, we will demonstrate that these 4-point amplitudes indeed satisfy the constraints derived from the string side—specifically, the analytical structure in section \ref{Analyticstructure} and the higher equations of motion in section \ref{GRHEOM}. Furthermore, we argue that under the premise of these constraints (together with some mild assumptions), the amplitudes are uniquely determined. Finally, we numerically evaluate the moduli integral of $\mA^{(b)-}_{0,4,0}$ and find that the result matches that from the matrix model.  This further confirms both the proposed dual matrix models for $\SCLS$ and $\hSCLS$.

\subsection{Analytic continuation}
In this subsection, we consider the analytic continuation of the expressions for $\ma^{(b)}_{0,4,0}$ and $ \hma^{(b)}_{0,4,0}$ in \eqref{4pointamplitude} so that they can be defined on a more general region of $P_i \, (i=1,2,,3,4)$. These amplitudes involve  two types of infinite sums:  the single sum ($\sum_m...$) and the double sum ($\sum_{m_1,m_2}...$).  We analyze these two types in turn.  While the analysis of $\ma^{(b)}_{0,4,0}$    has already been presented in \cite{Collier:2024kwt}, we revisit this analysis here to facilitate a direct comparison with that of $ \hma^{(b)}_{0,4,0}$.

\subsection*{Single Sum}
  We denote the single sums in \eqref{4pointamplitude} respectively as $\ma^{(b)}_{0,4,0}(\boldsymbol{P})^{(1)}$  and $ \hma^{(b)}_{0,4,0}(\boldsymbol{P})^{(1)}$, and rewrite them as
 \begin{align}\label{singlesum}
 \sum_m&\frac{\prod_{j=1}^4\sin(2\pi mbP_j)}{\sin(\pi m b(b+b^{-1}))^2}
 =-\frac{1}{4}\sum_m \sum_{\sigma_1,...,\sigma_4=\pm}\frac{\sigma_1\sigma_2\sigma_3\sigma_4e^{2\pi i m b\sum^4_{j=1}\sigma_jP_j}}{(e^{\pi i m b(b+b^{-1})}-e^{-\pi i m b(b+b^{-1})})^2}\\
 &=-\frac{1}{4}\sum_m\sum_{\sigma_1,...,\sigma_4}\sum_{k=0}^\infty k\sigma_1\sigma_2\sigma_3\sigma_4 e^{2\pi i m (\sum_{j=1}^4\sigma_jP_j+k(b+b^{-1}))}\\
 &=\begin{cases}
  -\frac{1}{4}\sum_{k=0}^{\infty}\sum_{\sigma_1,...,\sigma_4=\pm}\frac{k\sigma_1\sigma_2\sigma_3\sigma_4}{e^{-2\pi i b(\sum_{j=1}^{4}\sigma_jP_j+kb)}-1},\quad &m\in \mathbb{Z}_{\geq 1}\\
  -\frac{1}{4}\sum_{k=0}^\infty\sum_{\sigma_1,...,\sigma_4=\pm}\frac{(-1)^k k \sigma_1\sigma_2\sigma_3\sigma_4e^{-\pi i b (\sum_{j=1}^4\sigma_jP_j+kb)}}{e^{-2\pi i b (\sum_{j=1}^4\sigma_jP_j+kb)}-1},\quad &m\in \mathbb{Z}_{\geq 1}-\frac{1}{2} 
 \end{cases}
 \end{align}
Here we assume that $\Im(b^2)>0$ to expand the sines in the denominator. Note that the summations in the last two lines of  \eqref{singlesum}  converge everywhere.

\subsection*{Double Sum}
We denote the double sums in \eqref{4pointamplitude} respectively as $\ma^{(b)}_{0,4,0}(\boldsymbol{P})^{(2)}$  and $ \hma^{(b)}_{0,4,0}(\boldsymbol{P})^{(2)}$, and rewrite them as
 \begin{align}\label{2sum}
  \sum_{m_1,m_2}&\frac{\prod_{j=1}^2\sin(2\pi m_1 bP_j)\prod_{j=3}^4\sin(2\pi m_2 bP_j)}{\sin(\pi m_1 b(b+b^{-1}))\sin(\pi m_2 b(b+b^{-1}))}\z(\frac{1}{(m_1+m_2)^2}-\frac{1-\delta_{m_1,m_2}}{(m_1-m_2)^2}\y)\nonumber\\
  =&-\frac{1}{4}\sum_{m_1,m_2}\sum_{\sigma_1,...,\sigma_4=\pm}\frac{\sigma_1\sigma_2\sigma_3\sigma_4e^{2\pi i b (m_1\sum_{j=1}^2\sigma_jp_j+m_2\sum_{j=3}^4\sigma_jp_j)}}{(e^{\pi im_1 b(b+b^{-1})}-e^{-\pi im_1 b(b+b^{-1})})(e^{\pi im_2 b(b+b^{-1})}-e^{-\pi im_2 b(b+b^{-1})})}\nonumber\\
  &\times \z(\frac{1}{(m_1+m_2)^2}-\frac{1-\delta_{m_1,m_2}}{(m_1-m_2)^2}\y)\nonumber\\
  =&-\frac{1}{4}\sum_{m_1,m_2}\sum_{k_1,k_2=0}^\infty\sum_{\sigma_1,...,\sigma_4=\pm}\sigma_1\sigma_2\sigma_3\sigma_4e^{2\pi i b m_1(\sum_{j=1}^2\sigma_jP_j+(k_1+\frac{1}{2})(b+b^{-1}))}\nonumber\\
  &\times e^{2\pi i b m_2(\sum_{j=3}^4\sigma_jP_j+(k_2+\frac{1}{2})(b+b^{-1}))}\z(\frac{1}{(m_1+m_2)^2}-\frac{1-\delta_{m_1,m_2}}{(m_1-m_2)^2}\y)\nonumber\\
  =&\frac{1}{4}\sum_{k_1,k_2=0}^\infty \sum_{\sigma,\sigma_1,...,\sigma_4=\pm}\sigma_1\sigma_2\sigma_3\sigma_4\nonumber\\
  \times&\begin{cases}
   \frac{\sigma\Li_2(-e^{2\pi i b(\sigma\sum^2_{j=1}\sigma_jP_j+(k_1+\frac{1}{2})b)})}{e^{-2\pi i b(\sum^4_{j=1}\sigma_jP_j+(\sigma(k_1+\frac{1}{2})+k_2+\frac{1}{2})b)}-1} &m\in \mathbb{Z}_{\geq 1}\\
   \frac{(-1)^{k_1+k_2+1}e^{-\pi i b(\sum^4_{j=1}\sigma_jP_j+(\sigma(k_1+\frac{1}{2})+k_2+\frac{1}{2})b)}\Li_2(-e^{2\pi i b(\sigma\sum^2_{j=1}\sigma_jP_j+(k_1+\frac{1}{2})b)})}{e^{-2\pi i b(\sum^4_{j=1}\sigma_jP_j+(\sigma(k_1+\frac{1}{2})+k_2+\frac{1}{2})b)}-1} &m\in \mathbb{Z}_{\geq 1}-\frac{1}{2} 
  \end{cases}\nonumber\\
  +&(\{1,2\}\leftrightarrow \{3,4\})
 \end{align}
Here we also assume $\Im(b^2)>0$. In the final equality, we employ equations \eqref{Liinteger} and \eqref{Lihalfinteger} to carry out the summation over 
 $m_1$ and $m_2$, yielding the dilogarithm function. As anticipated, this function introduces the expected branch cuts. The summations in \eqref{2sum} converge for arbitrary complex momenta, thereby providing an analytic continuation. Then the  4-point amplitude of $\SCLS$ can be expressed as
\begin{equation}
 \begin{aligned}
  &\ma^{(b)}_{0,4,0}(\boldsymbol{P})=-\frac{b^2}{2}\sum_{k=0}^\infty\sum_{\sigma_1,...,\sigma_4=\pm}\frac{k\sigma_1\sigma_2\sigma_3\sigma_4\mathsf{V}^{(b)}_{0,4}(iP_1,iP_2,iP_3,iP_4)}{e^{-2\pi i b(\sum_{j=1}^4\sigma_j P_j+kb)}-1}\\
  &-\frac{1}{4\pi^2}\sum_{1\leq j<\ell\leq 4}\sum_{k_1,k_2=0}^\infty\sum_{\sigma\sigma_1,...,\sigma_4=\pm}\frac{\sigma\sigma_1\sigma_2\sigma_3\sigma_4\Li_2(-e^{2\pi i b(\sigma(\sigma_jP_j+\sigma_\ell P_\ell)+(k_1+\frac{1}{2})b)})}{e^{-2\pi i b(\sum^4_{j=1}\sigma_jP_j+(\sigma(k_1+\frac{1}{2})+k_2+\frac{1}{2})b)}-1},
 \end{aligned}
\end{equation}
and the  4-point amplitude of  $\hSCLS$ can be expressed as
\begin{equation}
 \begin{aligned}
  \hma^{(b)}_{0,4,0}(\boldsymbol{P})=&-\frac{b^2}{2}\sum_{k=0}^\infty\sum_{\sigma_1,...,\sigma_4=\pm}(-1)^ke^{-\pi i b(\sum_{j=1}^4\sigma_jP_j+kb)} \frac{k\sigma_1\sigma_2\sigma_3\sigma_4\mathsf{V}^{(b)}_{0,4}(iP_1,iP_2,iP_3,iP_4)}{e^{-2\pi i b(\sum_{j=1}^4\sigma_j P_j+kb)}-1}\\
  &-\frac{1}{4\pi^2}\sum_{1\leq j<\ell\leq 4}\sum_{k_1,k_2=0}^\infty\sum_{\sigma,\sigma_1,...,\sigma_4=\pm}(-1)^{k_1+k_2+1}e^{-\pi i b (\sum_{j=1}^4\sigma_jP_j+(\sigma(k_1+\frac{1}{2})+k_2+\frac{1}{2})b)}\\
  &\times \frac{\sigma_1\sigma_2\sigma_3\sigma_4\Li_2(-e^{2\pi i b(\sigma(\sigma_jP_j+\sigma_\ell P_\ell)+(k_1+\frac{1}{2})b)})}{e^{-2\pi i b(\sum^4_{j=1}\sigma_jP_j+(\sigma(k_1+\frac{1}{2})+k_2+\frac{1}{2})b)}-1}
 \end{aligned}
\end{equation}
These expressions define the analytic continuation of the two amplitudes to arbitrary complex momenta.

\subsection*{Discontinuities}
The dilogarithm gives rise to the discontinuities in the amplitudes. To see this, we  rewrite $\ma^{(b)}_{0,4,0}(\boldsymbol{P})^{(2)}$  and $\hma^{(b)}_{0,4,0}(\boldsymbol{P})^{(2)}$ by first summing over $k_2$
 \begin{align}\label{A(2)}
  \ma^{(b)}_{0,4,0}(\boldsymbol{P})^{(2)}=-\frac{1}{2\pi^2 b }\sum_{1\leq j<\ell\leq 4}\sum_{k=0}^\infty &\sum_{\sigma_j,\sigma_\ell=\pm}\sigma_j\sigma_\ell\Li_2(-e^{2\pi i b (\sigma_jP_j+\sigma_\ell P_\ell+(k+\frac{1}{2})b)})\nonumber\\
  &\times \ma_{0,3}^{(b)}\z(\frac{b+b^{-1}}{2}+\sigma_jP_j+\sigma_\ell P_\ell,P_m,P_n\y)\nonumber\\
  \hma^{(b)}_{0,4,0}(\boldsymbol{P})^{(2)}=-\frac{1}{2\pi^2 b }\sum_{1\leq j<\ell\leq 4}\sum_{k=0}^\infty &\sum_{\sigma_j,\sigma_\ell=\pm}\sigma_j\sigma_\ell\Li_2(-e^{2\pi i b (\sigma_jP_j+\sigma_\ell P_\ell+(k+\frac{1}{2})b)})\\
  &\times \hma_{0,3}^{(b)}\z(\frac{b+b^{-1}}{2}+\sigma_jP_j+\sigma_\ell P_\ell,P_m,P_n\y)\nonumber
 \end{align}
Here $\{j,\ell,m,n\}=\{1,2,3,4\}$. Since the dilogarithm has a branch point at $z=1$,  the branch points of the 4-point amplitudes are located at
\begin{equation}
 P_*=\sigma_jP_j+\sigma_\ell P_\ell+(k+\frac{1}{2})+(s+\frac{1}{2})b^{-1}=0,\quad k, s\in \mathbb{Z}.
\end{equation}
Furthermore, the discontinuity of the dilogarithm is given by $\underset{z=1}{\text{Disc }} \Li_2(z)=-2\pi i \log(z)$, which thus gives the following discontinuities of the 4-point amplitudes:
\begin{equation}
 \begin{aligned}
  &\underset{p_*=0}{\text{Disc }}\ma^{(b)}_{0,4}(\boldsymbol{P})=8\pi i P_*\underset{p=p_*}{\Res}\z(\ma_{0,3,0}^{(b)}(P_1,P_2,P)\ma_{0,3,0}^{(b)}(P,P_3,P_4)+2\text{ perms}\y)\\
  &\underset{p_*=0}{\text{Disc }}\hma^{(b)}_{0,4}(\boldsymbol{P})=8\pi i P_*\underset{p=p_*}{\Res}\z(\hma_{0,3,0}^{(b)}(P_1,P_2,P)\hma_{0,3,0}^{(b)}(P,P_3,P_4)+2\text{ perms}\y)
 \end{aligned}
\end{equation}
They are in agreement with the result \eqref{DiscA} derived from the string side.

\subsection*{Poles}
For the pole structures, we examine those of 
 $\ma^{\pm}_{0,4,0}$. We first write down their explicit expressions as follows:
\begin{equation}\label{a4pm}
 \begin{aligned}
  \ma^{\pm}_{0,4,0}&(\boldsymbol{P})\\
  =&-\frac{b^2}{2}\sum^\infty_{k=0}\sum_{\sigma_1,...,\sigma_4=\pm}k\sigma_1\sigma_2\sigma_3\sigma_4\mV_{0,4}^{(b)}(iP_1,iP_2,iP_3,iP_4)\frac{1\pm e^{-\pi i b (\sum^{4}_{j=1}\sigma_jP_j+k(b+b^{-1}))}}{e^{-2\pi i b (\sum^{4}_{j=1}\sigma_jP_j+k(b+b^{-1}))}-1}\\
  &-\frac{1}{2\pi^2 b}\sum_{1\leq j<\ell\leq 4}\sum_{k=0}^\infty\sum_{\sigma_j,\sigma_\ell=\pm}\sigma_j\sigma_\ell\Li_2(-e^{2\pi i b (\sigma_jP_j+\sigma_\ell P_\ell+(k+\frac{1}{2})b)})\\
  &\times \ma^{\pm}_{0,3,0}\z(\frac{b+b^{-1}}{2}+\sigma_jP_j+\sigma_\ell P_\ell,P_m,P_n\y)
 \end{aligned}
\end{equation}
Similarly, we also denote the summations in the second line as $\ma^{(b)\pm}_{0,4,0}(\boldsymbol{P})^{(1)}$ and and those in the third and fourth lines as $\ma^{(b)\pm}_{0,4,0}(\boldsymbol{P})^{(2)}$. Then
for $\ma^{(b)\pm}_{0,4,0}(\boldsymbol{P})^{(1)}$, the poles are located at
\begin{equation}
 P_1\pm P_2 \pm P_3\pm P_4 =rb+sb^{-1},\quad \begin{cases}
  r+s\in2\mathbb{Z} &\text{ for }\ma^{(b)+}_{0,4,0}(\boldsymbol{P})^{(1)}\\
  r+s\in2\mathbb{Z}+1&\text{ for }\ma^{(b)-}_{0,4,0}(\boldsymbol{P})^{(1)}
 \end{cases},\quad r\in \mathbb{Z}_{\neq 0},s\in \mathbb{Z}
\end{equation}
 For  $\ma^{(b)\pm}_{0,4,0}(\boldsymbol{P})^{(2)}$, the poles are located at
\begin{equation}
  P_1\pm P_2 \pm P_3\pm P_4 =rb+sb^{-1},\quad \begin{cases}
  r+s\in2\mathbb{Z} &\text{ for }\ma^{(b)+}_{0,4,0}(\boldsymbol{P})^{(2)}\\
  r+s\in2\mathbb{Z}+1&\text{ for }\ma^{(b)-}_{0,4,0}(\boldsymbol{P})^{(2)}
 \end{cases},\quad r\in \mathbb{Z}_{\neq 0},s\in \mathbb{Z}
\end{equation}
Let's make a comment on these poles.
It seems that at $\sum_{j=1}^4\sigma_jP_j=sb^{-1}$ there is a pole in $\ma^{(b)\pm}_{0,4,0}(\boldsymbol{P})^{(2)}$. However,  from the expression \eqref{4pointamplitude} one can see that the summation  converges for $\sum_{j=1}^4\sigma_jP_j=sb^{-1}$ when $P_j$ are in the neighborhood of $\z\{bP_j\in \mathbb{R},\Im(b^2)>0\y\}$. It follows that    $\sum_{j=1}^4\sigma_jP_j=sb^{-1}$ are not poles of $ \ma^{(b)\pm}_{0,4,0}(\boldsymbol{P})$; from the $b\to b^{-1}$ duality,  $\sum_{j=1}^4\sigma_jP_j=rb$ are not poles of $ \ma^{(b)\pm}_{0,4,0}(\boldsymbol{P})$ as well. This fact, together  with the poles listed above for $\ma^{(b)\pm}_{0,4,0}(\boldsymbol{P})^{(1)}$ and $\ma^{(b)\pm}_{0,4,0}(\boldsymbol{P})^{(2)}$,  leads to the correct pole structure as in \eqref{polep+} and \eqref{polep-}.

\subsection{Duality and swap symmetries}
It is obvious that the amplitudes \eqref{4pointamplitude} satisfy both the $b\to-b$ duality and the $b\to -ib$ swap symmetry. Now we demonstrate that they also satisfy the $b\to b^{-1}$ duality. In \cite{Collier:2024kwt},  the $b\to b^{-1}$ duality of  $\ma_{0,4,0}^{(b)}$ is proven, using its (double) contour integral representation. Thus, for $\hma^{(b)}_{0,4,0}$, we also  express it  as a double contour integral
\begin{equation}
\hma^{(b)}_{0,4,0}(\boldsymbol{P})=\sum_{m=-\infty}^\infty\underset{P=\frac{(m+\frac{1}{2})b}{2}}{\Res}\widehat{f}_b(P)+\sum_{m_2=-\infty}^{\infty}\sum_{\substack{m_1 = -\infty \\ m_1 \neq m_2}}^\infty\underset{P=\frac{(m_1+\frac{1}{2})b}{2}}{\Res}\underset{P'=\frac{(m_2+\frac{1}{2})b}{2}}{\Res}\widehat{g}_b(P,P')
\end{equation}
where 
\begin{equation}
 \begin{aligned}
  \widehat{f}_b(P)&=-\frac{2\pi b \mathsf{V}_{0,4}^{(b)}(i\boldsymbol{P})\sin(2\pi b^{-1}P)\prod^4_{j=1}\sin(4\pi PP_j)}{\cos(2\pi bP)^2\cos(2\pi b^{-1}P)}\\
  \widehat{g}_{b}(P,P')&=\frac{\sin(4\pi PP_1)\sin(4\pi PP_2)\sin(4\pi P'P_3)\sin(4\pi P'P_4)}{2\cos(2\pi bP)\cos(2\pi b^{-1}P)\cos(2\pi bP')\cos(2\pi b^{-1}P')(P-P')^2}+2\text{ perms}
 \end{aligned}
\end{equation}
For the double sum term, 
the pole of $\widehat{g}_b(P,P')$ are 
\begin{equation}
 P=\frac{(m+\frac{1}{2})b}{2}, \frac{(m+\frac{1}{2})}{2b},\quad P'=\frac{(m+\frac{1}{2})b}{2},\frac{(m+\frac{1}{2})}{2b},\quad m\in\mathbb{Z}
\end{equation}
We can deform the integral contour and make the contour run around $P=\frac{(m_1+\frac{1}{2})b^{-1}}{2}$ and $P=\frac{(m_2+\frac{1}{2})b^{-1}}{2}$ to pick out the corresponding residues. Through this process, we obtain: 
 \begin{align}\label{hab}
  \hma_{0,4,0}^{(b)}(\boldsymbol{P})&=\hma^{(b^{-1})}_{0,4,0}(\boldsymbol{P})+\sum_{m=-\infty}^{\infty}\Bigg[\underset{P=\frac{(m+\frac{1}{2})}{2b}}{\Res}\underset{P'=\frac{(m+\frac{1}{2})}{2b}}{\Res}\widehat{g}_{b}(P,P')-\underset{P=\frac{(m+\frac{1}{2})b}{2}}{\Res}\underset{P'=\frac{(m+\frac{1}{2})b}{2}}{\Res}\widehat{g}_{b}(P,P')\nonumber\\
  &-\underset{P=\frac{(m+\frac{1}{2})}{2b}}{\Res}f_{b^{-1}}(P)+\underset{P=\frac{(m+\frac{1}{2})b}{2}}{\Res}f_{b}(P)\Bigg]\\
  &=\hma^{(b^{-1})}_{0,4,0}(\boldsymbol{P})\nonumber
 \end{align}
The reason for the last equality is as follows. One can check that: 
\begin{equation}
 \begin{aligned}
  &\underset{P=\frac{(m+\frac{1}{2})b}{2}}{\Res}f_{b}(P)-\frac{1}{2}\z(\underset{P=\frac{(m+\frac{1}{2})b}{2}}{\Res}\underset{P'=\frac{(m+\frac{1}{2})b}{2}}{\Res}+\underset{P'=\frac{(m+\frac{1}{2})b}{2}}{\Res}\underset{P=\frac{(m+\frac{1}{2})b}{2}}{\Res}\y)\widehat{g}_{b}(P,P')\\
  &=\underset{P=\frac{(m+\frac{1}{2})b}{2}}{\Res}\z(-\frac{\frac{\pi}{2}(b\tan(2\pi b P)-b^{-1}\tan(2\pi b^{-1} P))\prod_{j=1}^4\sin(2\pi PP_j)}{\cos(2\pi b P)^2\cos(2\pi b^{-1} P)^2}\y).
 \end{aligned}
\end{equation}
Thus, the infinite sum  in \eqref{hab} can be expressed  as the sum over all  residues of the above function (together with its analog with $b\leftrightarrow 1/b$), and it vanishes identically.

\subsection{The higher equations of motion}
As shown in the bosonic case \cite{Collier:2024kwt}, the 4-point amplitude $\ma_{0,4,0}^{(b)}$ indeed satisfies the corresponding higher equations of motion \eqref{HEOMmA}\footnote{Since both of them take the same form as the corresponding ones in the bosonic case. Note also that the HEOM \eqref{HEOMmA} requires $m+n\in 2\mathbb{Z}$, while in the bosonic case the HEOM does not have this restriction.}.  Thus,  
 we now show that the amplitude $\hma_{0,4,0}^{(b)}$ satisfies the higher equations of motion  \eqref{HEOMhmA}.

 We put one external momentum to be $P_4=\frac{mb}{2}-\frac{n}{2b}$, where $m+n\in 2\mathbb{Z}$. First, we analyse the  term  $\hma^{(b)}_{0,4,0}(\boldsymbol{P})^{(2)}$ in \eqref{A(2)}. Since $\hma_{0,3,0}^{(b)}$ vanishes when one external momentum degenerate, we have
\begin{equation}
\begin{aligned}
  \!\!\hma^{(b)}_{0,4,0}\z(P_1,P_2,P_3,\frac{mb}{2}-\frac{n}{2b}\y)^{(2)}=-&\sum_{j=1}^{3}\sum_{k=0}^{\infty}\sum_{\sigma_j,\sigma_4=\pm}\frac{\sigma_j\sigma_4}{2\pi^2b}\Li_2\z(-e^{2\pi i b (\sigma_jP_j+\sigma_4(\frac{mb}{2}-\frac{n}{2b})+(k+\frac{1}{2})b)}\y)\\
 &\times \hma_{0,3,0}^{(b)}\z(\frac{b+b^{-1}}{2}+\sigma_jP_j+\sigma_4\z(\frac{mb}{2}-\frac{n}{2b}\y),P_{j'},P_{j''}\y)
\end{aligned}
\end{equation}
where $\{j,j',j''\}=\{1,2,3\}$. Use the periodic conditions \eqref{NSb} and \eqref{NSRb} we can see 
\begin{equation}
\begin{aligned}
    \hma_{0,3,0}^{(b)}& \z(\frac{b+b^{-1}}{2}+\sigma_jP_j+\sigma_4\z(\frac{mb}{2}-\frac{n}{2b}\y),P_{j'},P_{j''}\y)\\
    &=(-1)^n\sigma_j\hma^{(b)}_{0,3,0}\z(P_1+\frac{(m-1)b}{2}+\frac{n-1}{2b},P_2,P_3\y).
\end{aligned}
\end{equation}
Next, we carry out the summation over $\sigma_4$ and $k$; most terms cancel out in this process, leading to the following result:
\begin{equation}
 \begin{aligned}
  \hma^{(b)}_{0,4,0}\z(P_1,P_2,P_3,\frac{mb}{2}-\frac{n}{2b}\y)^{(2)}=\frac{(-1)^n}{2\pi^2 b}&\sum_{j=1}^3\sum^{m-1}_{r\overset{2}{=}1-m}\sum_{\sigma_j=\pm}\Li_2\z(-e^{2\pi i b \sigma_j(P_j+\frac{rb}{2})+in\pi}\y)\\
 &\times \hma^{(b)}_{0,3,0}\z(P_1+\frac{(m-1)b}{2}+\frac{n-1}{2b},P_2,P_3\y)
 \end{aligned}
\end{equation}
For $\Re(P^2)>0$, the principal branch choice of the dilogarithm gives (see the identity \eqref{Dlogsum}),
\begin{equation}
 \Li_{2}(e^{2\pi i b P})+\Li_{2}(e^{-2\pi i b P})=\frac{\pi^2}{3}+2\pi^2b^2P^2-2\pi^2b\sqrt{p^2}
\end{equation}
From this equation, we have 
\begin{equation}\label{ha1}
 \begin{aligned}
  \hma^{(b)}_{0,4,0}\z(P_1,P_2,P_3,\frac{mb}{2}-\frac{n}{2b}\y)^{(2)}=\frac{(-1)^n}{2\pi^2 b}\sum_{j=1}^3\sum_{r\overset{2}{=}1-m}^{m-1}&\Big(\frac{3n^2\pi^2-\pi^2}{6}+2\pi^2b^2\z(P_j+\frac{rb}{2}\y)^2\\
 -2\pi^2b \sum^{n-1}_{s\overset{2}{=}1-n}\sqrt{\z(P_j+\frac{rb}{2}+\frac{s}{2b}\y)^2}\Big)&\hma^{(b)}_{0,3,0}\z(P_1+\frac{(m-1)b}{2}+\frac{n-1}{2b},P_2,P_3\y)
 \end{aligned}
\end{equation}
For the other term   $\hma^{(b)}_{0,4,0}(\boldsymbol{P})^{(1)}$, we  also perform the  summation over $\sigma_4$. Then the result can be rewritten as
\begin{equation}\label{ha2}
    \begin{aligned}
        &\hma_{0,4,0}^{(b)}\z(P_1,P_2,P_3,\frac{mb}{2}-\frac{n}{2b}\y)^{(1)}\\
        &=mb(-1)^n\mV^{(b)}_{0,4}\z(iP_1,iP_2,iP_3,i\z(\frac{mb}{2}-\frac{n}{2b}\y)\y)\hma^{(b)}_{0,3,0}\z(P_1+\frac{(m-1)b}{2}+\frac{(n-1)}{2b},P_2,P_3\y)
    \end{aligned}
\end{equation}
Taking \eqref{ha1} and \eqref{ha2} together we get 
\begin{equation}
\begin{aligned}
 \hma^{(b)}_{0,4,0}&\left(P_1,P_2,P_3,P_4=\frac{mb}{2}-\frac{n}{2b}\right)
 =(-1)^n\hma^{(b)}_{0,3,0}\left(P_1+\frac{(m-1)b}{2}+\frac{n-1}{2b},P_2,P_3\right)\\&\qquad\times\left[\frac{(mb+nb^{-1})mn}{2}-\sum_{i=1}^3\sum_{r\overset{2}{=}1-m}^{m-1}\sum_{s\overset{2}{=}1-n}^{n-1}\sqrt{\left(P_i+\frac{rb}{2}+\frac{s}{2b}\right)^2}\right]
\end{aligned}
\end{equation}
This matches precisely with the HEOM \eqref{HEOMhmA} of $\hSCLS$.

\subsection{Uniqueness}
We now turn to examining the potential differences between the string amplitudes and the amplitudes derived from the matrix integral. Focusing on the sphere four-point case, we define the difference between these two amplitudes as follows:
\begin{equation}
  \mathsf{B}^{(b)\pm}_{0,4,0}(\boldsymbol{P})\equiv\mA^{(b)\pm}_{0,4,0}(\boldsymbol{P})-\ma^{(b)\pm}_{0,4,0}(\boldsymbol{P}).
\end{equation}
Then $\mathsf{B}^{(b)\pm}_{0,4,0}(\boldsymbol{P})$ possess the following properties:
\begin{enumerate}
 \item $\mathsf{B}^{(b)\pm}_{0,4,0}(\boldsymbol{P})$ are meromorphic functions in the momenta $\boldsymbol{P}$. We have shown $\mA^{(b)\pm}_{0,4,0}$ and $\ma^{(b)\pm}_{0,4,0}$ have identical discontinuities (and poles), so the difference is meromorphic\footnote{Note that in the bosonic counterpart, the residues can be fixed by the so-called triality symmetry  \cite{Collier:2024kwt}. However, this property has not been proved in the supersymmetric case. It would be interesting to study whether SLFT has the triality symmetry, which would lead to the triality symmetry in the $\SCLS$ and the $\hSCLS$. }.
 
 \item $\mathsf{B}^{(b)\pm}_{0,4,0}(\boldsymbol{P})$  vanish at $P_i=\frac{rb}{2}+\frac{sb^{-1}}{2}$ for any $r,s\in \mathbb{Z}$ with $ r+s\in2\mathbb{Z}$. The reason is that $\mA^{(b)\pm}_{0,4,0}$ and $\ma^{(b)\pm}_{0,4,0}$ have the same trivial zeros and the same HEOMs\footnote{HEOMs of $\mA^{(b)\pm}_{0,4,0}$  ($\ma^{(b)\pm}_{0,4,0}$) can be directly read from those of $\mA^{(b)}_{0,4,0}$ and $\hmA^{(b)}_{0,4,0}$ ($\ma^{(b)}_{0,4,0}$ and $\hma^{(b)}_{0,4,0}$).}. 

 \item $\mathsf{B}^{(b)\pm}_{0,4,0}(\boldsymbol{P})$ are permutation symmetric and odd in all external momenta. In addition, they satisfy the dualities and swap symmetry with respect to the  super-Liouville parameter $b$.
\end{enumerate}
Functions that satisfy these constraints exist. We further write them as:
\begin{equation}
 \mathsf{B}^{(b)\pm}_{0,4,0}(\boldsymbol{P})=\prod_{i=1}^4b^4e^{i\pi P_i^2}\vartheta_1(bP_i|b^2)\vartheta_3(bP_i|b^2)\mathsf{B}^{(b)\pm\prime}_{0,4,0}(\boldsymbol{P})
\end{equation}
Then $\mathsf{B}^{(b)\pm\prime}_{0,4,0}(\boldsymbol{P})$ will be meromorphic functions in the momenta $\boldsymbol{P}$ and have the same poles as $\mA^{\pm}_{0,4,0}(\boldsymbol{P})$ and have no zeros. All the $b$-duality symmetries also hold for  $\mathsf{B}^{(b)\pm\prime}_{0,4,0}(\boldsymbol{P})$. Near infinity,  $\mathsf{B}_{0,4,0}^{(b)\pm}(\boldsymbol{P})$ behave like:
\begin{equation}
 \mathsf{B}_{0,4,0}^{(b)\pm}(\boldsymbol{P})\sim\mathsf{B}_{0,4,0}^{(b)\pm\prime}(\boldsymbol{P})e^{\pi \sum^4_{j=1}|P_j|^4}
\end{equation}
For possible solutions such as $\mathsf{B}_{0,4,0}^{(b)\pm\prime}(\boldsymbol{P})=1$, $\mathsf{A}_{0,4,0}^{(b)\pm}(\boldsymbol{P})$  will grow much faster at infinity than the solution we have found (our solution  grows polynomially fast). It seems  natural to assume that the physical amplitudes grow not too fast near infinity \cite{Collier:2024kwt}. In this sense, the amplitudes we get are unique.

\subsection{Numerical analysis}\label{Numerical}
In this subsection,  we perform (partial) numerical calculations of the moduli integrals for the 4-point amplitudes $\mA^{(b)}_{0,4,0}$ and $\hmA^{(b)}_{0,4,0}$, which provides further supports for our proposal of the dual matrix models. 

\subsection*{Recursion relation for superconformal blocks}
Before evaluating the moduli integral, we first  review the $c$-recursion relation for the superconformal blocks of the sphere 4-point function in the NS sector, as originally derived in \cite{Belavin:2007zz} and \cite{Hadasz:2006qb}.

Recall that in our convention, the relation between the  momentum of an NS superconformal primary operator and its conformal weight is:
\begin{equation}
 h = \frac{Q^2}{8} - \frac{P^2}{2}
\end{equation}
The idea of the $c$-recursion formula is to expand the conformal blocks as a sum of poles on the $c$-plane. The loci of the poles depend on the internal conformal weight $h$, and are parametrized by two integers $r,s$. It is useful to define the following functions
\begin{align}\label{usefulfunc}
\left(b_{r, s}(h)\right)^2 & =-\frac{1}{r^2-1}\left(4 h+r s-1+\sqrt{16 h^2+8(r s-1) h+(r-s)^2}\right) \nonumber\\
c_{r, s}(h) & =\frac{15}{2}+3 b_{r, s}(h)^2+3 b_{r, s}(h)^{-2} \nonumber\\
A_{r, s}(h) & =\frac{1}{2} \prod_{\substack{p=1-r \\
(p, q) \neq(0,0),(r, s);}}^r \prod_{\substack{q=1-s \\ p+q \in 2 \mathbb{Z}}
}^s \frac{\sqrt{2}}{ p b_{r, s}(h)+q b_{r, s}^{-1}(h)} \\
P_{r, s}\left(h_1, h_2\right) & =\prod_{\substack{p \in\{1-r, r-1 ; 2\} \\ q \in \{1-s,s-1;2\} \\
(p+q)-(r+s) \equiv 2 ~ \text{mod} ~4}} \frac{2 P_1-2 P_2-p b_{r, s}-q / b_{r, s}}{2 \sqrt{2}} \frac{2 P_1+2 P_2+p b_{r, s}+q / b_{r, s}}{2 \sqrt{2}}\nonumber\\ 
P_{r, s}\left(h_1, * h_2\right) & =\prod_{\substack{p \in\{1-r, r-1 ; 2\} \\ q \in \{1-s,s-1;2\} \\
(p+q)-(r+s) \equiv 0 ~ \text{mod} ~4}} \frac{2 P_1-2 P_2-p b_{r, s}-q / b_{r, s}}{2 \sqrt{2}} \frac{2 P_1+2 P_2+p b_{r, s}+q / b_{r, s}}{2 \sqrt{2}} \nonumber
\end{align}
Then, one can expand the superconformal blocks as a power series of the cross ratio $z$:
\begin{equation}\label{blockzexpansion}
\begin{aligned}
& \mc{F}^{1}_{h} \z[ \begin{smallmatrix}
  {} _{-} h_3 & {}_{-} h_2 \\
 h_4 & h_1
 \end{smallmatrix} \y](z)=z^{h-h_1-h_2}\left[1+\sum_{m=\in \mathbb{Z}_{+}} z^m F_m\left(h_4, {}_-h_3, {}_-h_2, h_1 ; h ; c\right)\right] \\
& \mc{F}^{\frac{1}{2}}_{h} \z[ \begin{smallmatrix}
  {} _{-} h_3 & {}_{-} h_2 \\
 h_4 & h_1
 \end{smallmatrix} \y](z)= z^{h-h_1-h_2} \sum_{k \in \mathbb{Z}_{+}-\frac{1}{2}} z^k F_k\left(h_4, {}_-h_3, {}_-h_2, h_1 ; h ; c\right),
\end{aligned}
\end{equation}
where the coefficients $F_m\left(h_4, {}_-h_3, {}_-h_2, h_1 ; h ; c\right)$ are functions of the external conformal weights $h_i$, the internal conformal weight $h$ and the central charge $c$. As we have introduced, these coefficients admit expansions of poles on the $c$-plane
\begin{equation}
 \begin{aligned}
  & F_0 \left(h_4, {}_-h_3, {}_-h_2, h_1 ; h ; c\right) = 1 \\
  & F_m \left(h_4, {}_-h_3, {}_-h_2, h_1 ; h ; c\right) = f_m\left(h_4, {}_-h_3, {}_-h_2, h_1 ; h ; c\right) \\
  &\quad + \sum_{\substack{r=2,3,\cd ,~ s = 1,2,\cd \\ 1<rs\leq 2m,~ r+s \in 2 \mathbb{Z}}} \frac{R^m_{r,s}\left(h_4, {}_-h_3, {}_-h_2, h_1 ; h \right)}{c-c_{r,s}(h)} F_{m-\frac{rs}{2}}\left(h_4, {}_-h_3, {}_-h_2, h_1 ; h+rs/2 ; c_{r,s}(h)\right)
 \end{aligned}
\end{equation}
where the ``seed'' block coefficients $f_m$ appearing on the RHS are the $c\to \infty$ limits of $F_m$, and have the following form
\begin{subequations}
\begin{align}
f_m\left(h_4, h_3, h_2, h_1 ; h\right) & = \begin{cases}\frac{1}{m!} \frac{\left(h+h_3-h_4\right)_m\left(h+h_2-h_1\right)_m}{(2 h)_m} &  m \in \mathbb{Z}_{>0}, \\
\frac{1}{(m-1 / 2)!} \frac{\left(h+h_3-h_4+1 / 2\right)_{m-1 / 2}\left(h+h_2-h_1+1 / 2\right)_{m-1 / 2}}{(2 h)_{m+1 / 2}} &  m \in \mathbb{Z}_{>0}-\frac{1}{2},\end{cases} \\
f_m\left(h_4, h_3, * h_2, h_1 ; h\right) & = \begin{cases}\frac{1}{m!} \frac{\left(h+h_3-h_4\right)_m\left(h+h_2-h_1+1 / 2\right)_m}{(2 h)_m} &  m \in \mathbb{Z}_{>0}, \\
\frac{1}{(m-1 / 2)!} \frac{\left(h+h_3-h_4+1 / 2\right)_{m-1 / 2}\left(h+h_2-h_1\right)_{m+1 / 2}}{(2 h)_{m+1 / 2}} &  m \in \mathbb{Z}_{>0}-\frac{1}{2},\end{cases} \\
f_m\left(h_4, * h_3, h_2, h_1 ; h\right) & = \begin{cases}\frac{1}{m!} \frac{\left(h+h_3-h_4+1 / 2\right)_m\left(h+h_2-h_1\right)_m}{(2 h)_m} &  m \in \mathbb{Z}_{>0}, \\
-\frac{1}{(m-1 / 2)!} \frac{\left(h+h_3-h_4\right)_{m+1 / 2}\left(h+h_2-h_1+1 / 2\right)_{m-1 / 2}}{(2 h)_{m+1 / 2}} &  m \in \mathbb{Z}_{>0}-\frac{1}{2},\end{cases} \\
f_m\left(h_4, * h_3, * h_2, h_1 ; h\right) & = \begin{cases}\frac{1}{m!} \frac{\left(h+h_3-h_4+1 / 2\right)_m\left(h+h_2-h_1+1 / 2\right)_m}{(2 h)_m} & m \in \mathbb{Z}_{>0}, \\
-\frac{1}{(m-1 / 2)!} \frac{\left(h+h_3-h_4\right)_{m+1 / 2}\left(h+h_2-h_1\right)_{m+1 / 2}}{(2 h)_{m+1 / 2}} & m \in \mathbb{Z}_{>0}-\frac{1}{2},\end{cases}
\end{align}
\end{subequations}
where $(\cdot)_m$ stands for the Pochhammer symbol. In addition, the residues of the poles are related to coefficients $R_{r,s}^m$, which take the following form
\begin{equation}
\begin{aligned}
     R_{r, s}^m&\left(h_4,{}_- h_3,{}_- h_2, h_1 ; h\right)\\
     &= \begin{cases}\sigma^{r s}\left({}_-h_3\right)\left(-\frac{\partial c_{r, s}(h)}{\partial h}\right) A_{r, s} P_{r, s}\left(h_1,{}_-h_2\right) P_{r, s}\left(h_4,{}_-h_3\right) & \text { for } m \in \mathbb{Z}_{+} \\ \sigma^{r s}\left({}_-h_3\right)\left(-\frac{\partial c_{r, s}(h)}{\partial h}\right) A_{r, s} P_{r, s} (h_1, \widetilde{{}_-h_2} ) P_{r, s} (h_4, \widetilde{{}_-h_3} ) & \text { for } m \in \mathbb{Z}_{+}-\frac{1}{2}\end{cases}
\end{aligned}
\end{equation}
Here, the fusion polynomials $P_{r,s}$ are defined as in \eqref{usefulfunc}. We have also introduced the notation $\sigma^{rs}(*h_3) = (-1)^{rs}$, $\sigma^{rs}(h_3)=1$, and $\widetilde{h}=*h,~ \widetilde{*h}=h$. 

The $z$-expansions \eqref{blockzexpansion} of the superconformal blocks are convergent within the unit disk $|z|<1$. In the actual  computation, it is useful to define the elliptic superconformal blocks \cite{Belavin:2007gz,Hadasz:2007nt,Suchanek:2010kq} as
\begin{equation}
\begin{aligned}
 \mc{F}^{1,\frac{1}{2}}_{h} \z[ \begin{smallmatrix}
  {} _{-} h_3 & {}_{-} h_2 \\
 h_4 & h_1
 \end{smallmatrix} \y](z) &= (16q)^{h-\frac{c-3/2}{24}} z^{\frac{c-3/2}{24}-h_1-{}_- h_2}(1-z)^{\frac{c-3/2}{24}- {}_- h_2 - {}_- h_3} \\
 &\quad \times \vartheta_3^{\frac{c-3/2}{2}-4(h_1 + {}_- h_2 + {}_- h_3 + h_4)} \mc{H}^{1,\frac{1}{2}}_{h} \z[ \begin{smallmatrix}
  {} _{-} h_3 & {}_{-} h_2 \\
 h_4 & h_1
 \end{smallmatrix} \y](q)
\end{aligned}
\end{equation}
where
\begin{equation}
 q = \exp(it),\quad t=i \pi \frac{K(1-z)}{K(z)},\quad K(z) = {}_2 F_1 \Big( \frac{1}{2},\frac{1}{2},1 \Big| z\Big).
\end{equation}
Expressed as a power series in $q$, the elliptic block $\mc{H}^{1,\frac{1}{2}}_{h}$ converges within the unit $q$-disk, which covers the whole $z$-plane.

\subsection*{The  computation of $\mA_{0,4,0}^{(b)-}(\boldsymbol{P})$}
In this subsection, we present the numerical computation of  $\mA_{0,4,0}^{(b)-}(\boldsymbol{P})$, which is defined in  \eqref{defApAm}. We will follow the method introduced in \cite{Collier:2023cyw} and \cite{Collier:2024kwt}, as reviewed below. 

The idea is to divide the whole $z$-plane into six regions:
\begingroup
\renewcommand\labelenumi{(\theenumi)}
\begin{multicols}{2}
\begin{enumerate}
 \item $\operatorname{Re}z\leq \frac{1}{2}$, $|1-z|\leq 1$;
 \item $|z| \leq 1$, $|1 - z| \geq 1$;
 \item $\operatorname{Re}z\leq \frac{1}{2}$, $|z| \geq 1$;
 \item $\operatorname{Re}z\geq \frac{1}{2}$, $|z|\leq 1$;
 \item $|1-z|\leq 1$, $|z|\geq 1$;
 \item $\operatorname{Re}z \geq \frac{1}{2}$, $|1-z|\geq 1$.
\end{enumerate}
\end{multicols}
\endgroup
These six regions are related to each other by compositions of $T:z\mapsto1-z$ and $S:z\mapsto z^{-1}$. Under the transformation $S$, the four-point function $\langle O_1(0)O_2(z) O_3(1) O_4(\infty) \rangle=G(1234|z)$ transforms according to
\begin{equation}\label{prefactorG}
 S:~ G(1234|z) = (-1)^{\#} z^{(h_4-h_3-h_2-h_1)} \bar{z}^{(\bar h_4-\bar h_3-\bar h_2-\bar h_1)}G(1324|z^{-1}),
\end{equation}
where $h_i,\bar h_i$ are the conformal weights of the fields $O_i$ and the sign $(-1)^{\#}$ should be added according to the statistical property of fields. Specifically,  $(-1)^{\#} = (-1)$ if and only if both $O_2$ and $O_3$ are fermions. Under the transformation $T$, the four-point function transforms as:
\begin{equation}
 T:~ G(1234|z) = (-1)^{\#} G(3214|1-z).
\end{equation}
Now we begin to compute the moduli integral of the string amplitudes. For simplicity, let us denote
\begin{equation}
 {}_-h^\pm_2 = h^\pm_2 +\delta^\pm_2,\quad {}_-h^\pm_3 = h^\pm_3 +\delta^\pm_3
\end{equation}
and similarly for the anti-holomorphic part. An observation is that, due to the distribution of the PCOs, each term in the integrand of the superstring amplitude is a product of four-point functions with the following relation
\begin{equation}\label{constraintdelta}
 \begin{aligned}
  & \delta_2^++\delta_2^- = \delta_3^++\delta_3^- = \frac{1}{2} \\
  & \bar\delta_2^++\bar\delta_2^- = \bar\delta_3^++\bar\delta_3^- = \frac{1}{2}. \\
 \end{aligned}
\end{equation}
Besides, we  have the constraints from the  self-consistency of the superstring theory
\begin{equation}
 \begin{aligned}
  &c^+ + c^- = 15 \\
  &h_i^+ + h_i^- = \frac{1}{2}. \\
 \end{aligned}
\end{equation}
Using these identities, one can check that the prefactors in \eqref{prefactorG} of the two super-Liouville correlators are canceled exactly by the Jacobian from the measure. Thus, each integral $I_{ab}$ or $J_{ab}$ can be expressed as:
\begin{equation}
\begin{aligned}
     I_{ab} \text{ or } J_{ab} = C_{S^2}\prod_{i=1}^4\mathcal{N}(P_i) \int_{\text{reg (1)}} &d^2z \langle O_1^+O_2^+O_3^+O_4^+ \rangle\langle O_1^-O_2^-O_3^-O_4^- \rangle\\
    &+ \Big( \text{5 other perms. of \{123\}} \Big).
\end{aligned}
\end{equation}
Depending on the order of the external momenta, there are six different channels for each $I_{ab}$ and $J_{ab}$. We denote them as $1234,~1324,~2134,~2314,~3124,~3214$ respectively. In the $t$-plane,  region (1) (in the $z$-plane) is  mapped to the domain  $F_0=\{ t = t_1+it_2\in \mathbb{C}| -\frac{1}{2} \leq t_1 \leq \frac{1}{2},~ |t|\geq 1 \}$. To do the integral in $F_0$,   We also need the following  formulae
\begin{equation}
 \begin{aligned}
  z = \z( \frac{\vartheta_2(t)}{\vartheta_3(t)} \y)^4 ,\quad 1-z = \z(\frac{\theta_4(t)}{\theta_3(t)}\y)^4,\quad d^2z = \z| \pi i \z( \frac{\vartheta_2(q)\vartheta_4(q)}{\vartheta_3(q)} \y)^4 \y|^2 d^2 t
 \end{aligned}
\end{equation}
Then for the channel $1234,~1324$, we have 
\begin{equation}\label{channel1234}
\begin{aligned}
\text{1234 or 1324 of }~ I_{ab} \text{ or } J_{ab} &= C_{S^2}\prod_{i=1}^4\mathcal{N}(P_i) \pi^2 \sum_{\star}\int_{F_0} d^2t \sum_{p_+ ,p_-} |16q|^{p^2_+ + p_-^2} |\theta_4(q)|^{-4} \\
 & \qquad \times \text{\stackon[0.5pt]{$C^{\mathrlap{\scriptstyle +}}$}{$\scriptstyle\star$}}_{\ \ }\text{\stackon[0.5pt]{$C^{\mathrlap{\scriptstyle +}}$}{$\scriptstyle\star$}}_{\ \ }\text{\stackon[0.5pt]{$C^{\mathrlap{\scriptstyle -}}$}{$\scriptstyle\star$}}_{\ \ }\text{\stackon[0.5pt]{$C^{\mathrlap{\scriptstyle -}}$}{$\scriptstyle\star$}}_{\ \ }\overset{\star}{\mathcal{H}}_+ (q)\overset{\star}{\mathcal{H}}_+ (\bar q)\overset{\star}{\mathcal{H}}_- (q)\overset{\star}{\mathcal{H}}_- (\bar q) 
\end{aligned}
\end{equation}
The summation $\sum_{\star}$ means to sum over all possible types of the 3-point coefficients and conformal blocks chosen according to \eqref{diagonalblcok} and \eqref{nondiagonalblock}. The summation $\sum_{p_+,p_-}$ should be understood as the integration with the measure \eqref{measurerho} for both $p_+$ and $p_-$.

For all other channels, namely 2134, 2314, 3124, 3214, one can use the following list of superconformal Ward identities to move the supercharges back to the middle two vertices
 \begin{align}
 &\langle \Lambda V\bar{\Lambda}V\rangle=-\langle V\Lambda\bar{\Lambda}V\rangle-\langle VVWV\rangle\nonumber\\
 &\langle \bar{\Lambda}V\Lambda V\rangle=-\langle V\bar{\Lambda}\Lambda V\rangle+\langle VVWV\rangle\nonumber\\
  &\langle \bar{\Lambda}\Lambda VV\rangle =\langle VWVV\rangle +\langle V\Lambda \bar{\Lambda}V\rangle \\
  &\langle \Lambda\bar{\Lambda}VV\rangle =-\langle VWVV\rangle +\langle V\bar{\Lambda}\Lambda V\rangle \nonumber\\
  & \langle{W VVV} \rangle = \langle{V WVV}\rangle + \langle{V VWV}\rangle + \langle{V \Lambda \bar{\Lambda}V}\rangle - \langle{V \bar{\Lambda} {\Lambda}V}\rangle \nonumber
 \end{align}
where the ordering of the fields is fixed, and we only change the location of the supercharges. After applying the superconformal Ward identities, one can check that for $\mA_{0,4,0}^{(b)-}(P_1, P_2, P_3, P_4)$, the loci of supercharges in the integrands are the same as those of the channels 1234 and 1324; the only difference is that the ordering of the external momenta is permuted.
This then means that the supercharges are acted upon in a way that satisfies   the rule \eqref{constraintdelta}. In fact, for a specific channel, say 2134, we can collect all the contributions from $I_{12}, I_{21}, J_{12}$ and $J_{21}$. It then turns out that the terms violating the rule \eqref{constraintdelta} cancel each other:
\begin{equation}
 \begin{aligned}
   &\langle{W VVV} \rangle^+ \langle{VVWV} \rangle^- + \langle \Lambda V\bar{\Lambda}V\rangle^+ \langle \bar \Lambda V {\Lambda}V\rangle^- +(+\leftrightarrow -) \\
   =& \langle{V WVV} \rangle^+ \langle{VVWV} \rangle^- + \langle V \Lambda \bar{\Lambda}V\rangle^+ \langle V\bar \Lambda {\Lambda}V\rangle^- +(+\leftrightarrow -)
 \end{aligned}
\end{equation}
where we have used  superscripts $+$ and $-$ as abbreviations to denote the corresponding correlators in the theories super-Liouville$^+$ and super-Liouville$^-$  respectively. 
 The symbol  ``$(+\leftrightarrow -)$'' means to swap the  $+$ and $-$ superscripts of all explicitly written terms. 
A fully analogous structure also arises in the remaining channels. Thus, we have
\begin{equation}\label{NumericalAm}
\begin{aligned}
 \mA_{0,4,0}^{(b)-}(P_1, P_2, P_3, P_4) &= C_{S^2} \prod_{i=1}^4\mathcal{N}(P_i) \int_{\mathbb{C}}d^2 z ~( I_{12}+I_{21}+J_{12} +J_{21}) \\
 &= C_{S^2}\prod_{i=1}^4\mathcal{N}(P_i) \pi^2 \sum_{\star} \int_{F_0} d^2t \sum_{p_+ ,~p_-} |16q|^{p^2_+ + p_-^2} |\theta_4(q)|^{-4} \\
 & \qquad \times \text{\stackon[0.5pt]{$C^{\mathrlap{\scriptstyle +}}$}{$\scriptstyle\star$}}_{43p_+}\text{\stackon[0.5pt]{$C^{\mathrlap{\scriptstyle +}}$}{$\scriptstyle\star$}}_{p_+21}\text{\stackon[0.5pt]{$C^{\mathrlap{\scriptstyle -}}$}{$\scriptstyle\star$}}_{43p_-}\text{\stackon[0.5pt]{$C^{\mathrlap{\scriptstyle -}}$}{$\scriptstyle\star$}}_{p_-21} \overset{\star}{\mathcal{H}}_+ (q)\overset{\star}{\mathcal{H}}_+  (\bar q)\overset{\star}{\mathcal{H}}_-(q)\overset{\star}{\mathcal{H}}_- (\bar q) \\
 & \qquad\qquad \qquad \qquad \qquad + \Big( \text{5 other perms. of \{123\}} \Big)
\end{aligned}
\end{equation}
Note that before using the superconformal Ward identities, the integrals in the channels 2314 and 3214 are actually divergent near $z\to 0$ or $\operatorname{Im} t \to \infty$, due to the collision of the PCOs. Fixing a channel, the superconformal Ward identities indicate that the singularities in $I_{12},~I_{21},~J_{12}$ and $J_{21}$ actually cancel with each other. By employing this trick, we avoid the need to deal with the singularity. 

In the actual computation, we divide the integral on $F_0$ into two parts: $F_1=F_0 \cap \{\operatorname{Im}t \leq t_{2\text{,max}}=5\}$ and $F_2=F_0 \cap \{\operatorname{Im}t \geq t_{2\text{,max}}=5\}$. For the integral on $F_1$, we compute the elliptic blocks up to order $q^8$, and numerically evaluate the integrations  in \eqref{NumericalAm} directly. In region $F_2$, we note that when $\operatorname{Im}t>5$, $|q|=e^{- \pi t_2}<10^{-6}$ is small, leading to:
\begin{equation}\label{asympH}
 \begin{aligned}
  \mc{H}_h^{1} \z[ \begin{smallmatrix}
  {} _{-} h_3 & {}_{-} h_2 \\
 h_4 & h_1
 \end{smallmatrix} \y](q) &\to 1 \\
  \mc{H}_h^{\frac{1}{2}} \z[ \begin{smallmatrix}
  h_3 & h_2 \\
 h_4 & h_1
 \end{smallmatrix} \y](q) &\to 0, \\
 \mc{H}_h^{\frac{1}{2}} \z[ \begin{smallmatrix}
  *h_3 & h_2 \\
 h_4 & h_1
 \end{smallmatrix} \y](q) &\to 0, \\
 \mc{H}_h^{\frac{1}{2}} \z[ \begin{smallmatrix}
  h_3 & * h_2 \\
 h_4 & h_1
 \end{smallmatrix} \y](q) &\to 0, \\
 \mc{H}_h^{\frac{1}{2}} \z[ \begin{smallmatrix}
  * h_3 & * h_2 \\
 h_4 & h_1
 \end{smallmatrix} \y](q) &\to -\vartheta_2(q^2)h. \\
 \end{aligned}
\end{equation}
For $\mA_{0,4,0}^{(b)-}(\boldsymbol{P})$, we can safely take $\mathcal{H}^1 \simeq 1$ and $\mathcal{H}^{\frac{1}{2}} \simeq 0$, since the block in the last line of \eqref{asympH} does not show up in $\mA_{0,4,0}^{(b)-}(\boldsymbol{P})$. In addition, we truncate the function $|\theta_4(q)|^{-4}\simeq 1$ in \eqref{NumericalAm}. After taking the approximate values, the integrand is a polynomial function of $q$ and $\bar{q}$, so we can compute  the integral over $t$ on $F_2$ analytically, and then perform the remaining integrals over the internal momenta $p_\pm$ numerically. 

\subsection*{Results}
For the numerical calculation of $\mA^{(b)-}_{0,4,0}(\boldsymbol{P})$, we choose the following external momenta and the super-Liouville parameter $b$:
\begin{equation}
 \begin{aligned}
  &\text{(a) $b=\frac{e}{\pi}e^{\frac{i\pi}{4}}$, $\{P_1,P_2,P_3,P_4\}=\{\frac{1}{3},0.05n,\frac{1}{7},\frac{1}{4}\}\times e^{-\frac{i\pi}{4}}$}\\
  &\text{(b) $b=\frac{10}{\pi}e^{\frac{i3\pi}{13}}$, $\{P_1,P_2,P_3,P_4\}=\{\frac{1}{3}e^{-\frac{i\pi}{7}},0.05ne^{-\frac{i\pi}{5}},\frac{1}{7}e^{-\frac{i\pi}{5}},\frac{1}{4}e^{-\frac{i\pi}{4}}\} $}.
 \end{aligned}
\end{equation}
For choice (a), the amplitude is real. For choice (b), since the phase of 
$b$ and the momenta are generic and analytically continued, the amplitude is complex-valued.

\begin{figure}[htbp]
 \centering
 \begin{subfigure}{\textwidth}
  \centering
  \includegraphics[width=1\textwidth, height=10cm, keepaspectratio]{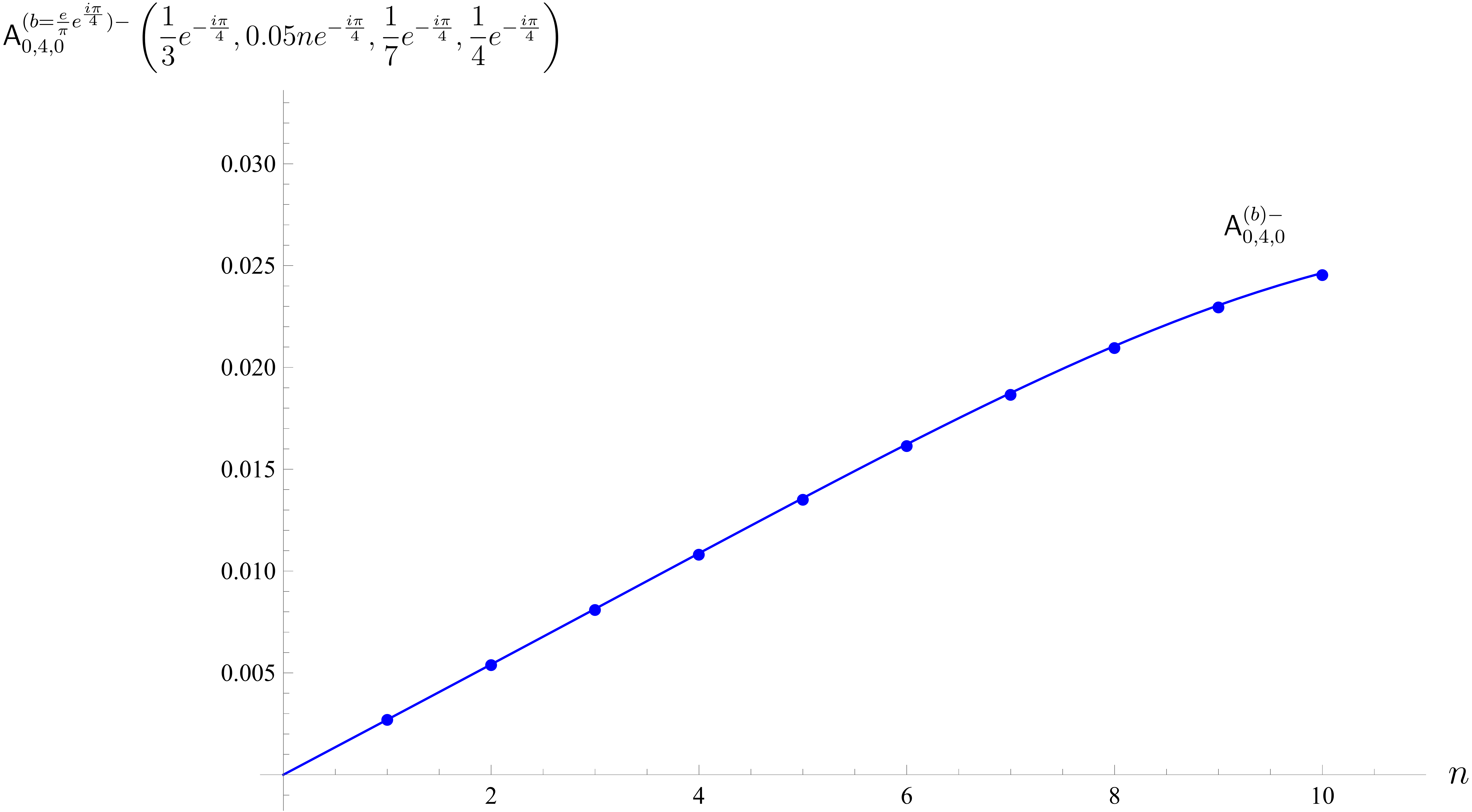}
  \caption{$\mA^{(b=\frac{e}{\pi}e^{\frac{i\pi}{4}})-}_{0,4,0}\z(\frac{1}{3}e^{-\frac{i\pi}{4}},0.05ne^{-\frac{i\pi}{4}},\frac{1}{7}e^{-\frac{i\pi}{4}},\frac{1}{4}e^{-\frac{i\pi}{4}}\y)$}
  \label{fig:sub1}
 \end{subfigure}
 
 \vspace{0.5cm}
 
 \begin{subfigure}{\textwidth}
  \centering
  \includegraphics[width=1\textwidth, height=10cm, keepaspectratio]{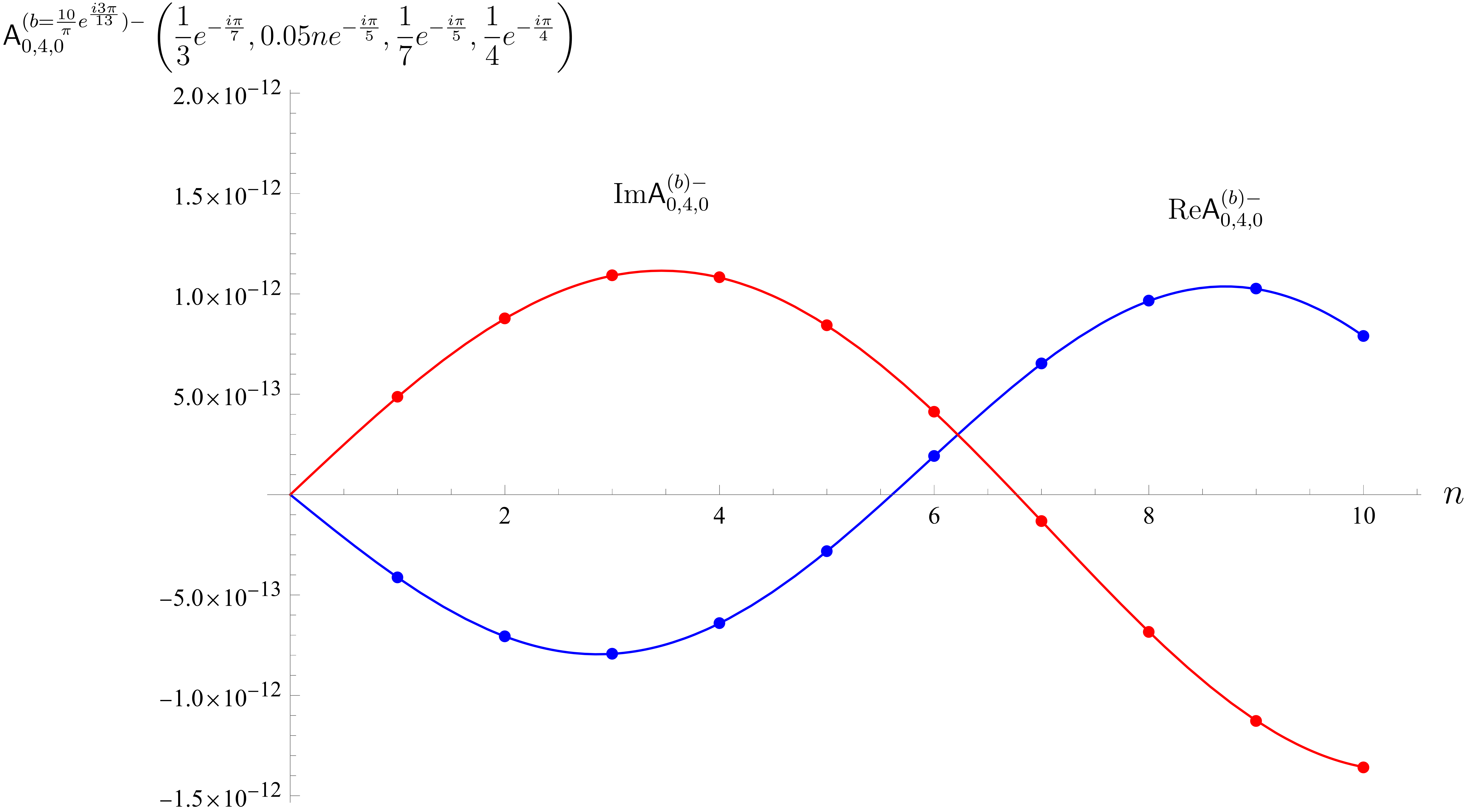}
  \caption{$\mA^{(b=\frac{10}{\pi}e^{\frac{i3\pi}{13}})-}_{0,4,0}\z(\frac{1}{3}e^{-\frac{i\pi}{7}},0.05ne^{-\frac{i\pi}{5}},\frac{1}{7}e^{-\frac{i\pi}{5}},\frac{1}{4}e^{-\frac{i\pi}{4}}\y)$}
  \label{fig:sub2}
 \end{subfigure}
 \caption{The dots represent the numerical results for $\mA^{(b)-}_{0,4,0}(\boldsymbol{P})$ under specific choices of external momenta, while the solid curves correspond to the exact results.}
 \label{fig:main}
\end{figure}

Figure \ref{fig:main} presents the numerical results for  $\mA^{(b)-}_{0,4,0}(\boldsymbol{P})$, which show good agreement with its exact form. The maximum discrepancy is $0.8\%$ for dataset (a) and $0.3\%$ for dataset (b).

\subsection*{Comments on the computation of $\mA_{0,4,0}^{(b)+}(\boldsymbol{P})$}
Unfortunately, we are currently unable to numerically reproduce the amplitude  $\mA_{0,4,0}^{(b)+}(\boldsymbol{P})$. Let us address why computing $\mA_{0,4,0}^{(b)+}(\boldsymbol{P})$ is more challenging. 

The computation for channels 1234 and 1324 is the same as in \eqref{channel1234}; however, that for the remaining channels is more difficult. For these, a naive adaptation of the previous method requires the use of the following superconformal Ward identities:
 \begin{align}
  &\langle \Lambda\Lambda VV\rangle=-\langle V\p VVV\rangle+\langle V\Lambda\Lambda V\rangle\nonumber\\
  &\langle \bar{\Lambda}\bar{\Lambda}VV\rangle=-\langle V\bp VVV\rangle+\langle V\bar \Lambda\bar \Lambda V\rangle\nonumber\\
  &\langle \Lambda V\Lambda V\rangle=-\langle VV\p VV\rangle-\langle V\Lambda\Lambda V\rangle \\
  &\langle \bar{\Lambda} V\bar{\Lambda} V\rangle=-\langle VV\bp VV\rangle-\langle V\bar{\Lambda}\bar{\Lambda}V\rangle \nonumber\\
  & \langle{WVWV} \rangle = \langle VWWV \rangle - \langle V \Lambda \bp\Lambda V \rangle - \langle V \bar{\Lambda} \p\bar{\Lambda} V \rangle - \langle VV\p \bp VV \rangle \nonumber\\
  &\langle WWVV\rangle=-\langle V\p\bp VVV\rangle+\langle V\p\bar{\Lambda}\bar{\Lambda}V\rangle+\langle V\bp\Lambda\Lambda V\rangle+\langle VWWV\rangle\nonumber
 \end{align}
Observe that several of the derivatives act on the third vertex operators; we may then use the formula below to move the derivative to the second vertex operators:
\begin{equation}
\begin{aligned}
  \langle O_1(0)O_2(z)\p O_3(1)O_4(\infty)\rangle &=(h_4-h_1-h_2-h_3)\langle O_1(0)O_2(z) O_3(1)O_4(\infty)\rangle \\ 
  & \qquad -z \p_z \langle O_1(0)O_2(z) O_3(1)O_4(\infty)\rangle
\end{aligned}
\end{equation}
Now, for the channel 3214 (or 2314), we can collect all the contributions of $I_{11}, I_{22}, J_{11}$ and $J_{22}$ as follows
\begin{equation}\label{Ap2314}
 \begin{aligned}
  &\langle WWVV\rangle^+ \langle VVVV\rangle^--\langle \Lambda\Lambda VV\rangle^+ \langle\bar{\Lambda}\bar{\Lambda} VV \rangle^-+(+\leftrightarrow -) \\
  &=\langle VWWV\rangle^+\langle VVVV\rangle^--\langle V\Lambda\Lambda V\rangle^+ \langle V\bar{\Lambda}\bar{\Lambda} V \rangle^-\\
  &+\bp\Big(\langle VVVV\rangle^+\langle V\Lambda\Lambda V\rangle^-\Big)
  +\p\Big(\langle VVVV\rangle^+\langle V\bar{\Lambda}\bar{\Lambda} V\rangle^-\Big)
-\frac{1}{2}\p\bp\Big(\langle VVVV\rangle^+\langle VVVV\rangle^-\Big)\\
&+(+\leftrightarrow -)
 \end{aligned} 
\end{equation}  
The second line (and its counterpart $(+\leftrightarrow -)$) obeys the rule \eqref{constraintdelta}, and it corresponds to permuting \{123\} as in the computation of $\mA_{0,4,0}^{(b)-}(\boldsymbol{P})$ in the previous subsection. Note that the total derivatives in \eqref{Ap2314} cannot be dropped out, as the integration domain $F_0$ has a boundary. Similarly, for the channels 2134 or 3124, we have 
\begin{equation}
 \begin{aligned}
  &\langle WVWV\rangle^+ \langle VVVV\rangle^--\langle \Lambda V\Lambda V\rangle^+ \langle\bar{\Lambda} V\bar{\Lambda} V \rangle^-+(+\leftrightarrow -) \\
  &=\langle VWWV\rangle^+\langle VVVV\rangle^--\langle V\Lambda\Lambda V\rangle^+ \langle V\bar{\Lambda}\bar{\Lambda} V \rangle^-\\
  &\!+\!\bp\Big(\bar{z}\langle VVVV\rangle^+\langle V\Lambda\Lambda V\rangle^-\Big)
  \!+\!\p\Big(z\langle VVVV\rangle^+\langle V\bar{\Lambda}\bar{\Lambda} V\rangle^-\Big)\!
-\!\frac{1}{2}\p\bp\Big(z\bar{z}\langle VVVV\rangle^+\langle VVVV\rangle^-\Big)\\
&+(+\leftrightarrow -)
 \end{aligned} 
\end{equation}
When $z\to 0$ (or Im $t\to \infty$), the most singular term is the last term in the third line of \eqref{Ap2314} (and its counterpart $(+\leftrightarrow -)$), since the leading term in its $z$-series expansion is of order $|z|^{-3}$ (similar divergence appears in  both $c=1$ string and $\hat{c}=1$ superstring \cite{Balthazar:2017mxh,Balthazar:2022atu}). This singularity arises once more from the collision of the PCOs,  yet it cannot be removed using the trick employed for $\mA_{0,4,0}^{(b)-}(\boldsymbol{P})$. Thus, a counter term is necessary, and we leave this for future work. 

\section{Discussion}

We conclude this work by outlining several interesting  future directions.

First and foremost,  S$\mathbb{C}$LS  will very likely offer a stringy realization of ($\mathcal{N}=1$) de Sitter holography, analogous to its bosonic counterpart \cite{Collier:2024kmo,Collier:2025pbm,Collier:2025lux}; thus, completing the work to verify this connection would be highly interesting and meaningful for deepening our understanding of de Sitter holography in (super)string theory. At this stage, it would also be valuable to establish a connection between the S$\mathbb{C}$LS and the double-scaled supersymmetric SYK model (see \cite{Blommaert:2025eps} for recent progress for the bosonic case). 

Several promising directions remain for advancing the computation of super complex Liouville string amplitudes:  our analysis has so far focused on sphere  3-point and 4-point amplitudes (only the NS-NS-NS-NS type for the latter), so extending to higher-genus cases (e.g., torus  1-point amplitudes) and completing the remaining two sphere 4-point amplitude types (R-R-R-R and NS-NS-R-R) is desirable.  Additionally, in our numerical evaluation of the moduli integral  of the sphere 4-point amplitude  $\mA^{(b)}_{0,4,0}=\mA_{0,4,0}^{(b)+}+\mA_{0,4,0}^{(b)-}$ (and  $\hmA^{(b)}_{0,4,0}=\mA_{0,4,0}^{(b)+}-\mA_{0,4,0}^{(b)-}$), we only explicitly computed $\mA_{0,4,0}^{(b)-}$. Since the moduli integral of $\mA_{0,4,0}^{(b)+}$ is divergent and more complicated, developing a proper regularization for $\mA_{0,4,0}^{(b)+}$ and numerically verifying its proposed analytic form also represents a key task.

In this work, we find evidence  that the perturbative string  amplitudes of type 0B S$\mathbb{C}$LS are identical to (two copies of) those of the bosonic  $\mathbb{C}$LS; consequently, they can be described by the same matrix integral. 
One could further test the proposed   S$\mathbb{C}$LS/matrix model correspondence at the   non-perturbative level, as in the bosonic case \cite{Collier:2024mlg}. It is plausible that S$\mathbb{C}$LS  and $\mathbb{C}$LS differ  non-perturbatively.

Finally, all the above discussions on  S$\mathbb{C}$LS can be extended to $\widehat{\text{S}\mathbb{C}\text{LS}}$ in parallel.  In particular,  it would be interesting to investigate whether  
$\widehat{\text{S}\mathbb{C}\text{LS}}$ is related to de Sitter holography.

\section*{Ackowledgements }
We thank Scott Collier, Lorenz Eberhardt, Beatrix M\"{u}hlmann, Cheng Peng, Victor A. Rodriguez, Wei Song, Diandian Wang and Jianming Zheng for useful discussions. 
Z.D. and K.L. are supported by the NSFC special fund for theoretical physics No. 12447108 and the national key research and development program of China No. 2020YFA0713000. Z.Y. is supported by the Postdoctoral Fellowship Program of CPSF under Grant Number GZC20241685 and funds from the  UCAS program 
of Special Research Assistant, and partly supported by NSFC Grant No. 12175237. Z.D. would like to thank the support of the PhD student fund for short-term overseas visits and the Simons Foundation Collaboration on Celestial Holography.

\appendix

\section{Properties of the special functions}
This section collects several  properties of the theta functions, the double Gamma function and  the dilogarithm.

\subsection*{Theta Function}\label{thetafunctions}
The theta functions are defined as
 \begin{align}
 &\vartheta_1\z(\nu|\tau\y)=i\sum_{n=-\infty}^\infty(-1)^n e^{\frac{(n-\frac{1}{2})^2}{2}2\pi i\tau}e^{(n-\frac{1}{2})2\pi i \nu}\nonumber\\
 &\vartheta_2\z(\nu|\tau\y)=\sum_{n=-\infty}^\infty e^{\frac{\z(n-\frac{1}{2}\y)^2}{2}2\pi i\tau}e^{\z(n-\frac{1}{2}\y)2\pi i\nu}\\
 &\vartheta_3\z(\nu|\tau\y)=\sum_{n=-\infty}^\infty e^{\frac{n^2}{2}2\pi i\tau}e^{n2\pi i \nu}\nonumber\\
 &\vartheta_4\z(\nu|\tau\y)=\sum_{n=-\infty}^\infty(-1)^n e^{\frac{n^2}{2}2\pi i\tau}e^{n2\pi i\nu}\nonumber
 \end{align}
They also admit the  infinite product representations
 \begin{align}
 &\vartheta_1(\nu|\tau)=2e^{\frac{\pi i\tau}{4}}\sin(\pi \nu)\prod_{m=1}^{\infty} (1-q^m)(1-zq^m)(1-z^{-1}q^m)\nonumber\\
 &\vartheta_2(\nu|\tau)=2e^{\frac{\pi i\tau}{4}}\cos(\pi \nu)\prod_{m=1}^{\infty} (1-q^m)(1+zq^m)(1+z^{-1}q^m)\\
 &\vartheta_3(\nu|\tau)=\prod_{m=1}^{\infty} (1-q^m)(1+zq^{m-\frac{1}{2}})(1+z^{-1}q^{m-\frac{1}{2}})\nonumber\\
 &\vartheta_4(\nu|\tau)=\prod_{m=1}^{\infty} (1-q^m)(1-zq^{m-\frac{1}{2}})(1-z^{-1}q^{m-\frac{1}{2}})\nonumber
 \end{align}
where $q=e^{2\pi i \tau},z=e^{2\pi i\nu}$. They satisfy the periodic properties
\begin{equation}
 \begin{aligned}
 \vartheta_1(\nu+1|\tau)=-\vartheta_1(\nu|\tau)&,\quad \vartheta_1(\nu+\tau|\tau)=-e^{-\pi i\tau-2\pi i\nu}\vartheta_1(\nu|\tau)\\
 \vartheta_2(\nu+1|\tau)=-\vartheta_2(\nu|\tau)&,\quad \vartheta_2(\nu+\tau|\tau)=e^{-\pi i\tau-2\pi i\nu}\vartheta_2(\nu|\tau)\\
 \vartheta_3(\nu+1|\tau)=\vartheta_3(\nu|\tau)&,\quad \vartheta_3(\nu+\tau|\tau)=e^{-\pi i\tau-2\pi i\nu}\vartheta_3(\nu|\tau)\\
 \vartheta_4(\nu+1|\tau)=\vartheta_4(\nu|\tau)&,\quad \vartheta_4(\nu+\tau|\tau)=-e^{-\pi i\tau-2\pi i\nu}\vartheta_4(\nu|\tau),
 \end{aligned}
\end{equation}
and the half-periodic properties
\begin{equation}
 \begin{aligned}
 \vartheta_1\left(\nu + \tfrac{1}{2}\middle|\tau\right) = \vartheta_2(\nu|\tau)&,\quad \vartheta_1\left(\nu + \tfrac{\tau}{2}\middle|\tau\right) = i e^{-\pi i \tau/4} e^{-\pi i \nu} \vartheta_4(\nu|\tau)\\
 \vartheta_2\left(\nu + \tfrac{1}{2}\middle|\tau\right) = -\vartheta_1(\nu|\tau)&,\quad \vartheta_2\left(\nu + \tfrac{\tau}{2}\middle|\tau\right) = e^{-\pi i \tau/4} e^{-\pi i \nu} \vartheta_3(\nu|\tau)\\
 \vartheta_3\left(\nu + \tfrac{1}{2}\middle|\tau\right) = \vartheta_4(\nu|\tau)&,\quad \vartheta_3\left(\nu + \tfrac{\tau}{2}\middle|\tau\right) = e^{-\pi i \tau/4} e^{-\pi i \nu} \vartheta_2(\nu|\tau)\\
 \vartheta_4\left(\nu + \tfrac{1}{2}\middle|\tau\right) = \vartheta_3(\nu|\tau)&,\quad \vartheta_4\left(\nu + \tfrac{\tau}{2}\middle|\tau\right) = i e^{-\pi i \tau/4} e^{-\pi i \nu} \vartheta_1(\nu|\tau).
 \end{aligned}
\end{equation}
The modular properties of the theta functions are:
\begin{equation}
 \begin{aligned}
 &\vartheta_1(\nu/\tau|-1/\tau)=-i\sqrt{-i\tau}e^{\pi i \nu^2/\tau}\vartheta_1(\nu|\tau)\\
 &\vartheta_2(\nu/\tau|-1/\tau)=\sqrt{-i\tau}e^{\pi i \nu^2/\tau}\vartheta_4(\nu|\tau)\\
 &\vartheta_3(\nu/\tau|-1/\tau)=\sqrt{-i\tau}e^{\pi i \nu^2/\tau}\vartheta_3(\nu|\tau)\\
 &\vartheta_4(\nu/\tau|-1/\tau)=\sqrt{-i\tau}e^{\pi i \nu^2/\tau}\vartheta_2(\nu|\tau)
 \end{aligned}
\end{equation}
Note that $\vartheta_1(\nu|\tau)$ is an odd function with respect to $\nu$, which implies $\vartheta_1(0|\tau)=0$. There is a relation between the derivative of $\vartheta_1$ and the $\eta$ function
\begin{equation}
 \p_\nu\vartheta_1(0|\tau)=2\pi \eta(\tau)^3,\quad \theta_1(0|\tau)=0
\end{equation}
In fact, the zeros of these theta functions are
\begin{equation}
 \begin{aligned}
  &\vartheta_1(\nu|\tau)=0,\quad \nu=n+m\tau\\
  &\vartheta_2(\nu|\tau)=0,\quad \nu=\z(n+\frac{1}{2}\y)+m\tau\\
  &\vartheta_3(\nu|\tau)=0,\quad \nu=\z(n+\frac{1}{2}\y)+\z(m+\frac{1}{2}\y)\tau\\
  &\vartheta_4(\nu|\tau)=0,\quad \nu=n+\z(m+\frac{1}{2}\y)\tau
 \end{aligned}
\end{equation}

\subsection*{Double Gamma Function}\label{doublegamma}
The double Gamma function is a meromorphic function that can be defined as the unique function satisfying the functional equations
\begin{equation}\label{shiftGamma}
 \begin{aligned}
 \Gamma_b(z+b)&=\frac{\sqrt{2\pi }b^{bz-\frac{1}{2}}}{\Gamma(bz)} \Gamma_b(z)\\
\Gamma_b(z+b^{-1})&=\frac{\sqrt{2\pi }b^{-b^{-1}z+\frac{1}{2}}}{\Gamma(b^{-1}z)} \Gamma_b(z),
 \end{aligned}
\end{equation}
with the normalization given by $\Gamma_b\z(\frac{b+b^{-1}}{2}\y)=1$. It admits an explicit integral representation in the half-plane $\Re(z)>0$
\begin{equation}
 \log\Gamma_b(z)=\int^\infty_0\frac{dt}{t}\z(\frac{e^{\frac{t}{2}(Q-2z)}-1}{4\sinh\z(\frac{bt}{2}\y)\sinh\z(\frac{t}{2b}\y)}-\frac{1}{8}(Q-2z)^2e^{-t}-\frac{Q-2z}{2t}\y).
\end{equation}
 $\Gamma_b(z)$ has simple poles at
\begin{equation}
 z=-rb-sb^{-1},\quad r,s\in \mathbb{Z}_{\geq 0}
\end{equation}
In order to describe the structure constants of SLFT, one can introduce the NS- and R- double Gamma functions
\begin{equation}
 	\begin{aligned}
		&\Gamma_b^{NS}(z)\equiv\Gamma_b\z(\frac{z}{2}\y)\Gamma_b\z(\frac{z+b+b^{-1}}{2}\y)\\
		&\Gamma^{R}_b(z)\equiv\Gamma_b\z(\frac{z+b}{2}\y)\Gamma_b\z(\frac{z+b^{-1}}{2}\y).
	\end{aligned}
\end{equation}
In the literature, the function $\Upsilon_b$ is  also frequently used:
\begin{equation}
 \Upsilon_b(z)\equiv \frac{1}{\Gamma_b(z)\Gamma_b(Q-z)}
\end{equation}
One can also define the NS- and R-Upsilon function
\begin{equation}
 \begin{aligned}
 \Upsilon_b^{NS}(z)&\equiv\Upsilon_b\left(\frac{z}{2}\right)\Upsilon_b\left(\frac{z+Q}{2}\right)=\frac{1}{\Gamma_b^{NS}(z)\Gamma_b^{NS}(Q-z)}\\
 \Upsilon_b^{R}(z)&\equiv\Upsilon_b\left(\frac{z+b}{2}\right)\Upsilon_b\left(\frac{z+b^{-1}}{2}\right)=\frac{1}{\Gamma_b^{R}(z)\Gamma_b^{R}(Q-z)}\\
 \end{aligned}
\end{equation}
 We also have the following useful formula (see \cite{Zamolodchikov:2005fy})
\begin{equation}\label{TwoGamma}
 \Gamma_b\z(\frac{b+b^{-1}}{2}\pm z\y)\Gamma_{-ib}\z(\frac{-ib+(-ib)^{-1}}{2}\pm iz\y)=\frac{e^{-\frac{\pi iz^2}{2}}\vartheta_3(0|b^2)}{\vartheta_3(bz|b^2)}
\end{equation}
Using this formula, we obtain:
\begin{equation}
 \begin{aligned}
 &\Gamma_b(b+b^{-1}\pm 2P)\Gamma_{-ib}(-ib+ib^{-1}\pm 2iP)\\
 &=e^{-2\pi iP^2+\frac{3i\pi}{4}+\frac{i\pi}{8}(b^2-b^{-2})}\frac{2\sin(2\pi Pb)\sin(2\pi Pb^{-1})}{P}\frac{\vartheta_3(0|b^2)}{\vartheta_1(2Pb|b^2)}.
 \end{aligned}
\end{equation}
Letting $P\to0$ leads to:
\begin{equation}
 \begin{aligned}
 &\Gamma_b(b+b^{-1})^2\Gamma_{-ib}(-ib+(-ib)^{-1})^2\\
 &=e^{\frac{3i\pi}{4}+\frac{i\pi(b^2-b^{-2})}{8}}\frac{2\pi}{b \eta(b^2)^3}\vartheta_3(0|b^2)
 \end{aligned}
\end{equation}
Using \eqref{TwoGamma} once more, we find:
\begin{equation}
 \begin{aligned}
  &\Gamma_b^{NS}\z(\frac{b+b^{-1}}{2}\pm z\y)\Gamma_{-ib}^{R}\z(\frac{-ib+ib^{-1}}{2}\pm iz\y)=\frac{e^{-i\pi\z(\frac{z^2}{4}+\frac{b^2+b^{-2}+2}{16}\y)}\vartheta_3(0|b^2)^2}{\vartheta_3\z(\frac{bz}{2}-\frac{b^2+1}{4}|b^2\y)\vartheta_3\z(\frac{bz}{2}+\frac{b^2+1}{4}|b^2\y)}\\
  &\Gamma_b^{R}\z(\frac{b+b^{-1}}{2}\pm z\y)\Gamma_{-ib}^{NS}\z(\frac{-ib+ib^{-1}}{2}\pm iz\y)=\frac{e^{-i\pi\z(\frac{z^2}{4}+\frac{b^2+b^{-2}-2}{16}\y)}\vartheta_3(0|b^2)^2}{\vartheta_3\z(\frac{bz}{2}-\frac{b^2-1}{4}|b^2\y)\vartheta_3\z(\frac{bz}{2}+\frac{b^2-1}{4}|b^2\y)}.
 \end{aligned}
\end{equation}
and 
\begin{equation}\label{GammaQ}
 \begin{aligned}
  \Gamma^{NS}_b\z(b+b^{-1}\pm z\y)&\Gamma^{NS}_{-ib}\z(-ib+ib^{-1}\pm iz\y)\\
  &=\frac{8e^{-\frac{i\pi z^2}{4}+\frac{3i\pi}{4}+\frac{i\pi}{8}\z(b^2-b^{-2}\y)}\sin\z(\frac{\pi b z}{2}\y)\sin\z(\frac{\pi b^{-1} z}{2}\y)\vartheta_3(0|b^2)^2}{z\vartheta_3\z(\frac{bz}{2}|b^2\y)\vartheta_1\z(\frac{bz}{2}|b^2\y)}\\
  \Gamma^{R}_b\z(b+b^{-1}\pm z\y)&\Gamma^{R}_{-ib}\z(-ib+ib^{-1}\pm iz\y)\\
  &=\frac{4e^{-\frac{i\pi z^2}{4}+\frac{i\pi}{8}(b^2-b^{-2})}\cos\z(\frac{\pi b z}{2}\y)\cos\z(\frac{\pi b^{-1} z}{2}\y)\vartheta_3\z(0|b^2\y)^2}{\vartheta_2\z(\frac{bz}{2}|b^2\y)\vartheta_4\z(\frac{bz}{2}|b^2\y)}
 \end{aligned}
\end{equation}
Taking $z\to 0 $ in the first equation of \eqref{GammaQ},  we  get
\begin{equation}
 \Gamma^{NS}_b(Q)^2\Gamma^{NS}_{-ib}(-ib+ib^{-1})^2=\frac{2\pi e^{\frac{3\pi i}{4}+\frac{\pi i}{8}\z(b^2-b^{-2}\y)}\vartheta_3(0|b^2)}{b\eta(b^2)^3}
\end{equation}

\subsection*{Dilogarithm}
The dilogarithm is defined as:
\begin{equation}
 \Li_2(z)=-\int^z_{0}\frac{\log(1-u)}{u}du,\quad z\in \mathbb{C}
\end{equation}
For $|z|\leq 1$, there is an infinite series representation:
\begin{equation}
 \Li_2(z)=\sum_{k=1}^\infty \frac{z^k}{k^2}
\end{equation}
The dilogarithm  is analytic everywhere on the complex plane except at $z=1$. We can choose a branch cut along the positive real axis $(1,\infty)$. The dilogarithm is continuous at the branch point,
\begin{equation}
 \Li_2(1)=\frac{\pi^2}{6}
\end{equation}
 For $x\in (1,\infty)$, it has the following discontinuity
\begin{equation}
 \Li_2(x+i\epsilon)-\Li_2(x-i\epsilon)=2\pi i\log x.
\end{equation}
There is a useful relation 
\begin{equation}\label{Dlogsum}
 \Li_2(z)+\Li_2\z(\frac{1}{z}\y)=-\frac{\pi^2}{6}-\frac{(\log(-z))^2}{2}.
\end{equation}
In the derivation of \eqref{2sum}, we need the following formula:
\begin{equation}\label{Liplus}
 \begin{aligned}
  \sum_{m_1,m_2=1}^\infty \frac{e^{m_1x}e^{m_2y}}{(m_1+m_2)^2}&=\sum_{s=2}^\infty\frac{1}{s^2}\sum_{m_1=1}^{s-1} e^{m_1x}e^{(s-m_1)y}\\
  &=\frac{e^{x-y}}{1-e^{x-y}}\sum_{s=1}^\infty\z(\frac{e^{sy}}{s^2}-e^{-(x-y)}\frac{e^{sx}}{s^2}\y)\\
  &=\frac{e^{x-y}\Li_2(e^y)-\Li_2(e^x)}{1-e^{x-y}}
 \end{aligned}
\end{equation}
and
\begin{equation}\label{Liminus}
 \begin{aligned}
  \sum_{m_1\neq m_2=1}^\infty \frac{e^{m_1x}e^{m_2y}}{(m_1-m_2)^2}&=\sum_{m_1>m_2}\frac{e^{m_1x}e^{m_2y}}{(m_1-m_2)^2}+\sum_{m_1<m_2}\frac{e^{m_1x}e^{m_2y}}{(m_1-m_2)^2}\\
  &=\sum_{n=2}^\infty\sum_{k=1}^{n-1}\frac{e^{nx}e^{ky}+e^{kx}e^{ny}}{(n-k)^2}\\
  &=\sum_{d=1}^\infty\frac{(e^{dx}+e^{dy})}{d^2}\sum^\infty_{k=1}e^{k(x+y)}\\
  &=\frac{e^{x+y}(\Li_2(e^x)+\Li_2(e^y))}{1-e^{x+y}}
 \end{aligned}
\end{equation}
Summing them gives us:
\begin{equation}\label{Liinteger}
 \begin{aligned}
  &\sum_{m_1,m_2=1}^\infty e^{m_1 x}e^{m_2 y}\z(\frac{1}{(m_1+m_2)^2}-\frac{1-\delta_{m_1,m_2}}{(m_1-m_2)^2}\y)\\
  &=\z(\frac{\Li_2(e^{x})}{e^{x-y}-1}-\frac{\Li_2(e^{x})}{e^{-x-y}-1}\y)+\z(\frac{\Li_2(e^{y})}{e^{y-x}-1}-\frac{\Li_2(e^{y})}{e^{-x-y}-1}\y)
 \end{aligned}
\end{equation}
Similarly we have
\begin{equation}\label{Lihalfinteger}
 \begin{aligned}
  &\sum^\infty_{m_1,m_2=0}e^{(m_1+\frac{1}{2})x}e^{(m_2+\frac{1}{2})y}\z(\frac{1}{(m_1+m_2+1)^2}-\frac{1-\delta_{m_1,m_2}}{(m_1-m_2)^2}\y)\\
  &=\z(\frac{e^{\frac{x-y}{2}}\Li_2(e^x)}{e^{x-y}-1}-\frac{e^{-\frac{x+y}{2}}\Li_2(e^x)}{e^{-x-y}-1}\y)+\z(\frac{e^{\frac{y-x}{2}}\Li_2(e^y)}{e^{y-x}-1}-\frac{e^{-\frac{x+y}{2}}\Li_2(e^y)}{e^{-x-y}-1}\y)
 \end{aligned}
\end{equation}

\section{The factor $B_{m,n}$ in the higher equations of motion}\label{Bmn}
In this section, we give the explicit form of the factor $B_{m,n}$ in \eqref{HEOM}. Though this factor is obtained in \cite{Belavin:2006pv}, it has a different expression in our convention (its form depends on the convention of the structure constants). Thus, to obtain   $B_{m,n}$,  here we  give the relation between the standard expressions of the structure constants and the ones (eq. \eqref{SC}) we used in this paper.

The standard form of the NS structure constants are \cite{Rashkov:1996np,Poghossian:1996agj,Fukuda:2002bv} :
\small{
\begin{equation}
\begin{aligned}
  &C^{std}_b(P_1,P_2,P_3)=\left(\pi \mu\gamma\left(\frac{bQ}{2}\right)b^{1-b^2}\right)^{\left(-\frac{Q}{2}-P_1-P_2-P_3\right)/b}\times\\
  &\frac{\Upsilon_b'^{NS}(0)\Upsilon_b^{NS}(Q+2P_1)\Upsilon_b^{NS}(Q+2P_2)\Upsilon_b^{NS}(Q+2P_3)}{\Upsilon_b^{NS}(\frac{Q}{2}+P_1+P_2+P_3)\Upsilon_b^{NS}(\frac{Q}{2}-P_1+P_2+P_3)\Upsilon_b^{NS}(\frac{Q}{2}+P_1-P_2+P_3)\Upsilon_b^{NS}(\frac{Q}{2}+P_1+P_2-P_3)}\\
  &\widetilde{C}^{std}(P_1,P_2,P_3)_b=\left(\pi \mu\gamma\left(\frac{bQ}{2}\right)b^{1-b^2}\right)^{\left(-\frac{Q}{2}-P_1-P_2-P_3\right)/b}\times\\
  &\frac{2i\Upsilon_b'^{NS}(0)\Upsilon_b^{NS}(Q+2P_1)\Upsilon_b^{NS}(Q+2P_2)\Upsilon_b^{NS}(Q+2P_3)}{\Upsilon_b^{R}(\frac{Q}{2}+P_1+P_2+P_3)\Upsilon_b^{R}(\frac{Q}{2}-P_1+P_2+P_3)\Upsilon_b^{R}(\frac{Q}{2}+P_1-P_2+P_3)\Upsilon_b^{R}(\frac{Q}{2}+P_1+P_2-P_3)}
\end{aligned}
\end{equation}}
Compared with \eqref{SC},  we have:
\begin{equation}
\begin{aligned}
  C_b(P_1,P_2,P_3)=K_b\frac{C^{std}_b(P_1,P_2,P_3)}{\prod_{i=1}^3\sqrt{\rho_b^{NS}(P_i)S(P_i)}}\\
 \widetilde{C}_b(P_1,P_2,P_3)=K_b\frac{\widetilde{C}^{std}_b(P_1,P_2,P_3)}{\prod_{i=1}^3\sqrt{\rho_b^{NS}(P_i)S(P_i)}}
\end{aligned}
\end{equation}
where
\begin{equation}
 \rho_b^{NS}(P)=-4\text{sin}(\pi b P)\text{sin}(\pi b^{-1}P),\quad S(P)=\left(\pi \mu\gamma\left(\frac{bQ}{2}\right)\right)^{-2P/b}\frac{b^2\gamma\left(bP\right)}{\gamma\left(-\frac{P}{b}\right)}
\end{equation}
and
\begin{equation}
 K_b=\left(\pi \mu\gamma\left(\frac{bQ}{2}\right)b^{1-b^2}\right)^{\frac{Q}{2b}}\frac{\Gamma_b^{NS}(2Q)}{2\Gamma_b^{NS}(Q)^3}\frac{1}{\Upsilon_b'^{NS}(0)}=\frac{\left(\pi \mu\gamma\left(\frac{bQ}{2}\right)b^{1-b^2}\right)^{\frac{Q}{2b}}}{2\pi}\frac{\Gamma_b^{NS}(2Q)}{\Gamma_b^{NS}(Q)}
\end{equation}
From the above relation, the definition \eqref{HEOM} of $B_{m,n}$ and the factor $B^{std}_{m,n}$ in the standard convention in \cite{Belavin:2006pv}, we  obtain the factor $B_{m,n}$:
\begin{equation}
 B_{m,n}=\left(\frac{S\left(P=\frac{mb}{2}+\frac{n}{2b}\right)\rho_b^{NS}(P=\frac{mb}{2}+\frac{n}{2b})}{S\left(P=-\frac{mb}{2}+\frac{n}{2b}\right)\rho_b^{NS}(P=-\frac{mb}{2}+\frac{n}{2b})}\right)^{\frac{1}{2}}B^{std}_{m,n}
\end{equation}
Note that though there is a derivative $\partial_P$ in the definition \eqref{HEOM}, an additional term that is proportional to $\partial_P (S\rho_b^{NS})$ will not appear, since the degenerate value $P=\frac{mb}{2}+\frac{n}{2b}$ are zeros of the structure constant. The factor $B^{std}_{m,n}$ is given in \cite{Belavin:2006pv}:
\begin{equation}
 B^{std}_{m,n}=(-1)^{m}2^{mn}i^{mn-2[mn/2]}b^{m-n+1}\left(\pi\mu\gamma\left(\frac{bQ}{2}\right)\right)^m \gamma\left(\frac{n-mb^2}{2}\right)\prod_{(k,l)\in\langle m,n\rangle_{\text{NS}}}\frac{kb+lb^{-1}}{2}
\end{equation}
where
\begin{equation}
 \langle m,n\rangle_{\text{NS}}=\{1-m:2:m-1,1-n:2:n-1\}\cup\{2-m:2:m-2,2-n:2:n-2\}/\{0,0\}
\end{equation}
Notice that the $ B^{std}_{m,n}$ above has an extra factor $(-1)^{m}=(-1)^{n}$ regarding  its from given in \cite{Belavin:2006pv}. This is because compared to the definition in \cite{Belavin:2006pv}, ours for  $ B^{std}_{m,n}$ \eqref{HEOM} swaps the order of $D_{m,n}$ and $\bar{D}_{m,n}$'s actions..

\section{OPE of the ground ring and physical operators}\label{Thelegfactor}
The leg-factor $\mathcal{N}_b(P)$ could be an arbitrary function of $P$ and $b$. 
 In this section, we show that when it takes the form in \eqref{normalizations}, the coefficient arising in the operator product expansion of the ground ring operator and the physical vertex operator is exactly $1$.  To achieve this unit OPE coefficient, we should at the same time normalize the ground ring operator as in \eqref{ground ring normalization}, which also depends on the form of the leg-factor (though the factor $\mathcal{N}_b(\frac{mb}{2}-\frac{n}{2b})$). 
 
 As an illustration, we show the computation of the simplest case $(m,n)=(3,1)$. We normalize the  ground ring and physical operators as 
 \begin{equation}\label{Preoperators}
\begin{aligned}
 \mathcal{O}_{3,1}&\equiv \mathcal{N}_{3,1}(b)H_{3,1}\bar{H}_{3,1}V^+_{P=\frac{3b}{2}+\frac{1}{2b}}V^-_{P=i\left(\frac{3b}{2}-\frac{1}{2b}\right)}\\
 \mathscr{V}_P&\equiv \mathcal{N}_b(P)c\bar{c}e^{-\Phi-\bar{\Phi}}V^+_PV^-_{iP}
\end{aligned}
\end{equation}
 Then from the fusion of the degenerate value $(m,n)=(3,1)$, the OPE turns out to be:
\begin{equation}\label{pregroundringope}
 \mathcal{O}_{3,1}(z)\mathscr{V}_P(0)=\sum_{r=-2,0,2}\mathcal{C}_b^{(r)}(P)\mathscr{V}_{P+\frac{rb}{2}}(0)+\text{BRST exact},
\end{equation}
where $\mathcal{C}_b^{(r)}(P)$ $(r=-2,0,2)$ are the to be computed OPE coefficients.

 For the computation, we need the OPE of the degenerate operator with $(m,n)=(3,1)$\footnote{Both the 2 degenerate operators in $\mathcal{O}_{3,1}$ (in \eqref{Preoperators}) are the ones with $(m,n)=(3,1)$ in the ``$+$'' and ``$-$'' super-Liouville theory respectively (for the ``$-$'' theory, recall that we use the convention where $V_P=V_{-P}$).} and a super-primary in $\mathcal{N}=1$ super-Liouville theory:
\begin{equation}
\begin{aligned}
 V_{\frac{3b}{2}+\frac{1}{2b}}(z)V_P(0)=&(z\bar{z})^{\frac{b^2+1}{2}-Pb}C^{(+)}_b(P)[V_{P+b}]_{ee}(0)+(z\bar{z})^{b^2+\frac{1}{2}}C^{(0)}_b(P)[V_{P}]_{oo}(0)\\
 +&(z\bar{z})^{\frac{b^2+1}{2}+Pb}C^{(-)}_b(P)[V_{P-b}]_{ee}(0)
\end{aligned}
\end{equation}
where the subscript ``$ee$'' and ``$oo$'' denote the collections of ``even-even'' and ``odd-odd'' descendents respectively.
From the explicit form of $H_{3,1}$ $(\bar{H}_{3,1})$ in \eqref{H31}, one finds that only the 3 terms in \eqref{pregroundringope}  appear in the OPE, and the 3 OPE coefficients are 
\begin{equation}\label{OPEcoe}
\begin{aligned}
 \mathcal{C}_b^{(-2)}(P)=& (2Pb)^2\mathcal{N}_{3,1}(b)\frac{\mathcal{N}_b(P)}{\mathcal{N}_b(P-b)}C_b^{(-)}(P)C_{-ib}^{(+)}(iP)\\
 \mathcal{C}_b^{(0)}(P)=& -\mathcal{N}_{3,1}(b)\frac{\mathcal{N}_b(P)}{\mathcal{N}_b(P)}C_b^{(0)}(P)C_{-ib}^{(0)}(iP)\\
 \mathcal{C}_b^{(2)}(P)=& (2Pb)^2\mathcal{N}_{3,1}(b)\frac{\mathcal{N}_b(P)}{\mathcal{N}_b(P+b)}C_b^{(+)}(P)C_{-ib}^{(-)}(iP)
\end{aligned}
\end{equation}
where 
\begin{equation}
\begin{aligned}
 C_b^{(-)}(P)=2\pi C_b^*\left(\frac{3b}{2}+\frac{1}{2b},P,P-b\right)\rho^{NS}_b(P-b)\\
 C_b^{(0)}(P)=2\pi \widetilde{C}_b^*\left(\frac{3b}{2}+\frac{1}{2b},P,P\right)\rho^{NS}_b(P)\\
 C_b^{(+)}(P)=2\pi C_b^*\left(\frac{3b}{2}+\frac{1}{2b},P,P+b\right)\rho^{NS}_b(P+b)
\end{aligned}
\end{equation}
In the above, the functions $C_b^*$ and $\widetilde{C}_b^*$ are the residue of the structure constant $C_b$ and $\widetilde{C}_b$ in \eqref{SC} respectively (since one operator is degenerate). 
Finally, substitute these values and the  form of the leg factor $\mathcal{N}_b(P)$ and $\mathcal{N}_{3,1}(P)$:
\begin{equation}
\begin{aligned}
 \mathcal{N}_b(P)&=-\frac{i(b^2-b^{-2})\rho^{(b)}_{NS}(P)}{8\pi P\cos\z(\frac{\pi b^2}{2}\y)\cos\z(\frac{\pi b^{-2}}{2}\y)}\\
 \mathcal{N}_{3,1}(P)&=\frac{\pi\mathcal{N}_b\left(\frac{3b}{2}-\frac{1}{2b}\right)}{B_{3,1}}
\end{aligned}
\end{equation}
into the OPE coefficients in \eqref{OPEcoe}, one finds:
\begin{equation}
 \mathcal{C}_b^{(-2)}(P)= \mathcal{C}_b^{(0)}(P)= \mathcal{C}_b^{(2)}(P)=1
\end{equation}
Generally, using the normalizations \eqref{ground ring normalization}, \eqref{normalizations}  of the ground ring and physical operators, 
 one can verify the OPE \eqref{groundringOPE} for higher values of $(m,n)$ ($m+n\in2\mathbb{Z}$) with the help of $\mathtt{Mathematica}$.

\bibliographystyle{JHEP} 
\bibliography{ref}

\end{document}